\documentclass[namedreferences]{solarphysics}

\usepackage[hyperref,optionalrh,showbiblabels]{spr-sola-addons} 
\usepackage{graphicx}        
\usepackage{color}           
\usepackage{breakurl}        
\usepackage{amsmath}

\chardef\us=`\_

\begin{document}

\begin{article}
\begin{opening}

\title{The Instruments and Capabilities of the {\it Miniature X-ray Solar Spectrometer} (MinXSS) CubeSats}

\author[addressref={aff1,aff2,aff3},corref,email={christopher.moore-1@colorado.edu}]{\inits{C. S.}\fnm{Christopher S.}~\lnm{Moore}}
\author[addressref=aff4,email={amir@boulder.swri.edu}]{\inits{A.}\fnm{Amir}~\lnm{Caspi}}
\author[addressref=aff2,email={tom.woods@lasp.colorado.edu}]{\inits{T. N.}\fnm{Thomas N.}~\lnm{Woods}}
\author[addressref={aff2,aff5},email={phil.chamberlin@lasp.colorado.edu}]{\inits{P. C.}\fnm{Phillip C.}~\lnm{Chamberlin}}
\author[addressref=aff5,email={brian.r.dennis@nasa.gov}]{\inits{B. R.}\fnm{Brian R.}~\lnm{Dennis}}
\author[addressref=aff2,email={Andrew.Jones@lasp.colorado.edu}]{\inits{A. R.}\fnm{Andrew R.}~\lnm{Jones}}
\author[addressref={aff2,aff5},email={james.p.mason@nasa.gov}]{\inits{J. P.}\fnm{James P.}~\lnm{Mason}}
\author[addressref={aff5,aff6},email={richard.a.schwartz@nasa.gov}]{\inits{R. A.}\fnm{Richard A.}~\lnm{Schwartz}}
\author[addressref={aff5,aff6},email={kim.tolbert@nasa.gov}]{\inits{A. K.}\fnm{Anne K.}~\lnm{Tolbert}}


\address[id=aff1]{Department for Astrophysical and Planetary Science, University of Colorado, Boulder, CO, USA}
\address[id=aff2]{Laboratory for Atmospheric and Space Physics, University of Colorado, Boulder, CO, USA}
\address[id=aff3]{Author now at Harvard-Smithsonian Center for Astrophysics, Cambridge, MA, USA}
\address[id=aff4]{Southwest Research Institute, Boulder, CO, USA}
\address[id=aff5]{NASA Goddard Space Flight Center, Code 671.0, Greenbelt, MD, USA}
\address[id=aff6]{American University, Washington, DC, USA}

\runningauthor{C. S. Moore et al.}
\runningtitle{MinXSS Instruments}

\begin{abstract}
The {\it Miniature X-ray Solar Spectrometer} (MinXSS) CubeSat is the first solar science oriented CubeSat mission flown for the NASA Science Mission Directorate, with the main objective of measuring the solar soft X-ray (SXR) flux and a science goal of determining its influence on Earth's ionosphere and thermosphere. These observations can also be used to investigate solar quiescent, active region, and flare properties. The MinXSS X-ray instruments consist of a spectrometer, called X123,  with a nominal 0.15 keV full-width-half-maximum (FWHM) resolution at 5.9 keV and a broadband X-ray photometer, called XP. Both instruments are designed to obtain measurements from 0.5\,--\,30 keV at a nominal time cadence of 10 seconds. A description of the MinXSS instruments, performance capabilities, and relation to the {\it Geostationary Operational Environmental Satellite} (GOES) 0.1\,--\,0.8 nm flux are discussed in this article. Early MinXSS results demonstrate the capability to measure variations of the solar spectral SXR flux between 0.8\,--\,12 keV from at least GOES A5--M5 (5 $\times$ 10$^{-8}$\,--\,5 $\times$ 10$^{-5}$ W m$^{-2}$) levels and infer physical properties (temperature and emission measure) from the MinXSS data alone. Moreover, coronal elemental abundances can be inferred, specifically Fe, Ca, Si, Mg, S, Ar, and Ni, when there is sufficiently high count rate at each elemental spectral feature. Additionally, temperature response curves and emission measure loci demonstrate the MinXSS sensitivity to plasma emission at different temperatures. MinXSS observations coupled with those from other solar observatories can help address some of the most compelling questions in solar coronal physics. Finally, simultaneous observations by MinXSS and {\it Reuven Ramaty High Energy Solar Spectroscopic Imager} (RHESSI) can provide the most spectrally complete soft X-ray solar flare photon flux measurements to date.
\end{abstract}

\end{opening}

\section{Introduction} 
     \label{S-Introduction} 
     The objective of the {\it Miniature X-ray Solar Spectrometer} (MinXSS) CubeSats is to explore the highly variable solar soft X-ray (SXR) spectral distribution and reveal its impact on Earth's ionosphere and thermosphere. The MinXSS X-ray instruments consist of a spectrometer, called X123,  with a nominal 0.15 keV full-width-half-maximum (FWHM) resolution at 5.9 keV and a broadband X-ray photometer, called XP. Both instruments are designed to obtain measurements from 0.5 - 30 keV at a nominal time cadence of 10 seconds. Solar soft X-rays are strongly absorbed by Earth's atmosphere in the E-region at an altitude of about $\sim$80 - 150 km. This energy input can strongly affect the energetics and dynamics of the ionosphere and thermosphere. These solar measurements can also be used to directly investigate the properties of the solar corona, which is dominated by  magnetic field dynamics, resulting in tenuous, high temperature plasma of over 1 MK. The primary heating source or the relative contributions of the many components to coronal heating are still in question. While MinXSS data alone cannot address the root of this question, the spectrally resolved measurements from MinXSS combined with other solar observations can yield critical information on this and other compelling questions in solar physics. MinXSS is not the first spectrometer to conduct these type of measurements, but it has unique new capabilities that can be advantageous over previous full Sun flux (irradiance) measurements from spatially integrating spectrometers.
     
     Currently, there are no solar instruments continuously conducting spectrally resolved soft X-ray measurements over a relatively large energy range. Fairly recent spectrally resolved, spatially integrated measurements have been conducted by {\it CORONAS-PHOTON Solar Photometer in X-rays} (SphinX) \citep{Gburek2013} and {\it MErcury Surface, Space ENvironment, GEochemistry, and Ranging} (MESSENGER) {\it Solar Assembly for X-rays} (SAX) \citep{Schlemm2007}. SphinX conducted solar measurements over a time-frame of roughly 9 months in 2009 during a time of very low solar X-ray flux levels, including the lowest solar X-ray levels ever recorded. The SphinX designed spectral coverage was $\sim$1 - 15 keV at a nominal 0.4 keV spectral resolution. MESSENGER SAX performed solar measurements primarily at an orbit around Mercury from 1 - 10 keV at a nominal spectral resolution of 0.6 keV and has measured numerous solar flares \citep{Dennis2015} from 2011 March to 2015 April. Like the MinXSS solar X-ray measurements, both of these spectrometers generated spatially integrated spectra over the instruments' field of view (FOV). MinXSS is designed to greatly improve upon these measurements and enhance the ability to determine emission line features with an improved spectral resolution (nominally $\sim$0.15 keV at 5.9 keV, due to detector architecture and electronics), a lower energy threshold (E$_{ph} \gtrsim $ 0.8 keV, due to a slightly thinner Be window for MinXSS-2), and by providing near-continuous {\it dedicated} solar measurements over a $\sim$6 year period (MinXSS-1 1 year mission $+$ MinXSS-2 5 year mission) during the declining phase of Solar Cycle 24, throughout solar minimum, and into the rising phase of Solar Cycle 25. MinXSS-1 was deployed from the International Space Station on 2016 May 16 and performed solar measurements until re-entry into Earth's atmosphere on 2017 May 6. MinXSS-2 is scheduled to launch no-earlier-than (NET) 2018 February.

\subsection{Current Solar Soft X-ray Measurements} 
  \label{Ss-Current_Measurements}

The aforementioned solar soft X-ray measurements lack direct spatial information. Combining spatially integrated, spectrally resolved measurements of MinXSS with data from other solar X-ray observatories can provide information on solar conditions. The Hinode {\it X-ray Telescope} (XRT) \citep{Kosugi2007, Golub2007} uses filters to create spectrally separated images of the soft X-ray intensity, but lacks fine spectral knowledge. The Geostationary Operational Environmental Satellite (GOES) {\it Soft X-ray Imager} (SXI) \citep{Hill2005}, exhibits similar qualitative spectral capabilities as XRT, but also suffers from the same issue of spectrally convolved images. The GOES {\it X-ray Sensor} (XRS) conducts spectrally \citep{Garcia1994} and spatially integrated measurements  in two bands (0.1 - 0.8 nm and 0.05 - 0.4 nm). A ratio of these two bands can yield an isothermal approximation to the coronal plasma temperature at the time of measurement \citep{White2005}. The {\it Reuven Ramaty High Energy Solar Spectroscopic Imager} (RHESSI) \citep{Lin2002} has provided spectral and spatial information using a Fourier imaging technique. RHESSI's primary spectral coverage extends from 6 keV - 17 MeV, with systematics-limited sensitivity below 6 keV. RHESSI's best spectral resolution in the 3 - 100 keV bandpass is $\sim$1 keV FWHM coupled with a 2.3 arcsecond FWHM spatial resolution. Additionally, the astronomy based {\it Nuclear Spectroscopic Telescope ARray} (NuSTAR) \citep{Harrison2013} satellite has performed a series of solar measurements which have been summarized in \cite{Grefenstette2016} and \cite{Hannah2016}. The NuSTAR nominal 0.4 keV FWHM spectral resolution can produce spectral images with $\sim$18 arcsecond spatial resolution over its  $\sim$11 arcminute field of view.

\subsection{MinXSS Science Goals} 
  \label{Ss-Science_Goals}

MinXSS data combined with the soft X-ray instrument data mentioned earlier, can be used in conjunction with other UV-Visible space observatories such as the Solar Dynamics Observatory (SDO) \citep{Pesnell2012}, the Hinode EUV Imaging Spectrograph (EIS) \citep{Culhane2007} and the Solar Optical Telescope (SOT) \citep{Tsuneta2008}, the Interface Region Imaging Spectrometer (IRIS) \citep{DePontieu2014}, to mention a few, to address pertinent science questions about the solar atmosphere. Specifically, the corona's high temperature, low density, and magnetic conditions have been of keen interest for decades since the observations of `coronium' lines in the solar spectrum \citep{Rayet1868, Seechi1875, golubpasachoff2010solar}. Better understanding the solar soft X-ray energy distribution allow for improved inferences of plasma conditions present during various stages of the solar cycle and during solar flares. In addition understanding the solar soft X-ray influence on Earth's ionosphere and thermosphere, a few main solar science questions and tasks that MinXSS data will help address are: 

\flushleft
\begin{itemize}
\item  What is the solar soft X-ray energy distribution as a function of solar cycle phase (at least the falling and rising phases) for the following components?
	\begin{itemize}
	\item[-] flares	
	\item[-] active regions (AR)
	\item[-] the quiescent Sun (QS)
	\end{itemize} 
\end{itemize}

\begin{itemize}
\item What is the AR variation in temperature, density, emission measure, chemical composition and magnetic complexity as a function of AR age and solar cycle phases?
\end{itemize}

\begin{itemize}
\item How are processes different between eruptive and non-eruptive compact flares? 
\end{itemize}

\begin{itemize}
\item What is the soft X-ray spectral connection to magnetic complexity in the solar atmosphere?
\end{itemize}

In order to effectively include MinXSS data in any solar analysis, it is necessary to understand the performance capabilities of the MinXSS X-ray instruments. This paper describes the basic instrument characteristics and will be a reference for scientists interested in utilizing MinXSS. The remainder of this paper includes basic descriptions of the MinXSS CubeSat mission in Section~\ref{S-Mission}, an overview of the instruments in Section~\ref{S-Instruments}, capabilities of these instruments in Section~\ref{S-Capabilities}, followed by examples of MinXSS-1 measurements from low solar levels (GOES A5) to an M5 flare and plasma inferences in Section~\ref{S-MinXSS_GOES}. Additional references for MinXSS include an overview of the MinXSS CubeSat and its subsystems by \cite{Mason2016}, pre-flight calibration results by \cite{Moore2016}, and early mission results by \cite{Woods2017}.


\section{The MinXSS CubeSats Mission} 
     \label{S-Mission} 

\begin{figure}    
   \centerline{\includegraphics[width=1.0\textwidth,clip=]{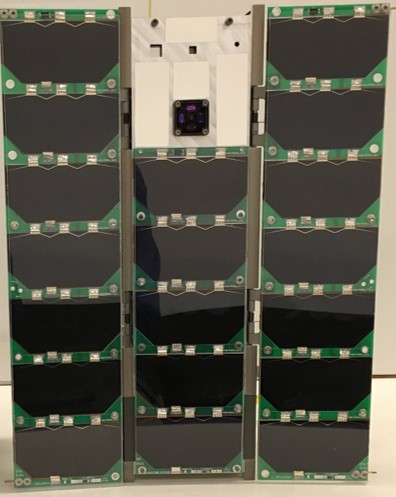}
              }
                            \caption{Picture of one of the twin Miniature X-ray Solar Spectrometer (MinXSS) 3U CubeSats. The CubeSat is oriented so that the solar panels and instrument apertures are facing the viewer (desired Sun facing side on-orbit). The MinXSS CubeSats are {\it designed} to measure the solar X-ray flux from 0.5 - 30 keV using the X-ray Photometer (XP) for spectrally integrated measurements across the entire energy band and the X123 spectrometer for energy resolved photon-counting measurements. The X123 spectrometer has a nominal spectral resolution of 0.15 keV full-width-half-maximum (FWHM). MinXSS-1 was deployed from the International Space Station on 2016 May 16 with an initial LEO altitude of $\sim$402 km for an anticipated mission lifetime of $\sim$12 months, depending on solar activity. MinXSS-2 is to be launched to a Sun-synchronous orbit of $\sim$500 km NET February of 2018 for an anticipated 5 year mission.      
             }
   \label{Figure1}
   \end{figure}

\begin{figure}    
   \centerline{\includegraphics[width=0.8\textwidth,clip=]{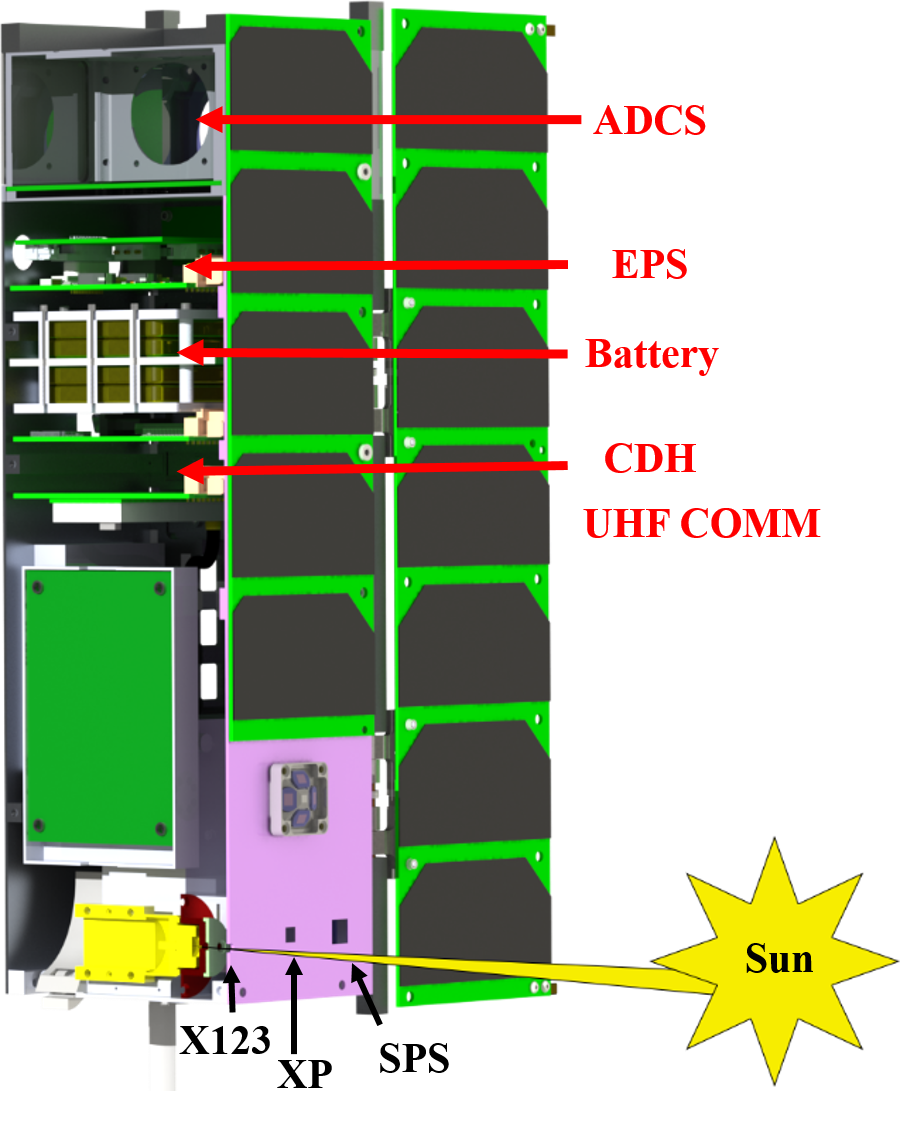}
              }
              \caption{Cut out of the MinXSS CubeSat to demonstrate the optical path of the Sun (not to scale) through the field of view limiting aperture and housing for the X123 spectrometer. The X123, XP and SPS apertures, and other subsystem locations are labeled for clarity. 
                      }
   \label{Figure2}
   \end{figure}

The MinXSS CubeSats development and testing involved extensive graduate student involvement in collaboration with the University of Colorado-Boulder's Laboratory for Atmospheric and Space Physics (LASP), and the Aerospace Engineering Sciences Department, with assistance from professors and professionals. The first of the twin satellites, MinXSS-1 was ferried to the International Space Station (ISS) from the Kennedy Space Center on 2015 December 6. MinXSS-1 was deployed from the ISS on 2016 May 16 to a Low Earth Orbit (LEO) with an initial altitude of $\sim$402 km. MinXSS-1 commenced science operations on 2016 June 9. The second CubeSat, MinXSS-2, is scheduled to launch to a Sun-synchronous orbit NET February of 2018. Figure~\ref{Figure1} is a picture of the MinXSS-1 3U CubeSat, noting that 1 Unit (1U) is 10 cm $\times$ 10 cm $\times$ 11.35 cm in size. Even though they are small relative to {\it traditional} solar observing satellites, the MinXSS CubeSats are fully functioning satellites. They include triple junction GaAs solar cells from AzurSpace, Li-polymer batteries, Electrical Power System (EPS), an Attitude and Determination Control System (ADCS) supplied by Blue Canyon Technologies, a tape measure for a radio antenna, a Li-1 radio for communication, Command and Data Handling (CDH) microcontroller, and science instruments. The positions of these subsystems in the MinXSS spacecraft are in Figure~\ref{Figure2}. An overview of these subsystems is described in \cite{Mason2016}. MinXSS-2 is an augmented version of MinXSS-1 with planned improvements in hardware, software and the implementations of lessons learned from the MinXSS-1 mission. Scientifically, one of the most important upgrades of MinXSS-2 is a newer version of the X-ray spectrometer. The MinXSS-1 instruments are listed in Table~\ref{T-Table1} and consist of a visible-light Sun Position Sensor (SPS), an X-ray photometer (XP), and an X-ray spectrometer called X123. Nominally the MinXSS science data are composed of 10 second integrations, that can be decreased to 3 seconds or increased to 1 minute depending on the scientific objectives. The next section describes the full set of MinXSS instruments and their capabilities.

  \begin{table} 
\caption{MinXSS satellite launch, orbit and mission lifetimes. A few instrument properties are also listed.}
  \rotatebox{90}{ 
 \label{T-Table1}
\begin{tabular}{lccccccc@{.}l c} 
  \hline
  \hline
Satellite & \multicolumn{3}{c}{MinXSS-1} &  & \multicolumn{3}{c}{MinXSS-2} \\  
\cline{2-4}  \cline{6-8}
Orbit Insertion Date &    & 16--May--2016 &   &  &     & NET\tabnote{NET = No Earlier Than} Feb--2018 &    \\
Anticipated Mission Lifetime &    & $\sim$1 year &  &   &     & $\sim$5 years &    \\
Initial Orbit Altitude (km) &    & $\sim$400 km &   &  &     & $\sim$500 km &    \\
\hline
Instrument & SPS & XP & X123 &  & SPS  & XP & X123 \\
\cline{2-4}  \cline{6-8}
Si Detector Depletion Depth ($\mu$m) & 55 & 55 & 500 & & 55  & 55 & 500 \\
Aperture Area (cm$^{2}$) & 4.0 $\times$ 10$^{-2}$ & 2.0 $\times$ 10$^{-1}$ &  2.5 $\times$ 10$^{-4}$  & &  4.0 $\times$ 10$^{-2}$ &  2 .0x 10$^{-1}$ &  2.5 $\times$ 10$^{-4}$ \\
Window Type-Material & ND7\tabnote{Neutral density filter - 10$^{7}$} & Be & Be & & ND7\tabnote{Neutral density filter - 10$^{7}$}  & Be & Be + Zn \\
Window Thickness ($\mu$m - uncertainty) & -- & 19.0 (0.1) & 24.5 (0.6) & & --  & 18.0 (0.1) & 11.2 (0.3) + 0.1 (0.1) \\
Full Field of View (FOV, $^{\circ}$) & 8 & 8 & 4 & & 8  & 8 & 4 \\
  \hline
  \hline
\end{tabular}
} 
\end{table}


\section{MinXSS Instrument Description} 
\label{S-Instruments} 

\subsection{Sun Position Sensor (SPS)} 
  \label{Ss-SPS}

	 SPS is a visible-light sensitive quad silicon-photodiode (Si-photodiode) arrangement behind an ND7 (neutral density filter with an attenuation factor of 10$^{7}$). The Si-photodiodes have a 55 $\mu$m thick depletion depth. The SPS square aperture is 2 $\times$ 2 mm and the layout is discussed in the Appendix in Section~\ref{S-appendix_SPS} and in Figure~\ref{FigureA1}. The relative solar illumination on each of the four diodes is used to calculate the solar position within the MinXSS X-ray instruments' field of view (FOV). The absolute position knowledge is accurate to within 10 arcseconds and the pointing is controlled to a precision of 10 arcseconds, limited by the capabilities of the Blue Canyon XACT ADCS system.

\subsection{X-ray Photometer (XP)} 
  \label{Ss-XP}

	The LASP designed XP unit consists of a Si-photodiode (55 $\mu$m thick depletion depth) with a beryllium (Be) window 19.0 $\mu$m thick for MinXSS-1 and 18.0 $\mu$m thick for MinXSS-2 to attenuate visible-light contamination. The XP aperture is 5 mm in diameter. The main purpose of XP is to serve as a consistency check to the detected photon flux and linearity of the X-ray spectrometer X123, assessing long term degradation trends, and comparison to the GOES XRS photometers. The Si-photodiodes in XP and SPS have been flown numerous times, such as on SDO \citep{Woods2012}, are known to be stable in space, and have linear response over the full range of solar flux levels after corrections to various influences. The XP photodiode response can have gain and dark current variations due to thermal fluctuations, and can possibly suffer noise from sources internal to the MinXSS spacecraft (e.g. emf, radio transmission, microphonics, etc.), which must be corrected for.  After gain and dark current corrections, the XP product returns one spectrally integrated value for the solar X-ray contribution over its expected spectral response from $\sim$0.5 - 30 keV. This type of information is valuable, but of similar class to the GOES XRS. The spectral information provided by the X-ray spectrometer, X123, can greatly enhance the interpretation of the XP data.

\subsection{X-ray Spectrometer (X123)} 
  \label{Ss-X123}

	The MinXSS main scientific instrument, X123, is an X-ray spectrometer purchased from Amptek (Amptek website: \url{http://amptek.com/}). The data returned from X123 will be the most important data product that MinXSS will provide to the scientific community. X123 is composed of a Silicon Drift Detector (SDD) with a 500 $\mu$m thick Si depletion depth behind a Be window 24.5 $\mu$m thick for MinXSS-1 and 11.2 $\mu$m thick for MinXSS-2. From pre-flight calibrations, there appears to be Zn contaminant in the MinXSS-2 Be filter with an approximate thickness of 0.1 $\mu$m for MinXSS-2. The X123 dead-layer is comprised mostly of SiO and has a nominal thickness around 0.15 $\mu$m. We did not measure the deadlayer thickness ourselves but include the nominal value into our detector modeling. This thickness of SiO does not strongly impact the detector high energy efficiency, but could contribute to the low energy ($<$ 1 keV) efficiency. This folds into our low energy efficiency.
	
	Amptek provides the X123 electronics for power regulation and X-ray photon energy detection and embedded software for reading the spectrometer output. Thus, in addition to X-ray characterization, the main task remaining is to integrate the X123 output to the MinXSS CDH processor that compresses the data and formats it into data packets. The power draw from X123 is nominally $\sim$2.8 W during normal operations and at max $\sim$5.0 W (normally at turn on). The X123 spectrometer is actively cooled with a thermoelectric cooler (also provided by Amptek) to $\sim$224 K. The cooling is necessary to minimize thermal noise which will contribute to the lower energy bins (E$_{bin} \leq $ 1.0 keV). The X123 sensor resides inside a stainless steel housing behind a tungsten FOV-limiting pinhole aperture with a $\sim$0.18 mm diameter to protect from energetic particles and hard X-rays.  

\subsection{Detector Operation} 
  \label{Ss-Detector_Operation}
  
	The X123 SDD, XP and SPS  operation relies, in principle, on electron-hole pair generation in the Si lattice by the incident photon flux. Photons and any other energy sources (these contribute to the noise) that have more than the Si electron-hole pair generation energy (E$_{e-h} \gtrapprox $ 3.65 eV) can create a number of electron-hole pairs proportional to the energy of the incident photon. For the SPS and XP detectors, the liberated charges contribute to a current which is measured, amplified and converted to a digital unit. The difference for the X123 spectrometer is that the electrons generated by this energy deposition process drift toward the readout anode and the resulting charge is integrated on a capacitor within the detector rise time (peaking time). These energy `impulses' are deemed events. The preamplifier, which consists of a Field Effect Transistor (JFET for MinXSS-1 and MOSFET for MinXSS-2), that amplifies the signal, converts it to a voltage, and this voltage ramp is shaped to a trapezoid. Each of these shaped signals are used to discern the energy deposited in the active detector volume by the incident photon `event'. The performance capabilities of the X-ray instruments are discussed in the next Section.

\section{MinXSS X-ray Instrument Capabilities} 
\label{S-Capabilities} 

	The instruments were characterized using radioactive lab sources for spectral resolution, optimization of electronic settings, determining the gain and energy offset of the X123 spectrometer. SPS, XP and X123 FOV, XP and X123 window thicknesses, spectral efficiency, and linearity were determined from the National Institute for Standards and Technology (NIST) Synchrotron Ultraviolet Radiation Facility (SURF) measurements. MinXSS X123 spectrometer basic performance properties and characterization methodology are described in \cite{Moore2016}. Accurate knowledge of the electron beam current, electron beam energy, magnetic field strength, and source distance allows for the precise calculation of the synchrotron light spectral intensity to enable calibration of the X123 and XP responsivities with about 10$\%$ accuracy \citep{Moore2016}.

\subsection{Field of View Sensitivity} 
  \label{Ss-FOV}

	The MinXSS X-ray instruments integrate the incident radiation across their FOVs to create their respective signals. While XP and X123 do not have spatial resolution per se, they do possess a sensitivity to light intensity across their respective FOVs. XP nominally has an angular response across an 8$^{\circ}$ FOV and X123 has a FOV of 4$^{\circ}$. An example of the alpha-beta ($\alpha$-$\beta$) coordinate system used for the SURF calibrations is given in the SPS image in Figure~\ref{FigureA1}. An extended source such as the Sun viewed from Earth will have an angular size of roughly 0.5$^{\circ}$, but the response can vary strongly depending on the actual X-ray coronal structure during observations. This is well within the FOV of XP and X123. For the most accurate photometric measurements from MinXSS, it is important to correct for the solar disc position within the MinXSS FOV. 

\begin{figure}    
   \centerline{\includegraphics[width=1.0\textwidth,clip=]{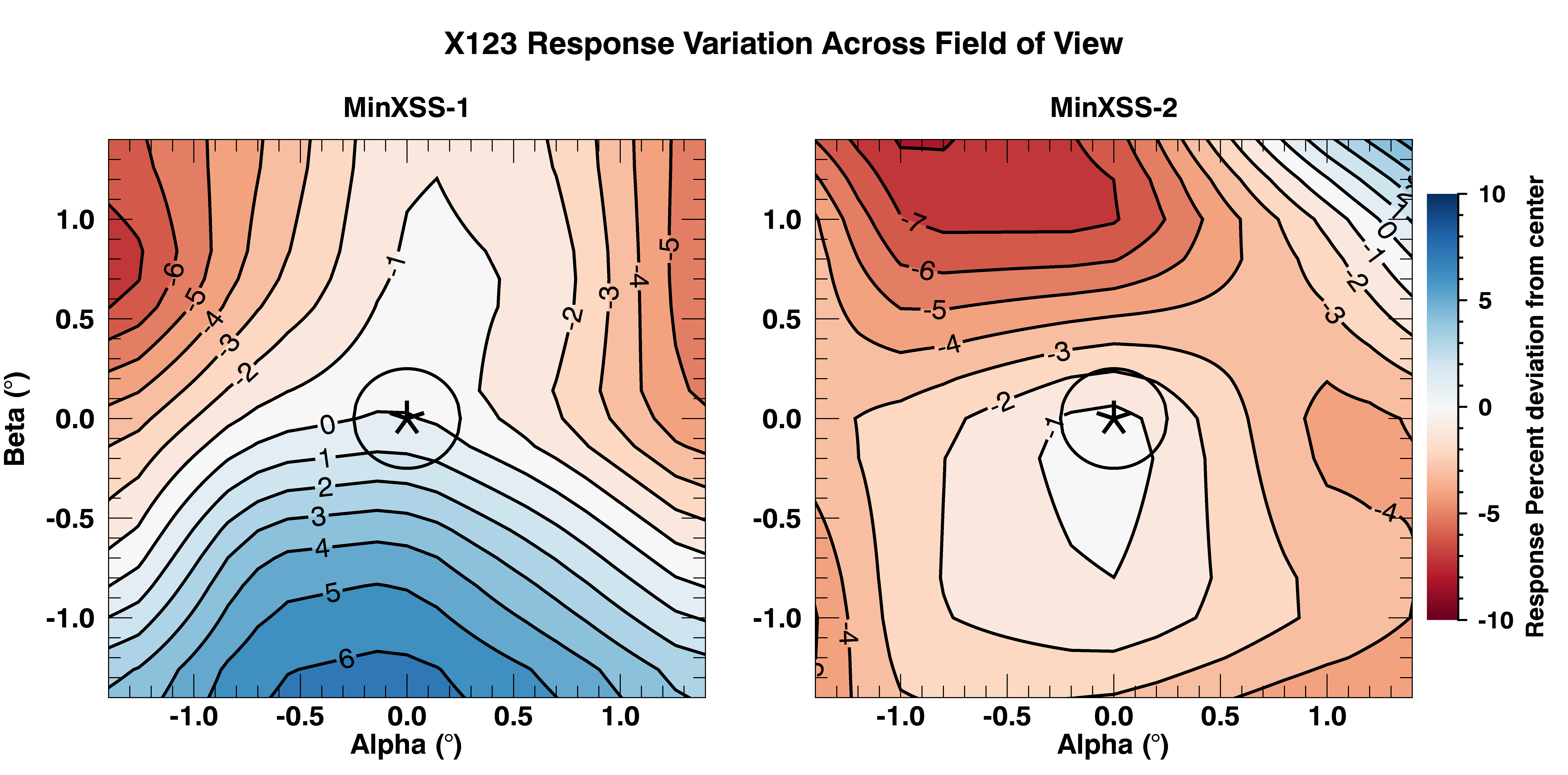}
              }
              \caption{X123 Field of View (FOV) sensitivity maps constructed from pre-launch data at the National Institute for Standards and Technology (NIST) Synchrotron Ultraviolet Radiation Facility (SURF) for each CubeSat (left: MinXSS-1 and right: MinXSS-2). The maps displayed are the MinXSS spectral response convolved with the apparent visible-light solar disc. The asterisk denotes the center of the spectrometer FOV which mechanically aligned to the boresight of the spacecraft. The black circle represents the size of the visible-light solar disk in the FOV. The contours and color map signify the percent difference in the X123 response from the center (asterisk), which is a few percent in magnitude.
                      }
   \label{Figure3}
   \end{figure}

	Figure~\ref{Figure3} shows FOV sensitivity maps of MinXSS-1 and MinXSS-2 X123. The maps are a 1.4$^{\circ}$ $\times$ 1.4$^{\circ}$ subset, centered on the FOV of X123. The XP map yields similar FOV sensitivities. The values are the raw NIST SURF determined MinXSS response maps, convolved with the visible-light solar disc, and expressed as a percent difference in the total signal summed over all energies at that position with respect to the center, to estimate the X-ray detection variation across the FOV. This pre-flight map serves as our baseline until on-orbit maps are created for comparison. The absolute variation across the centered 1.4$^{\circ}$ $\times$ 1.4$^{\circ}$ FOV is around 8$\%$. The MinXSS ADCS unit (XACT from Blue Canyon Technologies) has proven to keep the Sun within 0.3$^{\circ}$ of the center of the MinXSS FOV with a precision of 10 arcseconds ($\sim$0.028$^{\circ}$). Thus, the inner 1$^{\circ}$ diameter annular region is the most important for MinXSS on-orbit performance and the absolute variation within this region is no larger than about 5$\%$. We note that only mechanical alignment was implemented between the XACT ADCS unit and the instruments.

\subsection{Spectral Resolution} 
  \label{Ss-Resolution}

	The X123 spectral resolution, energy bin gain and offset have been estimated using radioactive line sources of $^{55}$Fe for the $\sim$5.90 and $\sim$6.49 keV line complexes, and $^{241}$Am for the $\sim$11.87,  $\sim$13.95, and $\sim$17.75 keV lines. The fit values for the both the MinXSS-1 X123 gain and MinXSS-2 X123 gain are 0.0297 keV/bin and are consistent with the nominal bin width (0.03 keV). The offset is dependent on the electrical grounding conditions at the time of operation, due to the fact that the pulse height analysis is computed on top of a baseline voltage, which can drift. Thus, the X123 energy offset in the lab at LASP for radioactive source measurements, at NIST SURF and on-orbit may not be exactly the same. Nominally, the value for MinXSS-1 energy bin offset is around -0.076 keV and the MinXSS-2 energy bin offset is close to -0.265 keV. 
	
	The X123 has a customizable peaking time, which dictates how long the photon liberated electron charge cloud is integrated on the readout capacitor. The longer the peaking time, the better the spectral resolution. MinXSS-1 has a fixed peaking time of 4.8 $\mu$s but we have tested the MinXSS-2 X123 detector for various peaking times between 0.6 - 9.6 $\mu$s. We will operate the MinXSS-2 X123 for peaking times between 1.2 - 4.8 $\mu$s due to the tradeoff between spectral resolution and effective maximum count rate that can be recorded before photon pile-up begins to occur. We only show data for these 1.2 - 4.8 $\mu$s peaking times in this paper. The longer the peaking time, the lower the maximum count rate that can be accurately measured. Details on the maximum count rate are discussed in Section~\ref{Ss-Linearity}.
	
\begin{figure}    
   \centerline{\includegraphics[width=1.0\textwidth,clip=]{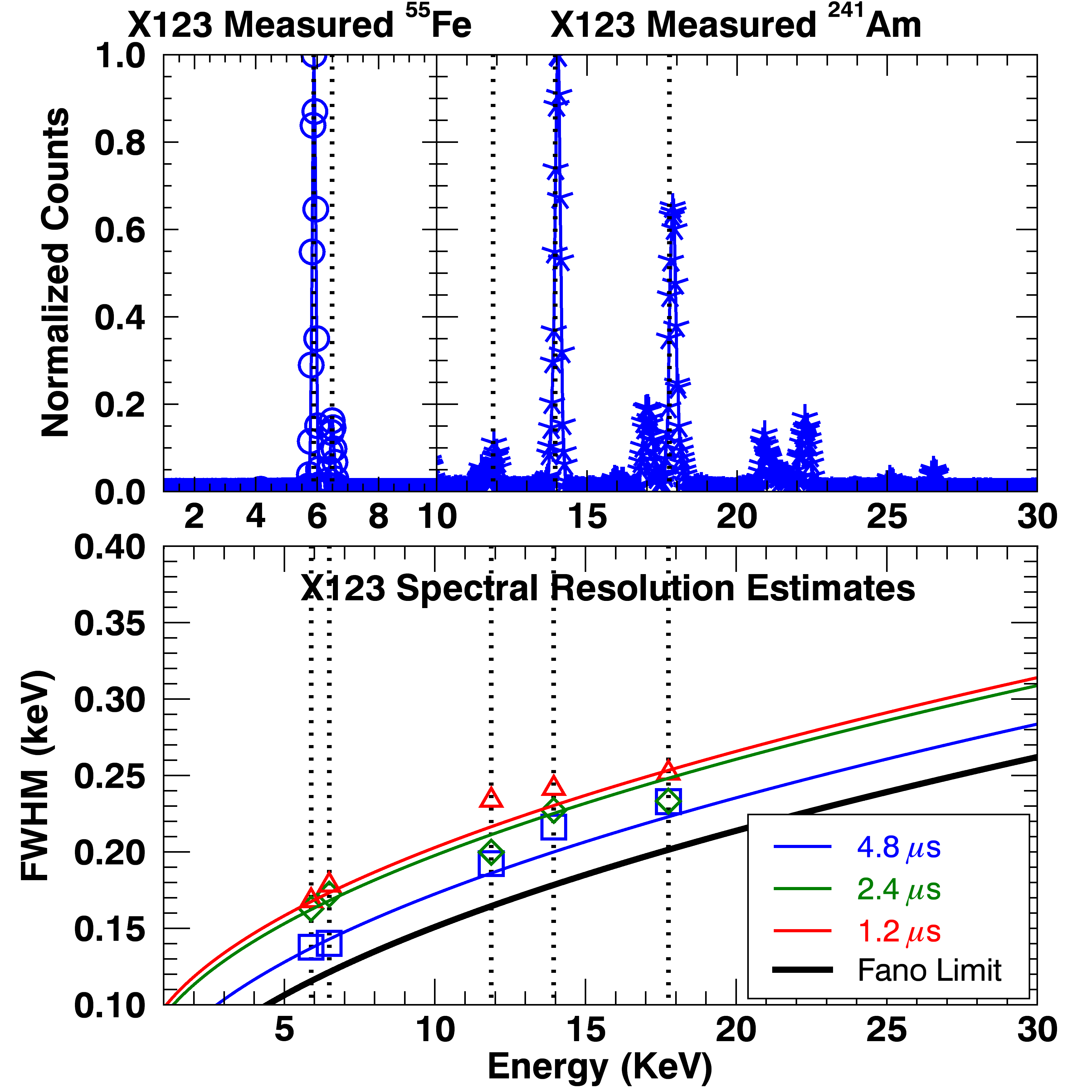}
              }
              \caption{Example plot of MinXSS-2 X123 spectral resolution estimates vs. photon energy using radioactive X-ray sources. The top split-plot shows the normalized counts from the $^{55}$Fe source from 1 - 10 keV and the $^{241}$Am source from 10 - 30 keV. The $^{55}$Fe $\sim$5.90 and $\sim$6.49 keV line complexes are easily detected. The $^{241}$Am $\sim$11.87,  $\sim$13.95, and $\sim$17.75 keV lines are used for spectral resolution estimates. The vertical dotted line emphasizes which spectral lines were used for FWHM resolution estimates for three different spectrometer peaking times. Longer peaking times yield better photon energy resolving power up to the combined electronic and Fano Limit. The Fano Limit is the intrinsic statistical limit of bulk Si semiconductor material to resolve energy differences, and is over-plotted with the black solid line. 
The colored lines are the Fano Limit (from Fano Noise) with an estimated electronic noise contribution from the $^{55}$Fe measurements. These estimates depicted by the color lines are used to extrapolate the spectral resolution to {\it higher} photon energies. The extrapolation to lower energies is not expected to adhere to these lines due to other noise sources (microphonics, thermal noise, and other uncharacterized sources). 
                      }
   \label{Figure4}
   \end{figure}

	Figure~\ref{Figure4} shows example MinXSS-2 X123 $^{55}$Fe measurements from 1 - 10 keV and $^{241}$Am measurements from 10 - 30 keV. The subset of lines used to assess the spectral resolution at specific energies are signified by the vertical dotted lines. An expanded in plot of the $^{55}$Fe for the $\sim$5.90 and $\sim$6.49 keV line complexes are in \cite{Moore2016}. The nominal spectral resolution near 5.9 keV was confirmed for the respective peaking times. MinXSS-1 X123 detector resolution measurements were not performed like the MinXSS-2 detector. Thus, we asses the on-orbit resolution which currently estimated to be $\sim$0.24 keV, which is much broader than the nominal 0.15 keV spectral resolution at 5.9 keV for a 4.8 $\mu$s peaking time. We are currently assessing the reasons for a degraded on-orbit resolution. MinXSS-2 has an improved version of the X123 spectrometer, the X123 Fast SDD. The X123 Fast SDD has an improved preamplifier, a MOSFET transistor instead of a JFET for the MinXSS-1 version. This has a lower effective capacitance, lower noise and results in improved spectral resolution for the same peaking times as the older version. An advantage is increased maximum count rate, which will be discussed in Section~\ref{Ss-Linearity}.  The spectral resolution for MinXSS-2 at 5.9 keV is 0.137 keV for 4.8 $\mu$s, 0.162 keV for 2.4 $\mu$s, and 0.168 keV for 1.2 $\mu$s.

	The resolving power at X-ray energies for Si based detectors is limited by Fano Noise \citep{knoll2010}. Fano Noise is the intrinsic statistical variation in the electron-hole pair generation rate per event and is represented by the solid black line in Figure~\ref{Figure4}. The Fano Factor (F) is essentially the ratio between the observed variance to the expected Poisson variance and for silicon is around 0.1. The Fano Noise in FWHM units is $\sigma_{Fano} = 2.35*\sqrt{\frac{F*E_{ph}}{E_{e-h}}}$, where $E_{ph}$ is the photon energy and $E_{e-h}$ is the electron-hole pair generation. The electronic noise contribution was estimated from the $^{55}$Fe for the $\sim$5.90 and $\sim$6.49 keV data and used to scale up the Fano Noise to provide an estimate of the spectral resolution at the higher energies. The extrapolation to lower energies is not expected to adhere to these lines due to other noise sources (microphonics, thermal noise, and other uncharacterized sources).

\subsection{Spectral Efficiency and Effective Area} 
  \label{Ss-Efficiency}

	The XP 55 $\mu$m and X123 SDD 500 $\mu$m thick depletion depths  for Si based X-ray detectors dictate the high energy cut-off efficiency. The window material and composition affect the high and low energy efficiency. For Be based windows only the lower energy photons are attenuated and thus the Be window thickness dictates the low energy photon response. Noise contributions, primarily electronic noise, limit the XP signal fidelity and the X123 lower energy contribution. The MinXSS instrument spectral efficiency was determined through a series of measurements conducted at NIST SURF and the details are listed in \cite{Moore2016}. The SURF photon spectrum is known to within 10$\%$ near 1 keV and produces statistically significant counts up to about 3 keV. Model uncertainties in the fitted detector window thicknesses, depletion depth thicknesses and atomic coefficients yield uncertainties near 20$\%$ for the higher energies. Since the SURF spectrum is a continuum, it is not very useful for the spectral resolution estimates, such as those discussed in Section~\ref{Ss-Resolution}, but the absolute synchrotron photon spectral distribution can be used to determine the detection efficiency of the MinXSS X-ray instruments. 

The MinXSS XP signal can be calculated  from the following expression in Equation~\ref{Eq-XP_Signal}, where C$_{XP}$ is the XP count rate with units of $\frac{DN}{s}$, s is seconds, DN is the Data Number, which is converted from the femtoCoulomb (fC) signal by the XP gain, $G_{XP}$. The bracketed term is the femtoamps (fA) generated from the detected photon flux, from the source of photons, $S(E_{ph}, \Omega)$ with photon energy, $E_{ph}$.
         
   \begin{equation}  \label{Eq-XP_Signal}  
    C_{XP} = G_{XP} \int_{0}^{\infty} \bigg[ \int^{\Omega_{\odot}} S(E_{ph}, \Omega) A_{XP} R_{XP}(E_{ph}, \Omega) d\Omega \bigg] dE_{ph}.
   \end{equation}

In the case of the Sun, there is a distribution of photons as a function of X-ray energy, and a function of position in the extended corona. The Sun appears as an extended object and its position can vary within the MinXSS FOV. The physical extent of the X-ray emission from the Sun is encompassed in the solid angle,  $\Omega$, which is not necessarily the simple conversion of $\pi sin^{2}(\theta_{\odot}) \approx \pi \big(\frac{R\odot}{1 AU}\big)^{2} \approx 6.8$ $\times$ $10^{-5}$ sterradians, because the X-ray emission is not confined to the visible-light solar disk. An example of the distribution of X-ray emission that MinXSS can detect is illustrated in the Hinode  XRT images in Figure~\ref{Figure11}. Thus, $S(E_{ph}, \Omega)$ is an intensity or radiance, in units of photons s$^{-1}$ keV$^{-1}$ cm$^{-2}$ ster$^{-1}$ in this formulation. $A_{XP}$ is the XP aperture geometric area, in units of cm$^{2}$. The XP detector response $R_{XP}(E_{ph}, \Omega)$ includes the XP detector efficiency, the conversion from photon energy to fC, $\epsilon_{ph}$, and the FOV sensitivity which is encompassed in the $\Omega$ dependence. In theory, the integral of the bracketed term can include all photon energies from 0 to $\infty$ incident on the XP area, but the actual high and low energy limits are set by the XP detector response.
       
\begin{figure}    
   \centerline{\includegraphics[width=1.0\textwidth,clip=]{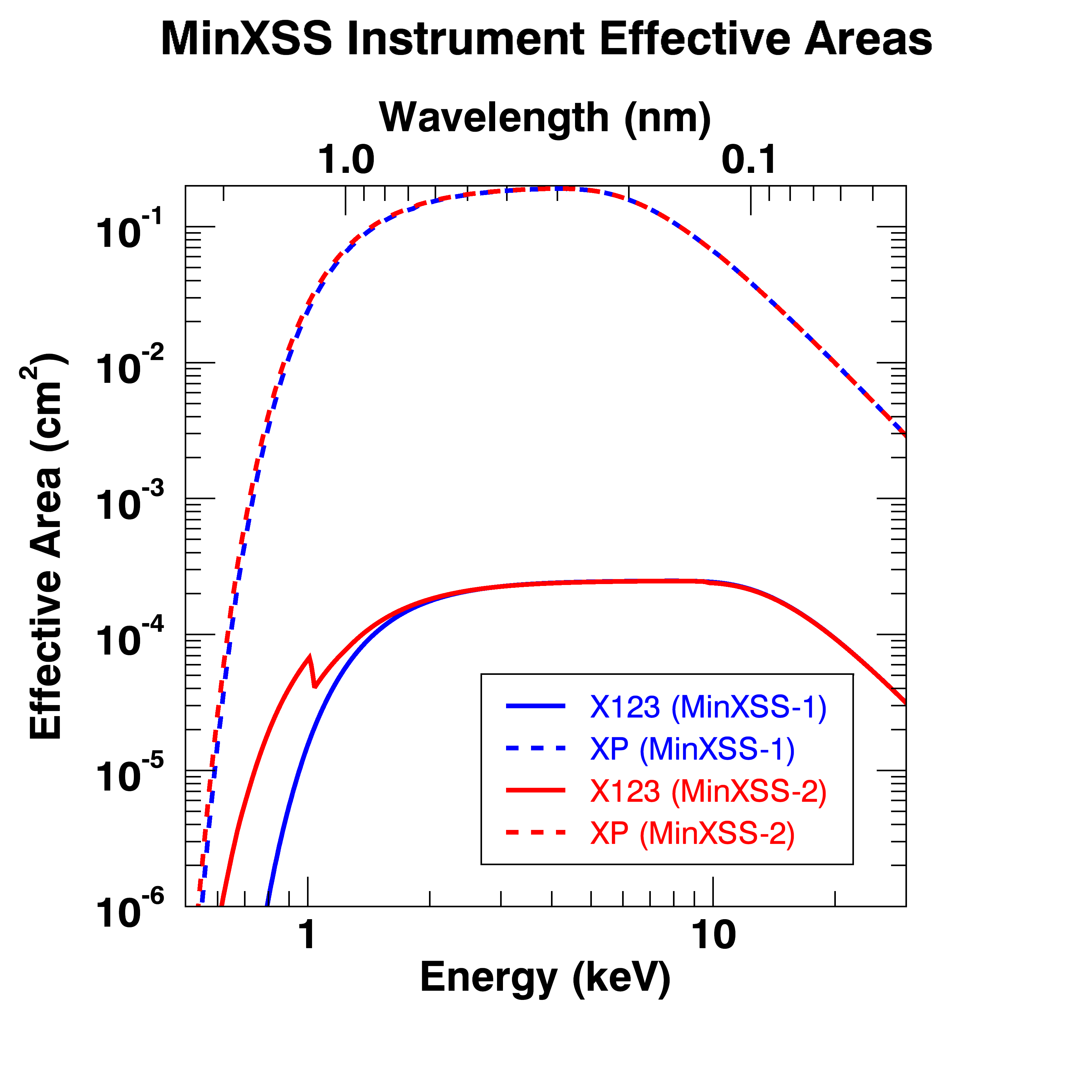}
              }
              \caption{MinXSS-1 and MinXSS-2 X-ray instrument effective area vs. photon energy. The main difference between the XP (dashed line) and X123 (solid line) is due to the geometric area of their respective apertures. The XP aperture diameter is $\sim$5 mm, while the X123 pinhole diameter is $\sim$0.18 mm. MinXSS-2 has an undesigned Zn contribution to the Be window, which results in an edge in the response near 1 keV.  
                      }
   \label{Figure5}
   \end{figure}

       The combination of the XP detector response and geometric area constitutes an `effective area'. The XP MinXSS-1 and MinXSS-2 effective area curves are displayed in Figure~\ref{Figure5}. The main XP photon response lies between 0.5 - 30 keV, with the actual count contribution depending on the solar X-ray spectrum, $S(E_{ph}, \Omega)$, during observations. Both MinXSS-1 and MinXSS-2 XP devices have similar responses as their Be windows are of similar thickness (19.0 $\mu$m for MinXSS-1 and 18.0 $\mu$m for MinXSS-2). 
       
       In a similar fashion to the XP calculation, the X123 count rate per energy bin denoted, j,  is \big(C$_{X123}$\big)$_{bin,j}$, in units of $\frac{Counts}{s}$ can be calculated by Equation~\ref{Eq-X123_Signal} and Equation~\ref{Eq-X123_Signal_Upsilon}

   \begin{equation}  \label{Eq-X123_Signal} 
    \big(C_{X123}\big)_{bin,j} = \int_{E_{min,j}}^{E_{max,j}} \bigg[ \Upsilon(E_{det}) \bigg] dE_{det}
  \end{equation}

   \begin{equation}  \label{Eq-X123_Signal_Upsilon} 
   \Upsilon(E_{det}) = \int_{0}^{\infty} \int^{\Omega_{\odot}} S(E_{ph}, \Omega) A_{X123}	\overline{\rm \Re}_{X123} (E_{ph}, \Omega, E_{det}) d\Omega dE_{ph}
  \end{equation}


 where $A_{X123}$ is the X123 aperture geometric area. The main difference between XP and X123 in terms of the signal calculation is the spectral binning of the data and the response function. $\overline{\rm \Re}_{X123} (E_{ph}, \Omega, E_{det})$, which is the photon $\Leftrightarrow$ detected energy bin {\it redistribution} function and includes the FOV dependence thru $\Omega$. The function $\overline{\rm \Re}_{X123} (E_{ph}, \Omega, E_{det})$ maps $E_{ph}$ to $E_{det}$ for forward modeling and vice-versa, for inverting detected counts to create an incident photon flux estimate. Thus, the bracketed term in Equation~\ref{Eq-X123_Signal} is only a function of $ dE_{det}$, because of the operation of $\overline{\rm \Re}_{X123} (E_{ph}, \Omega, E_{det})$ on the source intensity $S(E_{ph}, \Omega)$. Hence, the final integral over the individual energy bin energy limits, $E_{min,j}$ and $E_{max,j}$ to obtain a final count rate value in energy bin j. Again, in theory the integral over $E_{ph}$ in the bracket is over all photon energies, but in reality, this is limited to the X123 spectral efficiency, high and low photon limits.
 
 Similarly, for X123 the combination of the X123 spectral efficiency (the probability that a photon incident on the active area of the detector will be absorbed) and the aperture geometric area can be used to create the effective area curves for MinXSS-1 and MinXSS-2 X123 in Figure~\ref{Figure5}. The edge near 1 keV in the MinXSS-2 X123 detector effective area curve is due to a Zn contamination in the Amptek supplied Be window and has been included in the modeling. The primary spectral range is nominally 0.5 - 30 keV for both MinXSS spacecraft X123 detectors, but it is obvious in Figure~\ref{Figure5} that the MinXSS-2 detector has a higher efficiency for the lower photon energies  (E$_{ph} \leq $ 2 keV). This is due to the thinner 11.2 $\mu$m Be window on the MinXSS-2 detector as compared to the 24.5 $\mu$m thick window on MinXSS-1. The choice of the thinner Be window X123 detector being on the second MinXSS was intended, because MinXSS-2 has an anticipated longer misson of $\sim$5 years vs. MinXSS-1 expected maximum mission lifetime of $\sim$1 year. The higher efficiency for lower photon energies aids in extending the low energy limit of the MinXSS-2 X123 data product. Effective area curves are great for comparing the both the spectral and intensity sensitivities between various instruments on different observatories.

    \subsection{Detector Response Matrix} 
  \label{Ss-DRM}

\begin{figure}    
   \centerline{\includegraphics[width=1.0\textwidth,clip=]{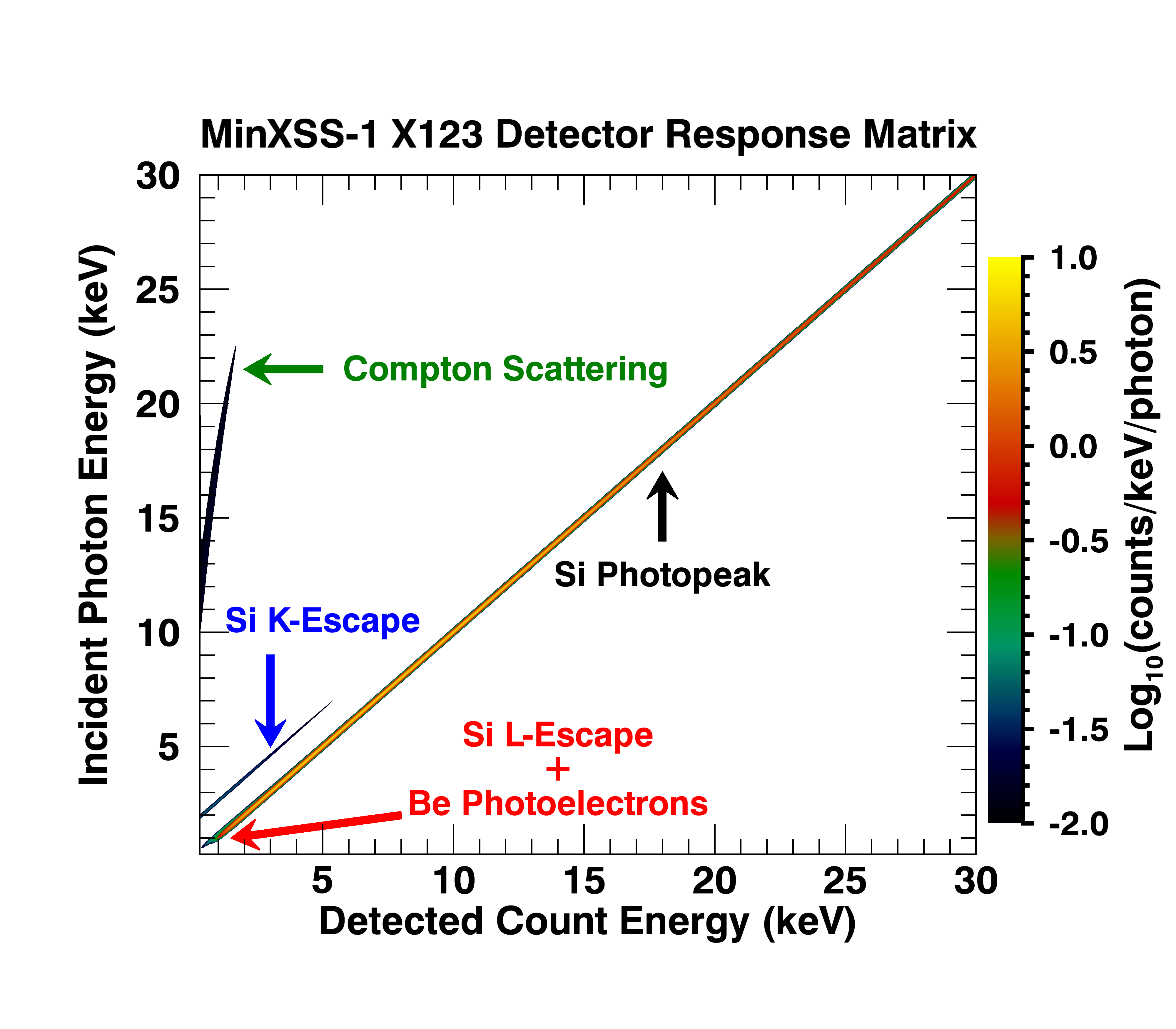}
              }
              \caption{Example of the X123 Detector Response Matrix (DRM) which includes Si photopeak, Compton scattering, Be window generated photoelectrons, and Si K and L (2s and 2p) escape processes. The DRM gives the probability that an incident photon of energy E$_{1}$ will deposit energy E$_{2}$ in the detector. 
                      }
   \label{Figure6}
   \end{figure}

     In reality, the resulting count space is discrete and not continuous after the integration of the bracketed quantity in Equation~\ref{Eq-X123_Signal} to create the binned data. Thus, one can interpret the photon-count {\it redistribution} function as a Detector Response Matrix (DRM) for the X123 spectrometer, with columns, k, that connect the incident photon energy to the row, j, which are the loses recorded in the X123 energy bins. This is listed in Equation~\ref{Eq-DRM}, where $\bf{\overrightarrow{C}_{j}}$ is the detected count rate spectrum, $\bf{\overline{\rm \Re}_{k,j}}$ is the X123 DRM and $\bf{\overrightarrow{S}_{k}}$ is the source intensity. An example of the MinXSS X123 DRM is displayed in Figure~\ref{Figure6}. The DRM incorporates the detection efficiency, the probability that a photon stopped in the detector by a photo-electric interaction will be recorded. The probability of the full-energy being deposited in the detector gives the diagonal response or photopeak efficiency. Additionally, there are loss processes such as fluorescence emission \citep{knoll2010}, which result in an event being recorded at a lower energy. These are the off-diagonal terms in the DRM and include loss processes such as Si K and L (2s and 2p) escape. These loss processes result in counts occurring at an energy bin that is the incident photon energy minus the Si edge energy ($\sim$1.8 keV for K, $\sim$0.15 keV for 2s, and $\sim$0.14 keV for 2p). These spectral detection shifts occur if the resulting Si escape photon actually leaves the active area of the detector without being absorbed, or else if the escape photon is reabsorbed by the detector, one retains the original photon energy in the detected energy bin.
     
        \begin{equation}  \label{Eq-DRM} 
       {\bf \overrightarrow{C_{j}}} = {\bf \overline{\rm \Re} _{k,j} \overrightarrow{S_{k}}}.
   \end{equation}
     
	Another process that we consider is the liberation of photoelectrons in the X123 Be window that can eventually be detected and appear in the spectrum. These events will occur at lower energies than the initial photon distribution that created the photoelectrons due to the energy loss processes in migrating to the window surface, the negative bias of the X123 detector front surface and transport through the detector deadlayer. The resulting contribution would be a continuum distribution at low energies (E$_{ph} \leq $ 1.5 keV) for typical solar fluxes (peak in photon distribution) and window thicknesses (absorption probability spectrum of window) on our X123 units. These model contributions are included in the functional form of $\overline{\rm \Re}_{X123}$ and are also discussed in \cite{Moore2016}.

      Another effect in MinXSS spectra is Compton scattering, but to a much smaller degree than Si-escape (both because of the lower probability and the lack of  {\it detectable} higher energy photons greater than 10 keV). Compton scattering is unlikely to be a major contaminant in the quality of MinXSS spectra due to the small effective area of X123 coupled with the steeply decreasing solar spectra at high energies. Only large X-class flares could produce large enough signal at photon energies greater than 10 keV to make Compton scattering important, but other issues such as photon pile-up and X123 dead-time will occur before that becomes relevant. We have used a flux linearity test at SURF to quantify when pile-up and detector dead-time will become severe and discuss this in Section~\ref{Ss-Linearity}

    \subsection{Temperature Response} 
  \label{Ss-Temperature_Response}

To connect the MinXSS instrument spectral response to the ability to detect plasma of differing temperatures, one can calculate a temperature response curve. The temperature response curve is the signal expected in an instrument from the plasma photon emission. This response curve is built as a function of plasma temperature iteratively, by using many isothermal emission models of differing temperatures. The temperature response curve is generated by using a spectral synthesis model to create an X-ray emission profile from physical parameters (temperature, density, plasma emission measure, elemental abundance, etc.) vs. photon energy, folding this through the MinXSS instrument response and totaling the counts over a specified number of energy bins (creating an effective energy bin width for this model) for X123 and the estimated fC for XP. In general, the spectral emission model used is computed for a range of isothermal plasma temperatures. This grid of input temperatures leads to a grid of MinXSS instrument counts per isothermal temperature.  Equation~\ref{Eq-XP_Temperature}, Equation~\ref{Eq-X123_Temperature}, and Equation~\ref{Eq-X123_Temperature_Upsilon} 
show the functional form of the calculation to compute the temperature responses for XP and X123 respectively.

   \begin{equation}  \label{Eq-XP_Temperature}  
    F(T)_{XP} = G_{XP} \int_{0}^{\infty} \bigg[ \int^{\Omega_{\odot}} S(E_{ph}, \Omega, T) A_{XP} R_{XP}(E_{ph}, \Omega) d\Omega \bigg] dE_{ph}.
   \end{equation}

  \begin{equation}  \label{Eq-X123_Temperature} 
 F(T)_{X123 \ bin,j} = \int_{E_{min,j}}^{E_{max,j}} \bigg[ \Upsilon(E_{det}, T) \bigg] dE_{det}
  \end{equation}

  \begin{equation}  \label{Eq-X123_Temperature_Upsilon} 
\Upsilon(E_{det}, T) =  \int_{0}^{\infty} \int^{\Omega_{\odot}} S(E_{ph}, \Omega, T) A_{X123}	\overline{\rm \Re}_{X123} (E_{ph}, \Omega, E_{det}) d\Omega dE_{ph} 
  \end{equation}

\begin{figure}    
   \centerline{\includegraphics[width=1.0\textwidth,clip=]{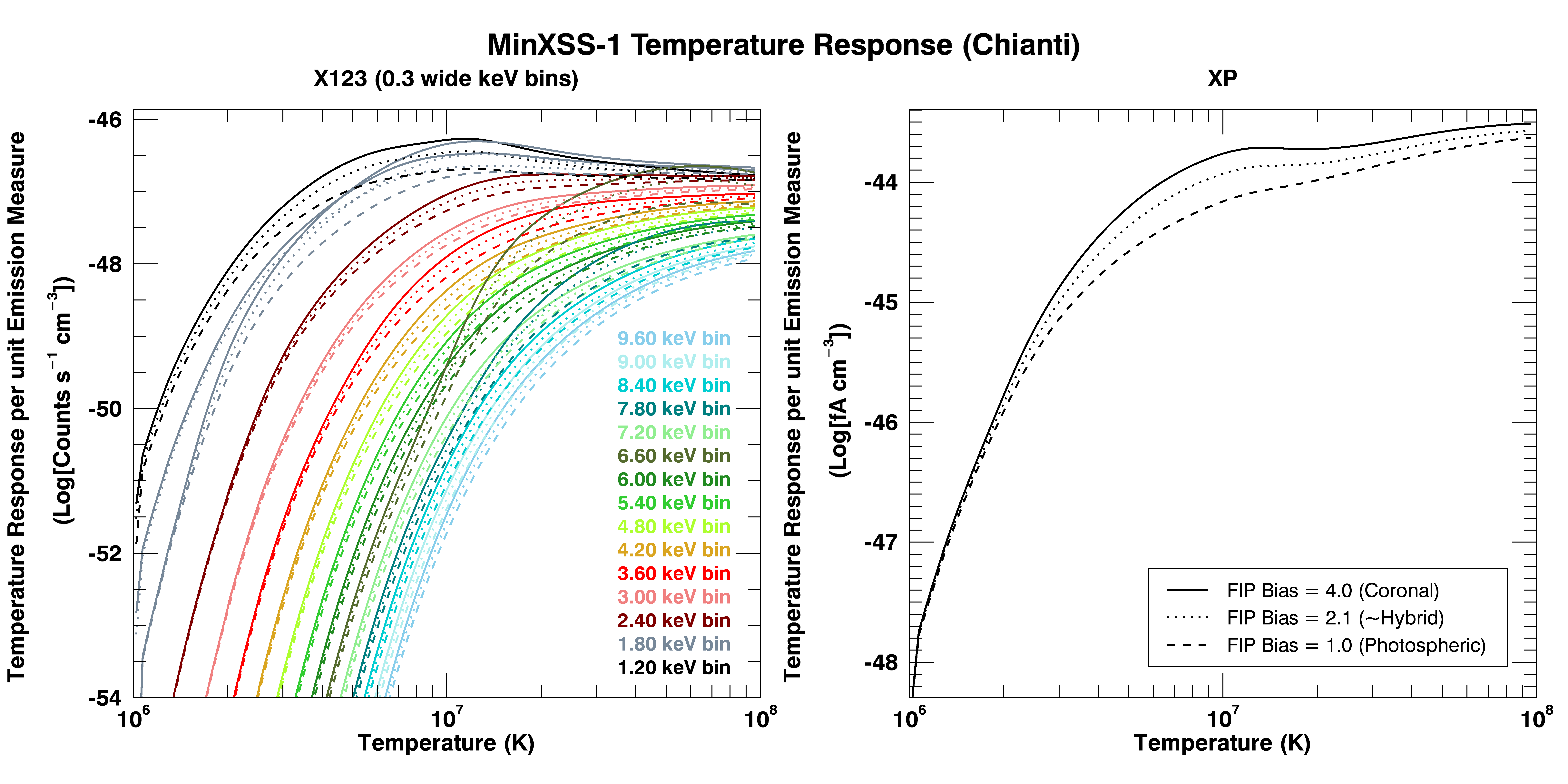}
              }
              \caption{An example of the X123 and XP temperature response functions for a spectrum summed to 0.3 keV wide bins for X123 (ten 0.03 native bins). The temperature response is in volume emission measure units of cm$^{-3}$. The isothermal spectral emission model used to compute the spectral response of the MinXSS instruments per plasma temperature is a spectrally extended version of the SolarSoftware (SSW) f$\_$vth function (which uses the Chianti Atomic Database). The temperature response in soft X-rays can vary due to differences in the abundance of the low-FIP elements of Fe, Mg, Si, Ca and the mid-FIP element S used in the spectral emission model. Thus, we display the temperature response for traditional `Coronal' \citep{Feldman1992} (4 times photospheric for the low-FIP elements), `Hybrid' \citep{Schmelz2012} ($\sim$2.1 times photospheric), and one of the latest photospheric \citep{Caffau2011} abundances. The MinXSS instrument temperature response begins to deviate for different abundances for plasma temperatures greater than 2 MK, due to the ions of the low-FIP elements. 
                      }
   \label{Figure7}
   \end{figure}

   An example of the MinXSS XP and X123 temperature response curves are in Figure~\ref{Figure7}. There are differences in the temperature response function depending on the abundances used in spectral emission model for the soft X-rays are primarily due to the variance in the low first ionization potential (low-FIP) elements of Fe, Mg, Si, Ca and the mid-FIP element S. Elements with a first ionization potential less than 10 eV have been measured to be overabundant with respect to the high-FIP elements in the solar corona when compared to photospheric values. This has become known as the FIP effect in the Sun. Summaries of the variations in Solar abundance are given by \cite{Laming2015} and \cite{Schmelz2012}. Thus, we calculate the temperature response for a range of abundances and display the abundance values corresponding to `common' reference values in literature. Results displayed are for traditional `Coronal' \citep{Feldman1992} (4 times photospheric for the low-FIP elements), `Hybrid' \citep{Schmelz2012} ($\sim$2.1 times photospheric), and one of the latest photosphere \citep{Caffau2011} abundances. The MinXSS instrument temperature response begins to deviate for plasma temperatures greater than 2 MK, primarily due to the ions of the low-FIP elements.

 Figure~\ref{Figure7} demonstrates the temperature range over which MinXSS X123 and XP can reliably extract information. X123 and XP inferences are biased towards the plasma temperatures greater than 2 MK. The dominant emission for non-large-flaring-times is expected to be around 2 - 4 MK. Thus, MinXSS will be able to infer non-large-flaring Sun properties between 1.5 - $\sim$4 MK with high confidence, but limited capabilities for temperatures below 1.5 MK. The temperature response is flattens for the X123 lower energy bins ($\leq$3 keV) for temperatures greater than $\sim$4.5 MK. Due to this flat nature and with limited significant counts from energy bins greater than 3 keV (which is not expected from the X123 relatively small effective area), X123 can only set upper limits on the emission measure, but cannot definitively constrain the temperature values for plasma hotter than $\sim$5 MK during non-large-flaring times. Plasma temperatures above $\sim$5 MK are expected for C, M and X class GOES flares. The higher energy bins ($\geq$ 3 keV) are mostly sensitive to plasma temperatures greater than $\sim$5 MK, but need substantial photon flux for statistically significant signals. All these attributes demonstrate that MinXSS has the greatest diagnostic capability for large flares on the Sun.

     \subsection{Linearity of Response} 
  \label{Ss-Linearity}
     
\begin{figure}    
   \centerline{\includegraphics[width=1.0\textwidth,clip=]{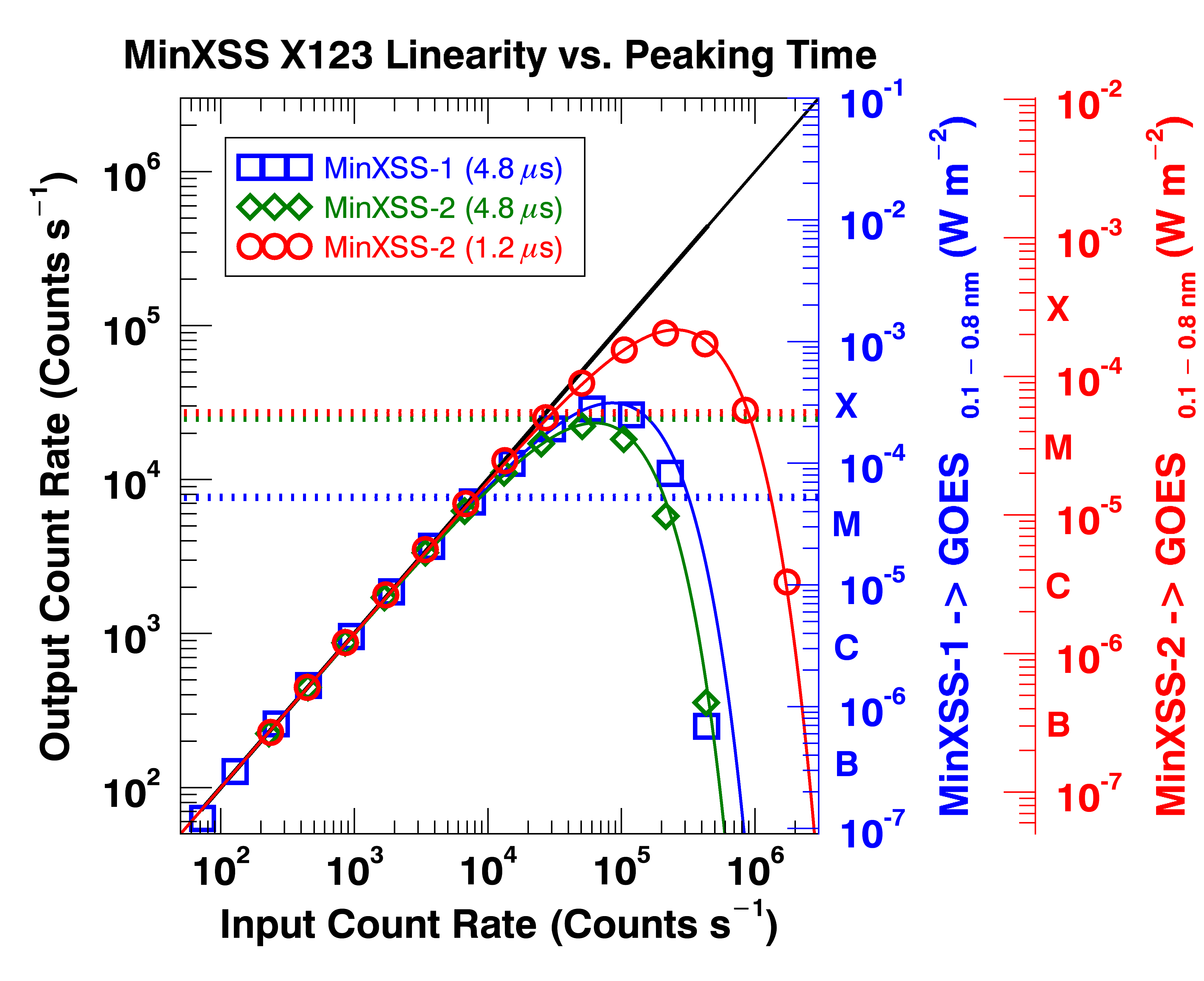}
              }
              \caption{X123 detected (output) count rate for an input (actual) count rate, for the MinXSS-1 (blue symbols) and MinXSS-2 (green and red symbols) expected operating peaking times. The lines indicates the dead-time model fit. Comparison of a MinXSS-1 observation based scaling to GOES flux levels (blue vertical axis) and model estimations for MinXSS-2 (red vertical axis). The horizontal dotted lines indicate the count rate where dead-time effects, pile-up effects and detector paralysis begin to occur. Spectra above these count rates must be heavily processed prior to analysis.
                      }
   \label{Figure8}
   \end{figure}

    	It is desired to have a linear response vs. light source intensity levels for the MinXSS instruments. The linearity of response for XP and X123 were assessed at NIST SURF plus early data from the MinXSS-1 mission and the data for the former are in Figure~\ref{Figure8}. The count rates discussed here are the measured counts summed over the spectrum during an accumulation divided by the accumulation time. MinXSS data products contain many types of time parameters, but it is best to use the actual measured count rates (counts per second; cps) to deduce the severity of dead-time and pile-up effects vs. directly using the time parameters in the data sets. The X123 spectral data can be corrected for dead-time losses up until a maximum input count rate for the slow counter, ${[{C}_{s}]}_{max}$, which depends on the slow counter peaking time, ${\tau}_{s}$. The slow counter peaking time directly effects the slow counter dead-time, ${\tau}_{ds}$, via the relation ${\tau}_{ds}$ = $B({\tau}_{s} + {\tau}_{flat})$, where B is a constant (over the mission) per detector, and ${\tau}_{flat}$ is the trapezoidal shaping flat top time. The maximum input count rate that the slow counter can be corrected for is ${[{C}_{s}]}_{max} = \textstyle \frac{1}{{\tau}_{ds}}$. This corresponds to $\sim$85 000 cps for MinXSS-1 (4.8 $\mu$s peaking time) and $\sim$255 000 cps for MinXSS-2 (1.2 $\mu$s peaking time). Below, we discuss the dead-time correction process with the NIST SURF data and the on-orbit specific correction process.

    	The MinXSS-1 X123 has a JFET preamplifier and MinXSS-2 with the X123 Fast SDD unit has a MOSFET preamplifier, allowing for lower noise and improved utility for shorter peaking times.  The deduced non-dead-time corrected maximum count rate for MinXSS-1 is $\sim$8 000 cps for its chosen 4.8 $\mu$s peaking time. The deduced non-dead-time corrected maximum count rate for MinXSS-2 is $\sim$7 000 cps for the 4.8 $\mu$s peaking time and $\sim$27 000 cps for the 1.2 $\mu$s peaking time. The latter will be the MinXSS-2 nominal operational peaking time. The horizontal thick dotted lines in Figure~\ref{Figure8} display these non-dead-time corrected spectrally integrated maximum count rates. These listed count rate values (${C}_{s}$) are from the slow counter, which creates the X123 spectrum. The `input' count rate (${C}_{in}$) is determined from the measured fast counter count rate (${C}_{f}$) corrected for fast counter dead-time (${\tau}_{df}$). The fast counter has a shorter peaking time of 100 ns (${\tau}_{pf}$) and an effective pair resolving time of $\sim$120 ns (${\tau}_{pair}$ = ${\tau}_{df}$ = 120 ns). The fast channel is used to determine if each event is `valid', if the X123 slow channel and hence the Digital Pulse Processor should include it in the spectrum. This helps minimize photon peak pile-up, where more than one photon is absorbed by the detector within the peaking time and the event is recorded as the sum of the photon energies. As a consequence of the much shorter peaking time, the fast channel has much lower spectral resolution and thus is not the preferred channel for accurate spectra accumulation. 
    	
    	All photon-counting X-ray detectors exhibit some form of count rate loss due to dead-time \citep{knoll2010}. At high count rates losses due to dead-time can become significant and expressions to approximate the true count rates for X123 are provided by \cite{Redus2008}. The linear black solid line in Figure~\ref{Figure8} displays the relation between the input and dead-time corrected fast counter output count rates. The dead-time correction for the fast counter count rate follows the non-paralyzable model, ${C}_{in} = \frac{{C}_{f}}{(1 - {C}_{f}{\tau}_{df})}$. ${C}_{in}$ resembles the `true' input count rate. This expression is accurate to within 5\% for true count rates $\leq$ 500 000 cps. For the NIST SURF calibrations we calculate ${C}_{in}$ directly from the fast counter and use ${C}_{model} =  {C}_{in}{e}^{-{C}_{in}{\tau}_{ds}}$, the paralyzable model, to directly estimate the count rate in the slow counter (${C}_{s}$). The expression for ${C}_{model}$ assumes a Poisson arrival probability for photons and registered events. The blue (MinXSS-1, ${\tau}_{s}$ = 4.8 $\mu$s) , green (MinXSS-2, ${\tau}_{s}$ = 4.8 $\mu$s) and red (MinXSS-2, ${\tau}_{s}$ = 1.2 $\mu$s) solid lines in Figure~\ref{Figure8} are the calculated dead-time suppressed slow counts (${C}_{model}$). The corresponding measured slow counts (${C}_{s}$) are the symbols.
    	
    	Overall, these predictions agree with the measured data until pile-up effects cause the measured count rate to lie below the calculations. The resultant pile-up effects depend on the shape of the photon flux spectrum, but in general, will be noticeable for input count rate values greater than the peak modeled count rate distribution ${C}_{model}$ (the solid lines in Figure~\ref{Figure8}). This occurs for ${C}_{in} > \textstyle \frac{1}{{\tau}_{ds}}$. ${\tau}_{ds}$ will be included in the MinXSS processing software release. Thus, in theory, if one has quality fast and slow counter measurements during an observation the true count rate can be deduced to within $5\%$ until 500 000 cps. Unfortunately, this NIST SURF procedure to correct the slow counter for dead-time effects is not directly applicable on-orbit because (1) the low energy thresholds for the fast and the slow counter may not be (and currently are not) at exactly the same energy, and (2) on-orbit the  MinXSS-1 fast counter has exhibited high noise (radio beacons, reaction wheel momentum changes, etc.) and thus cannot be used quantitatively, only qualitatively. MinXSS-2 ground testing does not exhibit the same noise characteristics and should have reduced fast counter noise on-orbit.
    	
    	As an alternative to estimate the true spectrum count rate on-orbit one can use the merit function in Equation~\ref{Eq-X123_Dead_Time_Correction}, $\mathsterling$, where n is the variable to estimate the spectrally integrated slow counter count rate without dead-time depression, in units of cps. The minimum of $\mathsterling$ with the restriction that, ${C}_{s} < n < \textstyle \frac{1}{{\tau}_{ds}}$ should yield the input count rate (${n}_{min}$), or `true' count rate that best estimates the measured spectrally integrated slow count rate, ${C}_{s}$. It is apparent from Figure~\ref{Figure8} that the model function ${C}_{model}$, is not monotonic nor uniquely defined for input count rates in the interval $0 < n < \infty$. Thus, to obtain a feasible result one must restrict the search domain to be between ${C}_{s} < n < \textstyle \frac{1}{{\tau}_{ds}}$. For $n > \textstyle \frac{1}{{\tau}_{ds}}$ other effects like pile-up must also be corrected for. Additionally, the quantity of the difference between the observed count rate and model count rate is squared to ensure positive concavity and thus the realization that ${n}_{min}$ will be the best fit result. The ratio of the best fit value to the actual spectrum summed count rate yields a correction factor P, (P = $\frac{{n}_{min}}{{C}_{s}}$) that can be multiplied by each binned count rate in the spectrum to adjust for dead-time during the respective integration. Using this technique, the MinXSS data utility range can be extended to counts rates determined by ${C}_{s} < \textstyle \frac{1}{{\tau}_{ds}}$. As a proxy to indicate when these correction will be needed, correlated XP data or the GOES XRS flux levels can be used.

      \begin{equation}  \label{Eq-X123_Dead_Time_Correction} 
 \begin{split}
 {\Big[ \mathsterling({C}_{s} < n < \textstyle \frac{1}{{\tau}_{ds}}, {C}_{s}, {\tau}_{ds}) \Big]}_{min}  & =  { \Big[ {\Big[{C}_{s} - {C}_{model}({C}_{s}< n < \textstyle \frac{1}{{\tau}_{ds}}, {\tau}_{ds}) \Big]}^{2} \Big]}_{min} \\
 & = { \Big[ {\Big[{C}_{s} - n{e}^{-n{\tau}_{ds}}\Big]}^{2} \Big]}_{min}
 \end{split}
  \end{equation}

 Early MinXSS-1 data from different GOES levels has been used to estimate the blue vertical axis in Figure~\ref{Figure8}, which roughly relates the GOES 0.1 - 0.8 nm (W m$^{-2}$) flux levels to the spectrally integrated MinXSS counts. The X123 counts are totaled across the {\it entire} operational spectral bins (E$_{ph} \gtrsim $ 0.8 keV,) and not limited to the corresponding GOES 0.1 - 0.8 nm band (1.55 - 12.4 keV). This relation is to serve as a general guide of what count rates one would expect for specific GOES levels and can plan for X123 effects as necessary (photon peak pile-up, dead-time, etc.). The MinXSS-2 estimates are {\it modeled} counts based on the response functions and process described in Section~\ref{Ss-Efficiency}, in Equation~\ref{Eq-X123_Signal}, where the photon source term, $S$ is generated from inverting the MinXSS-1 measured count spectrum. These rough estimates are displayed as the red vertical axis in Figure~\ref{Figure8}. Examples of the count rate, and estimated photon flux as a function of GOES levels from the early aspects of the MinXSS-1 mission are discussed in Section~\ref{S-MinXSS_GOES}.

\subsection{MinXSS Data Products} 
  \label{Ss-MinXSS_Data}
  
  The publicly available MinXSS data can be accessed on the MinXSS website \url{http://lasp.colorado.edu/home/minxss/}. These data will include Level0c, Level0d, Level1 - 5 products with one of the most relevant products being the Level1 spectral irradiance (photons s$^{-1}$ cm$^{-2}$ keV$^{-1}$). Measured count rates, spacecraft position, Sun-Earth distance, pointing information, etc. are also included. MinXSS data processing software to convert raw data to science quality data will be incorporated into SSW soon. 
              

\section{MinXSS-1 Solar Measurements - GOES A5 - M5 Levels} 
\label{S-MinXSS_GOES} 

  The MinXSS X123 spectrometer prototype was space-flight verified on two NASA Sounding Rocket flights for the calibration of the {\it SDO Extreme Variability Experiment} (EVE) \citep{Woods2012} and returned high quality science data for two 5 minute periods \citep{Caspi2015}. Thus, we have confidence in the MinXSS CubeSat versions of the X123 spectrometers to return high quality data. This was reaffirmed with the first few months of data downlinked from the MinXSS-1 mission. Over the early phases of the mission the X-ray flux has been as low as GOES $\sim$A5 and as high as GOES M5.0 during a flare. The times of the corresponding GOES levels are:
  
\begin{itemize}
\item {\bf A5} from 2016 June 29 10:29:32 - 2016 July 01 22:55:53,
\end{itemize}

\begin{itemize}
\item {\bf B5} from 2016 July 23 01:15:05 - 01:39:45,
\end{itemize}

\begin{itemize}
\item {\bf C2.7} from 2016 July 08 - 00:55:04 - 00:58:44,
\end{itemize}

\begin{itemize}
\item {\bf M1.2} from 2016 July 21 - 01:50:01 - 01:53:31,
\end{itemize}

\begin{itemize}
\item {\bf M5.0} from 2016 July 23 -  02:10:05 - 02:13:46
\end{itemize}

    These data have been filtered and only data that pass our `science quality' check (minimal background levels, particle events, non-SAA times, etc.) are analyzed below. These filter checks also isolate eclipse time data with only thermal noise apparent in the spectrum which result in $\sim$ 2 - 5 cps across the entire spectrum. Other external (astronomical) soft-X-ray flux contributions to the MinXSS X123 background are negligible due to the small instrument aperture. The solar flare times are centered on the flare peak total count rate in the MinXSS X123 spectrum. These solar fluxes have all been corrected for dead-time losses using Equation~\ref{Eq-X123_Dead_Time_Correction} and result in MinXSS-1 X123 count rate levels of 26 - 9 100 cps. The dead-time losses were $\sim$11$\%$ for the M5.0 flare, $\sim$3$\%$ for the M1.2 flare, $\sim$0.7$\%$ for the C2.7 flare, and less than $\sim$0.5$\%$ for B GOES levels and lower.
This demonstrates the MinXSS spectrometer capability to cover a wide range of solar X-ray flux levels. XP responds identically to the increasing GOES flux, with background subtracted signals in the range of 801 - 616 928 fC. This provides a confirmation of XP and X123 nominal operation.
    
        \begin{figure}    
   \centerline{\includegraphics[width=1.0\textwidth,clip=]{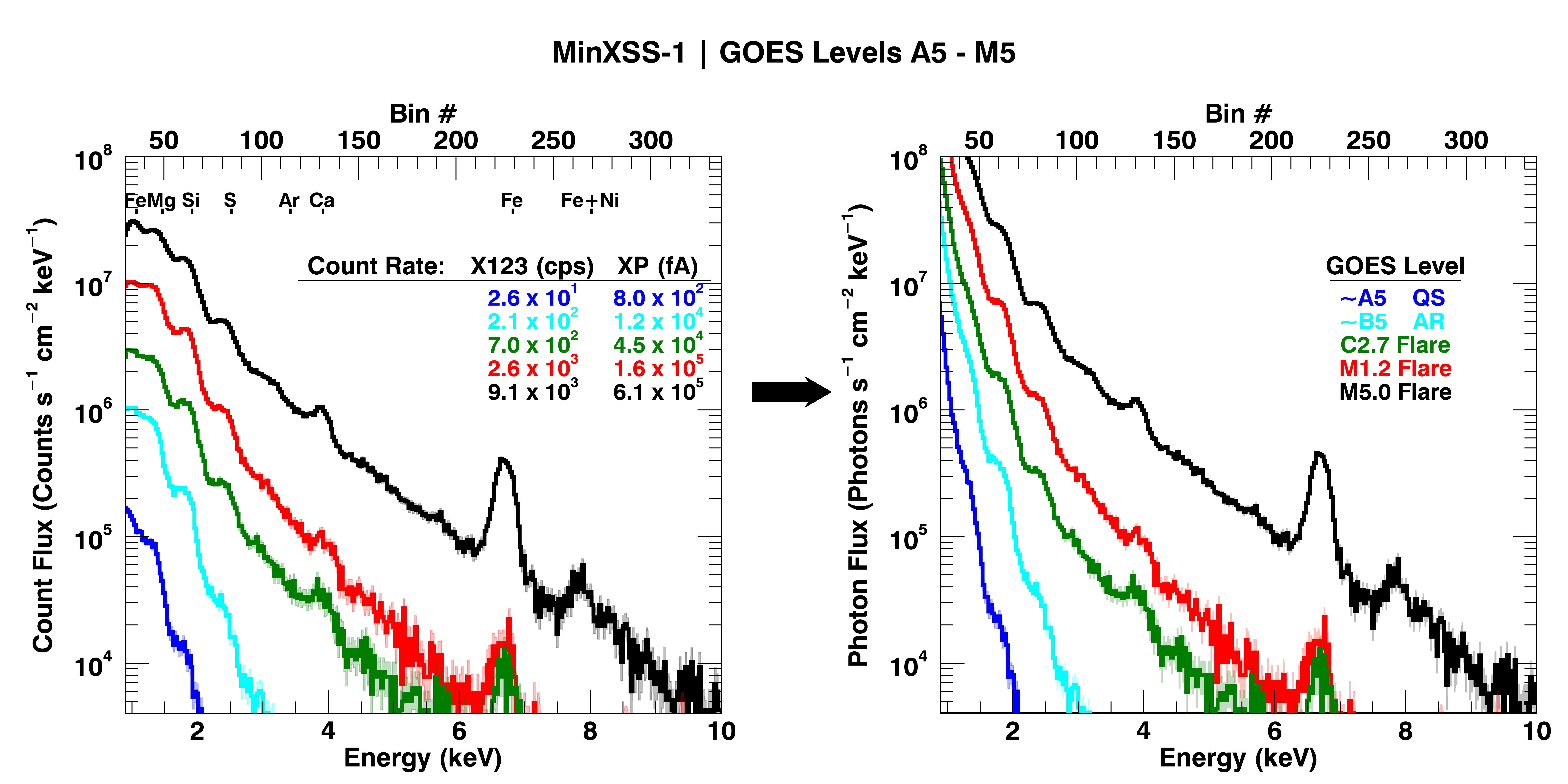}
              }
              \caption{MinXSS-1 X123 solar measurements (solid lines) from GOES A5 to M5 levels ($\sim$5 $\times$ 10$^{-8}$ - 5 $\times$ 10$^{-5}$ W m$^{-2}$). The left plot is the mean count flux and the right plot is the estimated photon flux. The uncertainties are depicted as the shaded region around the measurements. This demonstrates the dynamic range of the MinXSS-1 spectrometer, and the variation in spectral features for increasing solar flux levels. The `bumps' in the spectrum are due to groups of dominant emission lines from ionized Fe near 1.2 keV and 6.7 keV, Mg near 1.7 keV, Si around 2.1 keV, S by 2.7 keV, Ar (or lack thereof) near 3 keV, Ca by 4 keV, and the Fe+Ni complex at 8 keV. These features can be used as elemental abundance probes to assess deviations from the traditional `Coronal' abundance values during various solar conditions.
                      }
   \label{Figure9}
   \end{figure}

   Due to the X123 noise sources mentioned earlier, Be window thickness and uncertainties in the photon inversion process, and to the off-diagonal elements of the X123 response, the lower end of the MinXSS-1 X123 spectra valid for scientific analysis is $\sim$0.8 keV. The sharply decreasing solar flux for the higher energies limit and the MinXSS small aperture result in an effective high energy limit of $\sim$12 keV. Even for flares as large as M5.0, the flux is not large enough to produce a statistically significant count rate above 12 keV before effects such as detector dead-time and pile-up hinder the accuracy of the spectra. Thus, our MinXSS-1 X123 has an effective solar flux energy range of 0.8 - 12 keV and should return quality data up to low X-class flares if corrected for dead-time and pile-up effects. The MinXSS-2 X123 Fast SDD spectrometer with a nominal slow channel peaking time of 1.2 $\mu$s, is four times faster than the nominal 4.8 $\mu$s peaking time, thus one would expect roughly four times the count rate before inaccurate spectra. But the MinXSS-1 Be window is much thicker than the MinXSS-2 window (24.5 vs. 11.2 $\mu$m), making comparisons non-linear. Early estimates put the MinXSS-2 X123 maximum GOES level at X-class solar flares. Future data will reveal the full relation, keeping in mind that the axis on Figure~\ref{Figure8} are rough estimates and that the spectral distribution of photons is non-linear vs. GOES flux levels.

  \begin{table} 
\caption{MinXSS-1 count rate and photon energy flux of observations from GOES A5 - M5 levels. \newline Ratio$_{B5}$ is the count rate value of the corresponding row divided by the B5 count rate.}
  \rotatebox{90}{ 
\label{T-Table2}
\begin{tabular}{ccccccccccc}     
  \hline                   
  \hline
GOES Level & X123$_{measured}$ & Ratio$_{B5}$ X123 & X123 Total Photon Energy Flux & XP$_{measured}$ & Ratio$_{B5}$ XP & XP$_{X123-predicted}$ & 100*($\Delta_{XP}$/XP$_{measured}$)\\
  & (counts s$^{-1}$)  &  & $\geq$ 1 keV at 1 AU (W m$^{-2}$) &  (fA) &  & (fA) & \\
  \hline
QS $\sim$A5  & 2.63 $\times$ 10$^{1}$ & 1.25 $\times$ 10$^{-1}$ & 4.6 $\times$ 10$^{-6}$ & 8.01 $\times$ 10$^{2}$ & 6.67 $\times$ 10$^{-2}$ & 1.33 $\times$ 10$^{3}$ & -66.35\% \\
AR $\sim$B5  & 2.10 $\times$ 10$^{2}$ & 1.00 $\times$ 10$^{0}$ & 4.6 $\times$ 10$^{-5}$ & 1.20 $\times$ 10$^{4}$ & 1.00 $\times$ 10$^{0}$ & 1.15 $\times$ 10$^{4}$ & 4.2\% \\
C2.7 flare  & 7.01 $\times$ 10$^{2}$ & 3.33 $\times$ 10$^{0}$ & 1.5 $\times$ 10$^{-4}$ & 4.53 $\times$ 10$^{4}$ & 3.77 $\times$ 10$^{0}$ & 4.36 $\times$ 10$^{4}$ & 3.7\% \\
M1.2 flare & 2.60 $\times$ 10$^{3}$ & 1.23 $\times$ 10$^{1}$ & 5.5 $\times$ 10$^{-4}$ & 1.65 $\times$ 10$^{5}$ & 1.37 $\times$ 10$^{1}$ & 1.61 $\times$ 10$^{5}$ & 2.7\% \\
M5.0 flare & 9.1 $\times$ 10$^{3}$ & 4.38 $\times$ 10$^{1}$ & 2.1 $\times$ 10$^{-3}$ & 6.18 $\times$ 10$^{5}$ & 5.14 $\times$ 10$^{1}$ &  6.16 $\times$ 10$^{5}$ & 0.3\% \\
  \hline
  \hline
\end{tabular}
} 
\end{table}

The spectral photon flux estimates in Figure~\ref{Figure9} demonstrate the drastic change in the soft-X-ray spectra. The X123 soft X-ray spectrum measured change by orders of magnitude over a few GOES level changes. Additionally, the XP fA signal (converted from the measured DN signal) scales with the GOES flux levels of A5 to M5. This provides a consistency check for the X123 spectral signal. While the qualitative nature of this change is not new, the quantitative determination of the magnitude of this change is relevant. Table~\ref{T-Table2} lists the X123 and XP count rates as a function of GOES class and is plotted in Figure~\ref{Figure9}. The measured XP signal was compared to the X123 estimated XP signal, by taking the X123 estimated photon flux and computed the XP signal from this flux. The resultant `X123 modeled' XP signal is then compared to the measured XP signal. The XP measured and `X123 modeled' XP signal agree to within $\sim$4$\%$ except for the GOES A5 measurement. At low GOES levels the XP signal becomes comparable to the thermal photodiode noise and leads to underestimated signals. This is apparent in the QS emission measure inferences discussed in detail in Section~\ref{Ss-EM_Loci}.

A particularly interesting feature in the X123 spectrum is the presence of two `humps' near 1.7 keV due to Mg XI and Mg XII line groups, and 2.1 keV from Si XIII and Si XIV lines. These hump features persist for all GOES levels and provide useful diagnostics for element abundance estimations. The Fe XXIV and Fe XXV line complex near 6.7 keV is prominent for the GOES flares of levels C2.7 and higher. This line complex is well suited to estimate the Fe abundance modification during solar flares vs. the QS (non-large-flaring Sun). Only for the brightest flares (M5.0 and M1.2 in this paper) is the Fe-Ni complex near 8.0 keV pronounced and suitable for analysis. For smaller flares the signal is not statistically significant to infer physical properties from this feature. These measurements demonstrate the MinXSS dynamic range of its solar measurements. 

  \subsection{Soft X-ray Energy Content and Correlation to GOES XRS} 
\label{Ss-SXR_Energy}  
      
\begin{figure}    
   \centerline{\includegraphics[width=1.0\textwidth,clip=]{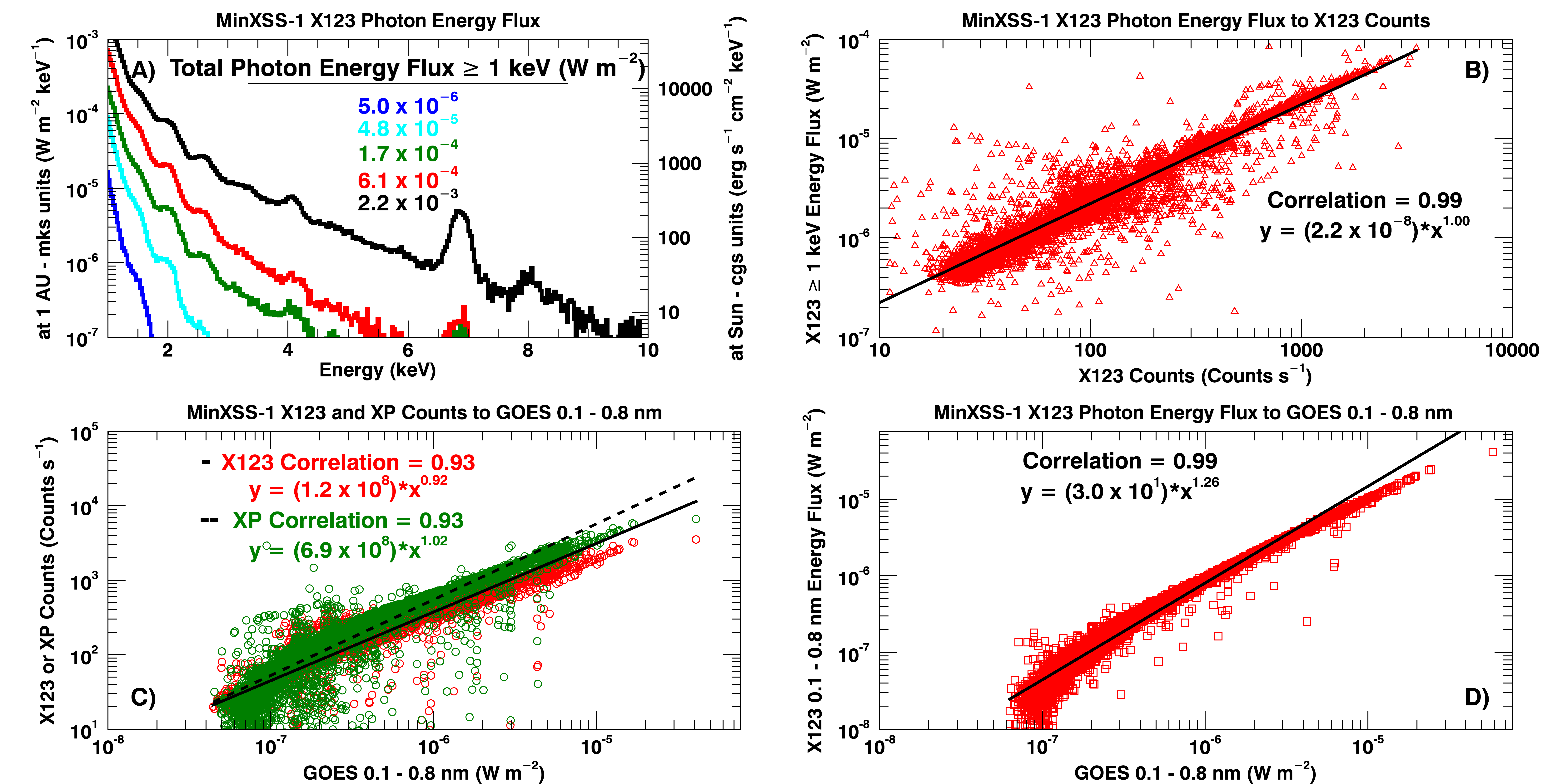}
              }
              \caption{Panel A shows the MinXSS-1 X123 derived photon energy flux at 1 AU (in mks units - W m$^{-2}$ keV$^{-1}$) and scaled back to the Solar surface (in cgs units - erg s$^{-1}$ cm$^{-2}$ keV$^{-1}$). The total energy flux at 1 AU as measured by MinXSS-1 for photon energies $\geq$ 1 keV is listed for the GOES $\sim$A5 (blue), B5 (cyan), C2.7 (green), M1.2 (red) and M5.0 (black) class observations. Panels B - D contain scatter plots, correlation coefficients and linear fit of MinXSS-1 X123 photon energy $\geq$ 1 keV to count rate (Panel B), MinXSS-1 X123/XP count rate to GOES 0.1 - 0.8 nm flux (Panel C) and MinXSS-1 X123 photon energy flux integrated from 0.1 - 0.8 nm ($\sim$1.55 - 12.4 keV) to GOES XRS 0.1 - 0.8 nm flux (Panel D) all show very strong correlations, validating the MinXSS data.  
                  }
   \label{Figure10}
   \end{figure}

         A main benefit of conducting routine spectrally resolved soft X-ray measurements is the ability to calculate and track the spectral distribution of energy flux. This information is important for models of Earth's atmospheric response to solar radiative forcing, solar flare analyses and comparison of the solar X-ray flux to other stars. Table~\ref{T-Table2} lists the integrated soft X-ray energy flux (multiplying the energy flux by the X123 bin width and summing over photon energy bins) above 1 keV estimated at 1 AU from MinXSS-1 X123 measurements for the main five GOES levels. This information is also in Panel A of Figure~\ref{Figure10}.  Panel B displays the fitted power law relationship, y = a(x$^{b}$), between the MinXSS-1 X123 total counts and the total radiative energy greater than 1 keV energy, which is not quite linear. The data in the scatter plots (Panels B, C and D) are from 2016 June 10 - 2016 November 30. The coefficient of 2.2 $\times$ 10$^{-8}$ J counts$^{-1}$ m$^{-2}$ can be used to obtain a rough estimate of the energy content from the X123 total counts alone. A large portion of this energy (up to an order of magnitude) resides below 2 keV and this is apparent when comparing to the corrected (divided by 0.7) GOES XRS 0.1 - 0.8 nm flux, see Panel A in Figure~\ref{Figure10}. 
         
    There is a correlation between the 0.1 - 0.8 nm ($\sim$1.55 - 12.4 keV) energy flux calculated from MinXSS-1 X123 and GOES XRS (Panel D of Figure~\ref{Figure10}). The XP count rate (DN s$^{-1}$) has a near linear relationship to the GOES XRS 0.1 - 0.8 nm flux, especially above GOES B1 levels and is useful as a proxy to the GOES XRS measurements. This is expected as XP tracks the total soft X-ray energy incident on the MinXSS-1 aperture (number of electron-hole pairs generated is proportional to E$_{ph}$). Exact linearity is not expected between X123 count rate and GOES XRS energy flux because X123 is photon counting. These results validate the dynamic response of the MinXSS-1 X123 and XP to the solar flux. The next section discusses the feasibility of extracting physical information from model fits of the MinXSS X123 data to estimate plasma temperature, density, emission measure and elemental abundances.

\subsection{Spectral Parametric Fits} 
\label{Ss-Parameter_Fits} 
    
\begin{figure}    
   \centerline{\includegraphics[width=1.0\textwidth,clip=]{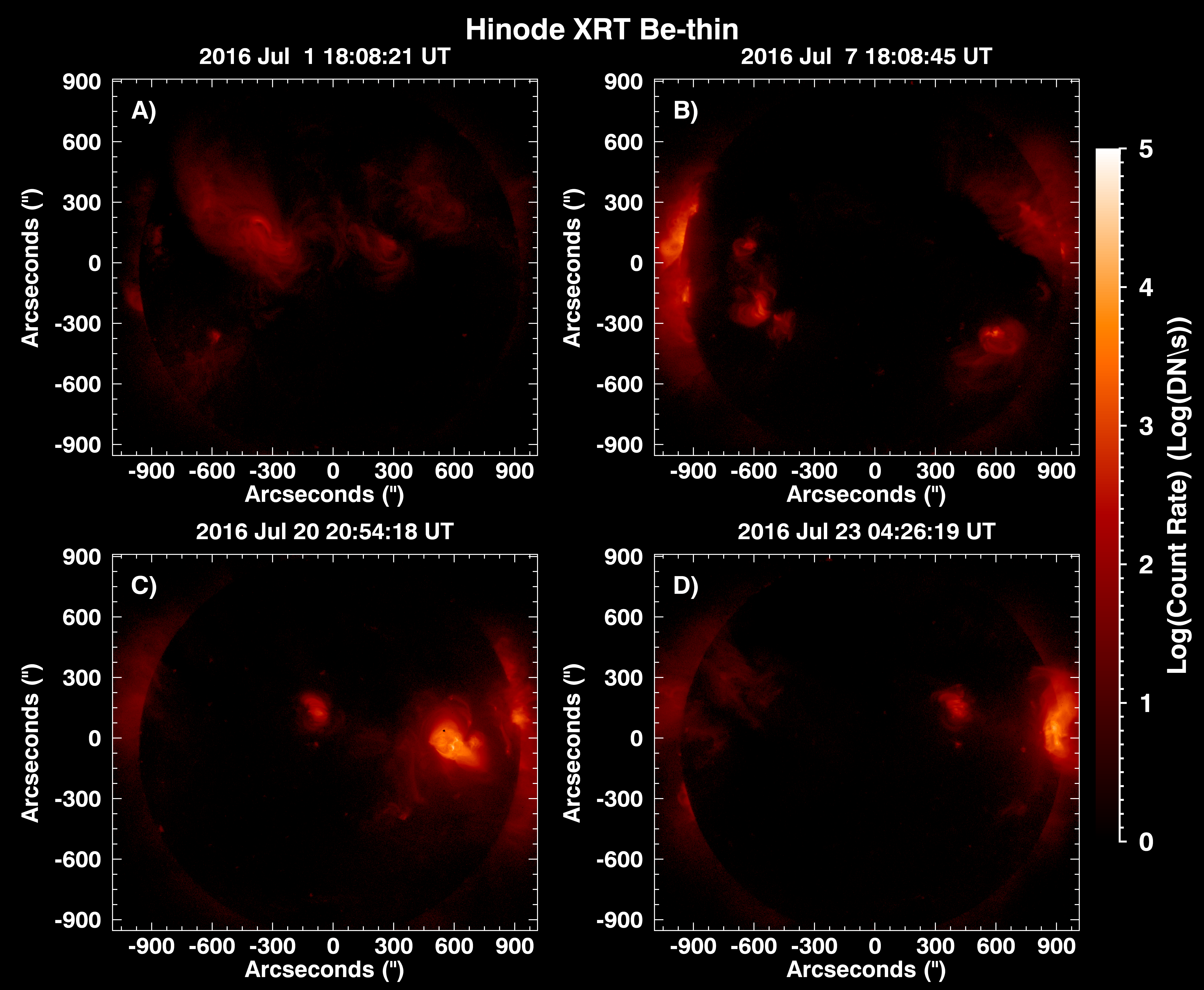}
              }
              \caption{Log-scaled count rate Hinode X-ray Telescope (XRT) Be-thin full Sun images near the time of the MinXSS-1 observed QS (Panel A), Pre-flare times for the C2.7 (Panel B), M1.2 (Panel C) and M5.0 (Panel D) flares. The XRT images provide information on the spatial distribution of the soft X-ray emitting plasma, since MinXSS measurements are integrated across the entire FOV.
                      }
   \label{Figure11}
   \end{figure}

    Hinode XRT Be-thin images are displayed in Figure~\ref{Figure11} to provide qualitative spatial context for the X-ray emission that MinXSS detects for non-large-flaring times. The XRT Be-thin serves as the closest XRT synoptic filter analog to the MinXSS spectral response. The four panels in Figure~\ref{Figure11} are the closest XRT full-Sun synoptic images in time to the QS (Panel A), pre-flare C2.7 (Panel B), pre-flare M1.2 (Panel C) and pre-flare M5.0 (Panel D) times. It is clear that active regions are present, their emission dominate the MinXSS count rates (due to the hotter plasma content). We are currently working to cross-calibrate the two instruments to develop quantitative relationships. With this knowledge we proceed to demonstrate an example of how MinXSS data can be used to extract physical information on the X-ray emitting regions. 
    
The unique spectrally resolved measurements of MinXSS are suitable for parametric fits of single temperature (1T), two temperature (2T), multi-temperature (multi-T) and differential emission measure (DEM) models to estimate the plasma conditions. QS, AR and flare data over the possible total 6 year MinXSS mission combined with data from the observatories mentioned in Section~\ref{Ss-Current_Measurements} can be used to address current questions in solar physics. We have performed a series of parametric fits using the OSPEX (\url{https://hesperia.gsfc.nasa.gov/ssw/packages/spex/doc/ospex_explanation.htm}) programming suite in Solar Software, which utilizes the Chianti Atomic Database \citep{DelZanna2015, Young2016}, on the seven sets (includes pre-flare times) of data described earlier from GOES A5 to M5 levels. The uncertainties calculated on the fit parameters in this paper are basic OSPEX returned fits uncertainties, based on the curvature matrix, which assumes that the curvature has a local Gaussian shape. 

\begin{figure}    
   \centerline{\includegraphics[width=1.0\textwidth,clip=]{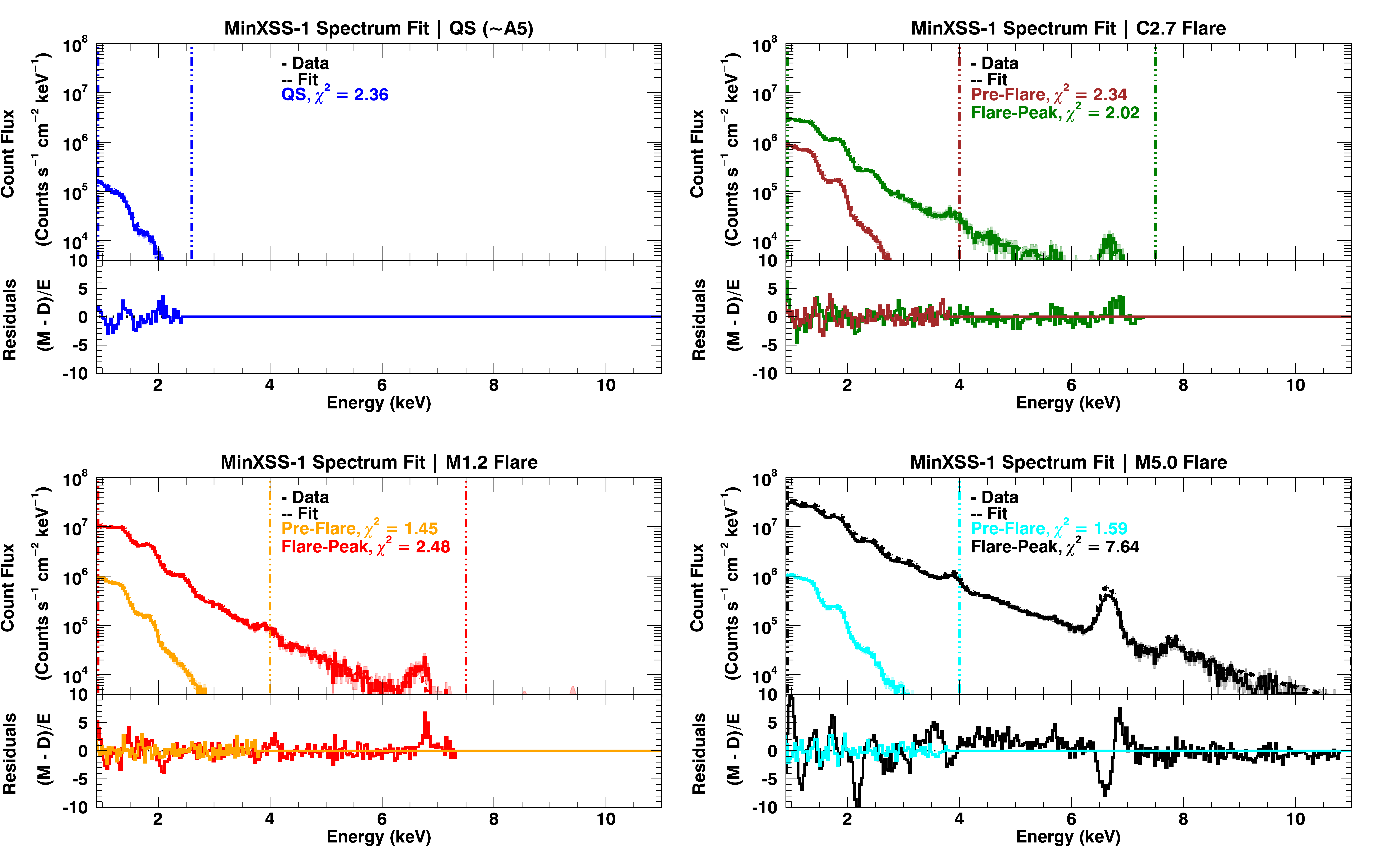}
              }
              \caption{MinXSS-1 X123 count flux solar measurements (solid lines) with the best fit spectra overlaid (dashed lines), for temperature and emission measures derived using the OSPEX suite. The residuals are listed also (M = model, D = data, and E = uncertainty). The shaded regions indicate the uncertainties in the count flux. A 2T model with select elemental abundance fit separately (2T-AllFree). The best fit parameters with their uncertainties are listed in Table~\ref{T-Table4} and Table~\ref{T-Table5}. There is a 2T model used for non-large-flaring times (QS and pre-flare) and an additional 2T model is added to compensate for the radiative enhancement during the flare-peak times. The vertical dash-dot-dot-dot lines show the high and low  energy limits for the spectral fits. 
                      }
   \label{Figure12}
   \end{figure}

We have used four models to fit the respective observations for comparisons between models. The best fit values are listed in Table~\ref{T-Table3}, Table~\ref{T-Table4}, Table~\ref{T-Table5},  Figure~\ref{Figure12}, Figure~\ref{Figure13} and Figure~\ref{Figure14}. In this analysis we denote a FIP-Bias value of 1 equal to photospheric which would have values similar to those in \cite{Caffau2011} and FIP-Bias of 4 equal to the traditional `Coronal' abundance and are those of \cite{Feldman1992}. The first model is a simple one temperature, fixed coronal abundance (1T-Coronal; A$_{FIP}$ = 4) model, which did not fit the data suitably for any of the seven data sets. There were large discrepancies for the elemental features labeled in Figure~\ref{Figure9}. The second model is a single temperature model in which the abundance of low FIP elements is allowed to vary through a single multiplicative factor, called the FIP-Bias. This model is called 1T-Free, but it underestimates the pre-flare flux for photon energies greater than $\sim$2.5 keV. The same 1T-Free model did not fit the flare components well neither. For the times where there is not substantial counts for X123 energy bins $>$ 2.5 keV, like the QS flux (GOES $\sim$A5) the 1T-Free model provides an adequate fit. 

The third model is a two temperature component with a single multiplicative factor for the FIP-Bias (2T-Free). These fits consistently produce satisfactory results with reduced $\chi^{2}$ between 1 - 4, except for the M5.0 flare. All pre-flare spectra in this paper are fit over the times listed in Table~\ref{T-Table3} with a 2T-Free model. The flare-peak fits include an additional 2T-Free model that is fit to account for the additional radiation on top of the fixed model pre-flare values. In this methodology the fixed pre-flare model during the peak-flare times serves as a background estimate. Thus, the result is a separate FIP-Bias for the 2T-Free pre-flare and the flare-peak functions. Values near 4 for this fit class resemble traditional coronal values and values near 1 are photospheric.

The abundance models are sensitive to the `humps' from emission line groups of Fe, Mg, Si, S, Ar, Ca, and Ni (where applicable) around $\sim$1.2 6.7 and 8.1, 1.7, 2.1, 2.7, 3.0, 4.0, and 8.1 keV respectively. For the 2T-Free model, a FIP bias value of 3.48 is found for the QS spectra, for the pre-flare times values between 2.0 - 3.5 are obtained and lower values between $\sim$1 - 1.41 are obtained for the peak-flare spectral fits. It is clear that there is a difference in the estimated abundance from pre-flare to flare-peak and the lower abundance during the flare is consistent with recent literature \citep{Schmelz2012, Dennis2015, Woods2017}. The availability of the Fe and Fe+Ni complexes at 6.7 and 8.1 keV respectively, provide clear diagnostics for the abundance factors and in turn are weighted more heavily for the flare abundances. The lower abundance is also in line with the theory of plasma from the lower atmosphere flowing up to the higher layers of the atmosphere and radiating in X-rays and UV.

The forth model type is a two temperature component where each of the elements, Fe, Ca, S, Mg, Si, Ar and Ni are allowed to vary (2T-AllFree) as long as there are sufficient counts in the X123 energy bins for the respective element's line group features in the spectrum. It is important to note that the Ni abundance scale factor is coupled to the Fe abundance scale factor. Thus, the Ni abundance is not a true independent inference. This is heritage in the fitting routine for the Fe+Ni complex near 8.0 keV and does not strongly skew the fits. The 2T-AllFree temperature and emission results are in Table~\ref{T-Table4}, the elemental abundance results in Table~\ref{T-Table5}, the spectral fits in Figure~\ref{Figure12} and the comparison to em loci in Figure~\ref{Figure14}. The separate abundance values are in abundance ratio units of coronal/photospheric, where the coronal values are from \cite{Feldman1992} and the photospheric values are from \cite{Caffau2011}.

For the QS spectra, there are not enough counts for energies $\geqq$ 2.5 keV and not `strong enough' features for a consistent fit and thus the individual element abundance parameters are poorly constrained. But for the pre-flare spectra and flares statistically significant abundance values can be deduced. Again the trend is clear that there is a decrease in the abundance majority of the low-FIP elements (Fe, Si, Mg, Ni) during the flare-peak vs. the pre-flare fits. Additionally, the individual element abundance variation observed provides further evidence for a more complicated fractionation process than a simple single FIP-Bias scaling for all the low-FIP elements.  Similar conclusions about abundances being more complex than a FIP-Bias scaling have been expressed in recent studies, such as in \cite{Schmelz2012} and \cite{Dennis2015}.

This is further complicated by the postulated decrease in abundance from the nominal coronal/photospheric ratio of Ar for the three flares here and Ca for two of the three flares. The abundances of Ar and Ca have been of recent interest in Hinode EIS spectrum \citep{Doschek2015, DoschekWarren2016} and MinXSS can provide an additional diagnostic in investing any anomalous behavior. A more rigorous analysis of elemental abundance variations, solar flares, quiescent conditions and active region evolution comprising DEM fits and will occur in the future.

    \begin{table} 
\caption{MinXSS-1 2T-Free (one FIP-Bias scale factor) spectral fits of observations from \newline GOES A5 - M5 levels. The uncertainties in the fit parameters are in parenthesis.  \newline ** highlights that the pre-flare data inferred dimmer and hotter second component is near the \newline limit of the MinXSS plasma diagnostic capabilities and thus not as well constrained.}
\label{T-Table3}
  \rotatebox{90}{ 
\begin{tabular}{ccccccccccc}     
  \hline                   
  \hline
GOES Level & Observation Times &  EM$_{1}$  & T$_{1}$ & EM$_{2}$  & T$_{2}$ & FIP-Bias$_{12}$ \\
     &   & (${10}^{49}$ cm$^{-3}$)  & (MK) & (${10}^{49}$ cm$^{-3}$)  & (MK) &  (1 = photospheric) \\
     &  &   &  &   &  &  (4 = coronal) \\
  \hline
QS $\sim$A5  & 2016-Jun-29 10:29:32 - 2016-Jul-01 22:55:53 & 19.9 (9.2) & 1.17 (0.08) & 0.21 (0.07) & 2.43 (0.12) & 3.48 (0.46) \\
C2.7 PreFlare ($\sim$B3)  & 2016-Jul-07 23:41:03 - 2016-Jul-08 00:04:24 & 10.8 (0.75) & 1.70 (0.02) & *0.12 (0.01)* & *4.58 (0.08)* & 2.89 (0.10) \\
C2.7 flare & 2016-Jul-08 00:55:04 - 2016-Jul-08 00:58:44 & 1.21 (0.04) & 4.28 (0.07) & 0.10 (0.005) & 14.91 (0.27) & 1.41 (0.06) \\
M1.2 PreFlare ($\sim$B3)  & 2016-Jul-20 18:04:13 - 2016-Jul-20 18:28:53 & 8.73 (0.81) & 1.82 (0.04) & *0.13 (0.02)* & *4.68 (0.14)* & 2.17 (0.09) \\
M1.2 flare & 2016-Jul-21 01:50:01 - 2016-Jul-21 01:53:31 & 6.21 (0.11) & 4.05 (0.03) & 0.46 (0.01) & 12.97 (0.12) & 1.41 (0.03) \\
M5.0 PreFlare ($\sim$B5) & 2016-Jul-23 01:15:05 - 2016-Jul-23 01:39:45 & 7.68 (0.80) & 1.86 (0.05) & *0.18 (0.02)* & *4.80 (0.12)* & 2.21 (0.09) \\
M5.0 flare & 2016-Jul-23 02:10:05 - 2016-Jul-23 02:13:46 & 17.10 (0.13) & 4.86 (0.02) & 2.22 (0.02) & 19.67 (0.08) & 0.98 (0.01) \\
  \hline
  \hline
\end{tabular}
} 
\end{table}

    \begin{table} 
\caption{Temperature and emission measure values from MinXSS-1 2T-AllFree (separate elemental abundance scale factors) spectral fits of observations from GOES A5 - M5 levels and plotted in Figure~\ref{Figure12}. The best fit abundances are listed in Table~\ref{T-Table5}. The uncertainties in the fit parameters are in parenthesis. ** highlights that the pre-flare data inferred dimmer and hotter second component is near the limit of the MinXSS plasma diagnostic capabilities and thus not as well constrained.}
\label{T-Table4}
\begin{tabular}{ccccccccccc}     
  \hline                   
  \hline
GOES Level &  EM$_{1}$  & T$_{1}$ & EM$_{2}$  & T$_{2}$ \\
     &  (${10}^{49}$ cm$^{-3}$)  & (MK) & (${10}^{49}$ cm$^{-3}$)  & (MK) \\
  \hline
QS $\sim$A5 & 15.2 (0.21) & 1.93 (0.05) & -- & -- \\
C2.7 PreFlare ($\sim$B3) & 7.29 (0.59) & 1.91 (0.04) & *0.05 (0.01)* & *5.25 (0.23)* \\
C2.7 flare & 1.91 (0.12) & 3.82 (0.08) & 0.09 (0.006) & 15.92 (0.45) \\
M1.2 PreFlare ($\sim$B3) & 11.90 (2.5) & 1.69 (0.08) & *0.18 (0.03)* & *4.45 (0.16)* \\
M1.2 flare & 8.27 (0.41) & 3.61 (0.05) & 0.57 (0.02) & 12.38 (0.16) \\
M5.0 PreFlare ($\sim$B5) & 5.44 (0.78) & 2.04 (0.08) & *0.11 (0.03)* & *5.19 (0.26)* \\
M5.0 flare & 8.16 (0.11) & 7.18 (0.02) & 1.38 (0.02) & 20.78 (0.16) \\
  \hline
  \hline
\end{tabular}
\end{table}

       \begin{table} 
\caption{Separate abundance values are in abundance ratio units of coronal/photospheric, \newline where the coronal values are from \cite{Feldman1992} and the photospheric values are from \newline \cite{Caffau2011} from MinXSS-1 2T-AllFree spectral fits of observations from \newline GOES A5 - M5 levels that are plotted in Figure~\ref{Figure12}. Elemental abundances that were fixed \newline during fitting have a `fixed' in parenthesis in place of an uncertainty. These values were fixed during \newline fitting when there were not sufficient counts in the corresponding spectral feature to ascertain an  \newline abundance. The abundances of He, C, O, F, Ne, Na, Al and K were fixed at photospheric values. \newline The best fit temperatures and emission measures are listed in Table~\ref{T-Table4}.}
\label{T-Table5}
  \rotatebox{90}{ 
\begin{tabular}{ccccccccccc}     
  \hline                   
  \hline
Element & Fe & Ca & S & Mg & Si & Ar & Ni\tabnote{The Ni abundance was linked to the Fe abundance. So this is not a truly independent estimate of the Ni abundance.} \\
FIP (eV) & 7.90 & 6.11 & 10.36 & 7.65 & 8.15 & 15.76 & 7.64 \\
  \hline
QS $\sim$A5 & 0.06 (1.75) & 4.00 (fixed) & 1.29 (fixed) & 1.43 (0.18) & 3.90 (0.89) & 1.20 (fixed) & 0.06 (1.88) \\
C2.7 PreFlare ($\sim$B3) & 2.26 (0.28) & 4.00 (fixed) & 1.32 (0.64) & 2.60 (0.10) & 3.39 (0.20) & 1.20 (fixed) & 2.43 (0.30) \\
C2.7 flare & 0.78 (0.08) & 3.18 (0.47) & 1.11 (0.69) & 1.06 (0.07) & 1.38 (0.07) & 0.81 (0.28) & 0.84 (0.08) \\
M1.2 PreFlare ($\sim$B3) & 1.94 (0.23) & 4.00 (fixed) & 0.98 (0.74) & 2.30 (0.13) & 1.80 (0.14) & 1.20 (fixed) & 2.08 (0.25) \\
M1.2 flare & 1.00 (0.06) & 1.64 (0.26) & 1.05 (0.67) & 1.59 (0.06) & 1.34 (0.04) & 0.75 (0.16) & 1.08 (0.06) \\
M5.0 PreFlare ($\sim$B5) & 2.39 (0.28) & 4.00 (fixed) & 1.37 (0.64 & 2.26 (0.12) & 2.41 (0.17) & 1.20 (fixed) & 2.56 (0.29) \\
M5.0 flare & 1.85 (0.02) & 4.23 (0.11) & 0.90 (0.65) & 1.62 (0.03) & 0.80 (0.01) & 1.10 (0.01) & 1.98 (0.02) \\
  \hline
  \hline
\end{tabular}
} 
\end{table}

The MinXSS-1 X123 spectral fits of the non-flaring Sun are consistent with a dominant emission component between 1 - 3 MK with volume emission measure values near 10$^{49}$ - 10$^{50}$ cm$^{-3}$, which is not surprising. To highlight the limit of inference in these simple 2T models, we include the more uncertain hotter ($\gtrapprox$ 4.5 MK), dimmer ($\lessapprox$ 10$^{48}$ cm$^{-3}$) secondary component. With this secondary component, the model fits the data well with reasonable $\chi^{2}$ values. Without this secondary temperature component, the 1T models underestimate the measured count rate above 2.5 keV, albeit a DEM fit could reconcile this excess. The ability of MinXSS data alone to constrain this contribution is limited (due to the small effective area) and can only provide upper limits to the emission measure. We caution against the magnitude of  the emission measure of this hot-dim component derived from 2T fits from MinXSS data alone and signify this with ** in Table~\ref{T-Table3} and Table~\ref{T-Table4} on the fit parameters and a dash-dot delta function without a black outline in Figure~\ref{Figure14}). Conclusive evidence of its existence can be provided by simultaneous observations from other more sensitive instruments. The purpose of this article is to highlight the MinXSS capabilities, but this also encompasses uncovering the limitations.

We checked the estimated GOES XRS 0.05 - 0.4 nm flux for the MinXSS 2T inferred pre-flare secondary hot-dimmer components using the goes$\_$fluxes.pro IDL code. The measured fluxes from the GOES 0.05 - 0.4 nm channel would have to be between at least 2.2, 2.9, and 4.2 $\times$ 10$^{-9}$ W m$^{-2}$ (for photospheric abundances) for the pre-flare times of C2.7, M1.2 and M5.0 for this component to have the emission measure similar to the MinXSS 2T-AllFree fits. All measured GOES 0.05 - 0.4 nm fluxes were at about factor of 2 below the estimated values. Spectral fits to accommodate the measured count rate above 2.5 keV is most likely reconciled by DEM fits of MinXSS data coupled with other soft X-ray data. This will be done soon. 

There has been numerous literature discussing active region hot components (T $\geqq$ 5 MK) inferred from soft X-ray data (see \cite{Caspi2015, Miceli2012, Reale2009, Schmelz2009a, Schmelz2009b, Schmelz2015} to state a few and references therein). All of these aforementioned studies had their respective limitations. Future measurements from the MaGIXS \citep{Kobayashi2011} sounding rocket and continued measurements from FOXSI \citep{Ishikawa2014} sounding rockets, plus the NuSTAR satellite \citep{Hannah2016} will provide additional data to validate or further the case to rebuke the previous claims of the hot-dim plasma's presence, or at the very least, provide firm upper limits on its emission measure. The only substantial inkling seems to be from the Extreme Ultraviolet Normal Incidence Spectrograph (EUNIS-13) \citep{Brosius2014} and most recent FOXSI \citep{Ishikawa2017} rocket flights. MinXSS data combined with the other soft X-ray and UV observatories can help to further constrain the existence of hot-dim plasma and investigate the solar plasma temperature distribution. MinXSS data combined with the other soft X-ray and UV observatories can help to further constrain the existence of hot-dim and investigate the solar plasma temperature distribution.
                     
   The peak-flare emission in this paper is best described by a persistent cooler component between 3 - 7 MK plus a hotter contribution ($\geqq$ 13 MK) that dominates at the higher energy flux. All flare data analyzed in this paper cover about a 3 - 4 minute time-frame centered about the peak soft X-ray emission times. Refinement of the MinXSS-1 spectral responses is a continual endeavor, particularly for high flux times, such as the M5.0 flare (which is currently the highest flux observation that we have analyzed thoroughly). The deviations near values of 5 between the uncertainty normalized difference between the model and observations for the Si 2.1 and the Fe 6.7 keV features exhibit the need for further improvement. 
   
     A further check of the MinXSS-1 flare measurements is provided with a comparison to near simultaneous RHESSI measurements. RHESSI results are overlaid as `R' in Figure~\ref{Figure14} for the M1.2 and M5.0 flares, which were the only flares in this paper simultaneously observed by both MinXSS-1 and RHESSI. The RHESSI fits were performed in OSPEX, composed of a 1T thermal and a non-thermal (thick2) bremsstrahlung component. The spectra were fit from $\sim$6 keV to the maximum photon energy with signal above the background ($\sim$30 keV). The RHESSI thermal fit components yield values near GOES estimates \citep{White2005} and the MinXSS-1  em loci curves (discussed in Section~\ref{Ss-EM_Loci}). RHESSI estimates a temperature of 15 MK for both flares and emission measure values of 0.1 and 2.5 $\times$ 10$^{49}$ cm$^{-3}$ respectively for the M1.2 and M5.0 flare. RHESSI fits indicate an elemental abundance FIP-Bias factors of 2.1 and 1.3 for the M1.2 and M5.0 flares respectively, both below `traditional' coronal values. 

It is not expected for MinXSS-1 and RHESSI to return the exact same temperature and emission measure values due to the different spectral responses, spectral resolution, effective areas and consequential temperature sensitivities, but one expects consistencies in RHESSI plasma parameters with MinXSS em loci and inferred photon flux. MinXSS-1 and RHESSI photon flux in Figure~\ref{Figure15} provides one of the few spectrally complete flare measurements from 1 keV to at least 15 keV. The combined MinXSS-1 and RHESSI dynamic range spans nearly eight dex of photon flux and there is overlap near the 6.7 keV Fe complex. The $\sim$8 keV Ni feature is also apparent in the MinXSS-1 spectra. Similarly, RHESSI data can be used to extract non-thermal contributions to the MinXSS-1 spectra for large flares. The discussion of the instrument complexities, and cross calibration between MinXSS and RHESSI is reserved for a future paper (Amir et al., in prep).

\subsection{Emission Measure Loci} 
\label{Ss-EM_Loci}           
          
Emission measure loci (em loci) provide a powerful diagnostic and a useful visualization tool to understand a spectral instrument's capability to infer plasma temperature distribution from a specific set of observations \citep{Warren2009, Landi2010}. Em loci can be calculated by taking the measured data (summed counts over a specified energy range for X123 or the total signal for XP) and dividing by the instrument's temperature response calculated for that energy range, for the corresponding best fit elemental abundance and emission model, $em loci = \textstyle \frac{Counts}{F(T)}$. Figure~\ref{Figure14} displays the MinXSS-1 X123 em loci for the QS and three flare times, along with the best fit 2T-Free emission measure results. The different colors show the em loci of the various summed 0.3 keV wide energy bins (counts from $\sim$10 X123 bins summed) with a significant count rate ($\geqq$ 0.05 counts per second per nominal bin), which depict the energy ranges that best restrict the emission at specific temperatures. In general the higher X123 energies are better for inferring the hotter plasma conditions.

                         \begin{figure}    
   \centerline{\includegraphics[width=1.0\textwidth,clip=]{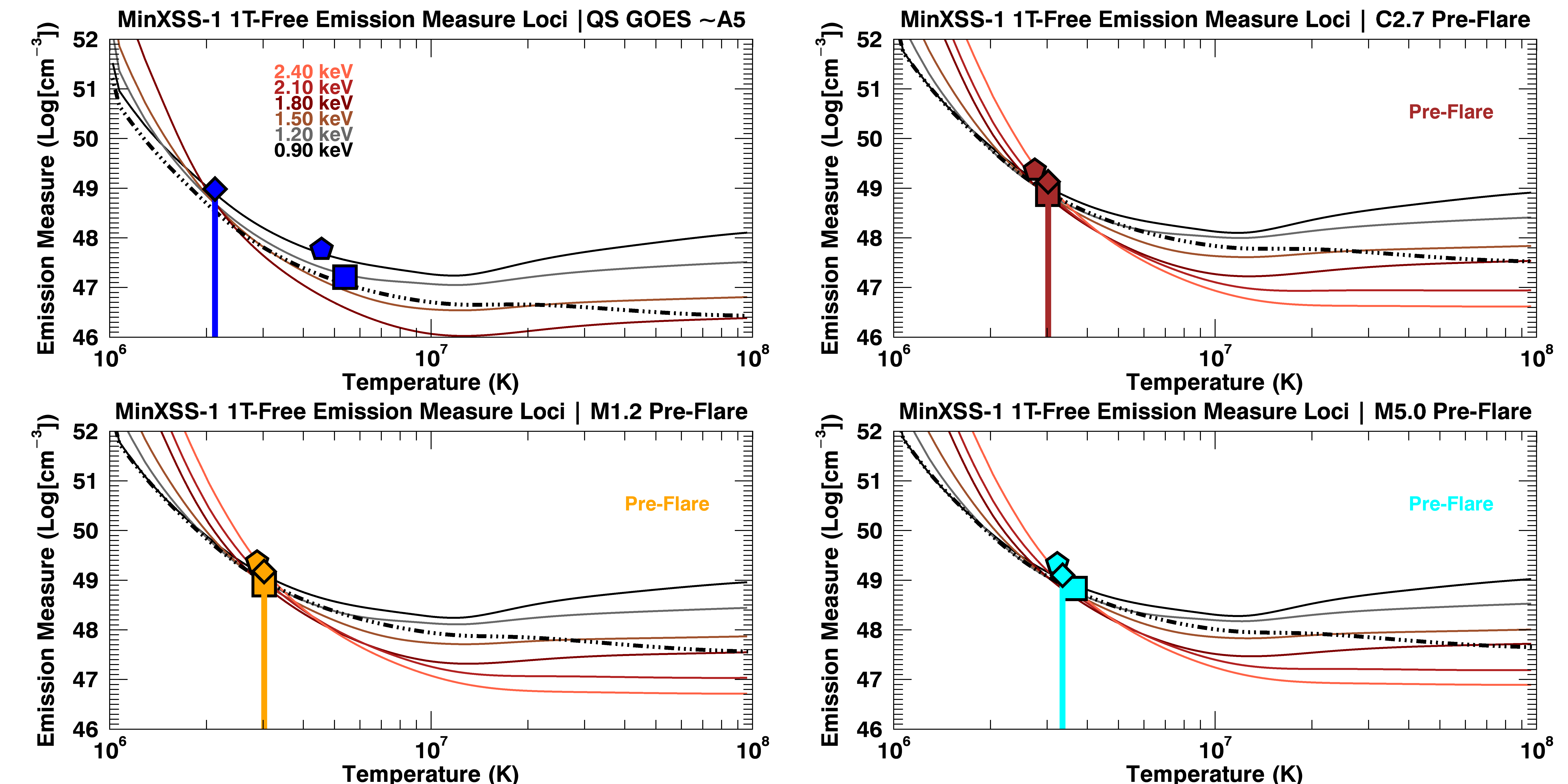}
              }
              \caption{Volume emission measure loci (em loci) plots with MinXSS-1 OSPEX 1T-Free fit parameters over-plotted as delta functions in temperature with filled diamonds indicating the emission measure value for the non-large-flaring sun (pre-flare). These MinXSS-1 em loci and fit parameters correspond to the spatial distribution captured by Hinode XRT in Figure~\ref{Figure11}. The solid colored lines correspond to X123 counts summed to 0.3 keV wide energy bins and the dash-dotted lines are the XP loci. The rainbow keV values in the top left plot indicate the color code for the minimum energy bin use for each X123 em loci. The em loci indicate the maximum emission if all the plasma was isothermal for each summed energy bins. GOES averaged values are listed for photospheric (pentagon) and coronal (square) abundances. The plot of X123 1T-Free fits are to demonstrate, 1) the agreement with the overlapping X123 em loci, 2) agreement with the overlapping XP em loci and 3) consistency with the GOES XRS isothermal estimation except for low GOES levels like the $\sim$A5 levels (due to the non-linearity of GOES for low flux levels). 
                      }
   \label{Figure13}
   \end{figure}

        \begin{figure}    
   \centerline{\includegraphics[width=1.0\textwidth,clip=]{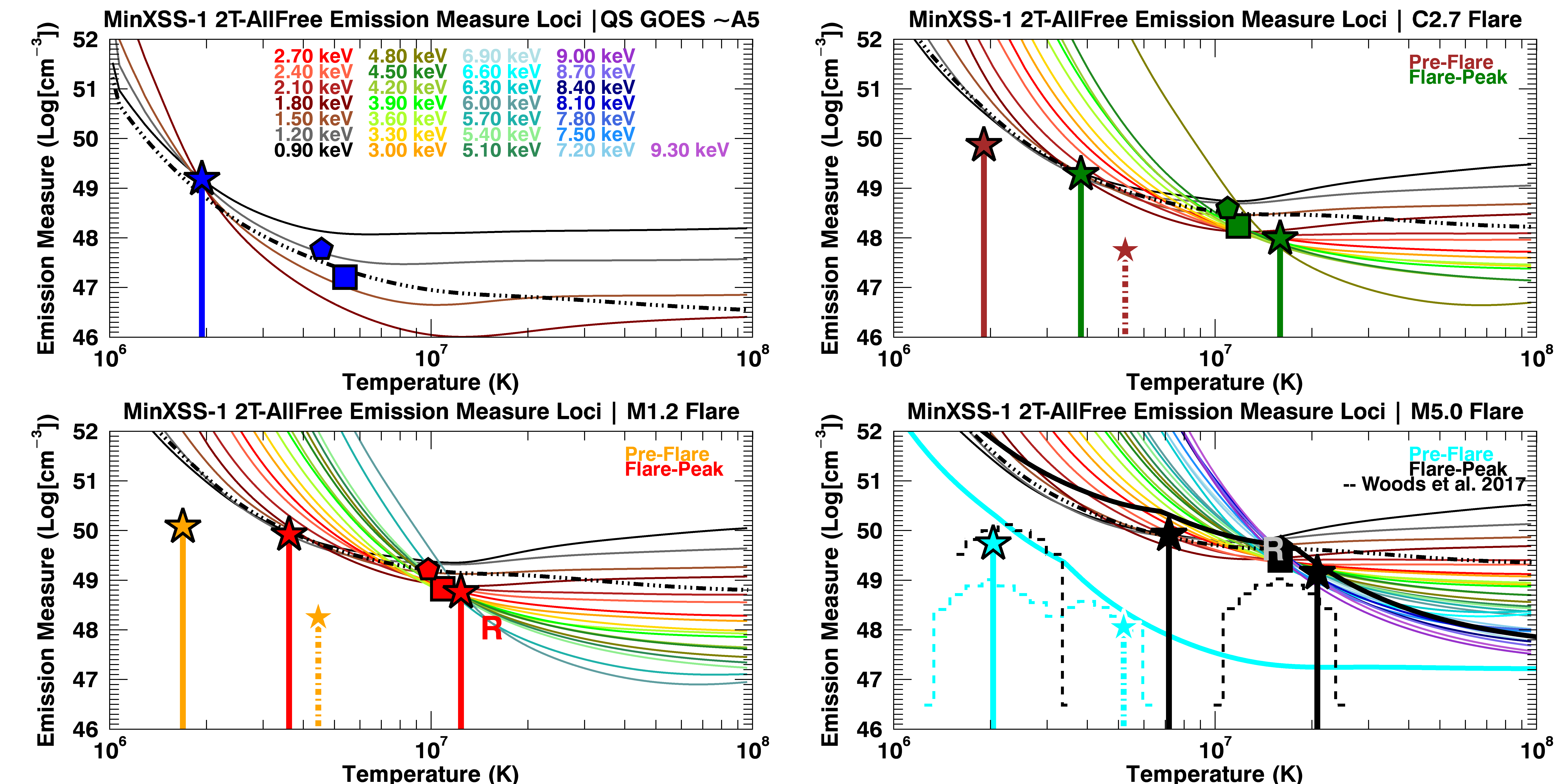}
              }
              \caption{Em loci plots with MinXSS-1 OSPEX 2T-AllFree fit flare-peak and QS parameters over-plotted as delta functions in temperature with filled stars indicating the emission emission measure value. The solid line delta functions are well constrained by the MinXSS data. The pre-flare dash-dot delta function without the black outline indicates that the hotter-dimmer component is less constrained by MinXSS data. The thin solid colored lines correspond to X123 counts summed to 0.3 keV wide energy bins and the dash-dotted lines are the XP loci. The rainbow keV values in the top left plot indicate the color code for the minimum energy bin use for each X123 em loci. The X123 and XP em loci are consistent. The \cite{Woods2017} M5.0 flare fit results are overlaid as the dashed histograms in the bottom right panel. The thick black em loci is for the M5.0 flare and the thick cyan em loci for the pre-flare, both are the minimum of all the individual energy bins corresponding to the spectral model used in the \cite{Woods2017} analysis. GOES averaged values are listed for photospheric (pentagon) and coronal (square) abundances. RHESSI values for the M1.2 and M5.0 flare are indicated by the `R'.
                      }
   \label{Figure14}
   \end{figure}

 If the plasma was isothermal, then one would expect an isothermal fit to `touch' (or overlap with) the em loci at the corresponding temperature value. Additionally, where the various em loci curves for different energy bins intersect is roughly the location that yield the isothermal temperature and emission measure that best describe the plasma. In the case of large flares during peak emission times, we expect the flaring plasma to dominate the solar spectral emission. Thus, a 1T model result would peak near this value and this is what was observed for the 1T models. The 1T-Free model for the QS and 3 pre-flare times are displayed in Figure~\ref{Figure13} to exemplify this  (the diamond symbols). The values for all the pre-flare times are near 3 MK and coincide with the GOES XRS isothermal estimates for coronal (squares) and photospheric (pentagons) abundances. Moreover, the X123 1T-Free values lie in between the GOES XRS coronal and photospheric estimates, as they should because the fit results for X123 reside in between coronal and photospheric FIP-Bias values. The 3 MK isothermal plasma temperature is consistent with the infered active region DEM peak from the combined Hinode EIS and SDO AIA study of \cite{DelZanna2013}. The 2T-AllFree em loci results are displayed in Figure~\ref{Figure14}. The best fit emission measure and temperatures are the delta functions with stars at the emission measure value. The \cite{Woods2017} M5.0 flare fit results are overlaid as the dashed histograms in the bottom right panel. The thick black em loci is for the M5.0 flare and the thick cyan em loci for the pre-flare, both are the minimum of all the individual energy bins corresponding to the spectral model used in the \cite{Woods2017} analysis. The GOES XRS average flare-peak time 1T emission measure and temperature results for photospheric and coronal elemental abundance are over-plotted for comparison. 
 
   The flare-peak GOES estimates are all lower temperature, higher emission measure and closer to the region where the MinXSS X123 em loci come close to intersecting for the three flare times. The X123 loci straddle the em loci both above and below the GOES values. 1T fits for X123 during the flare-peak times are similar to the GOES estimates, but are not shown because of the poor fit to the X123 flare spectra. The RHESSI fit results are also plotted for the M1.2 and M5.0 flare. The RHESSI estimates are hotter for the M1.2 flare than both MinXSS components and GOES, and the M5.0 flare is consistent with GOES estimates, but both RHESSI estimates are consistent with the MinXSS em loci.

        \begin{figure}    
   \centerline{\includegraphics[width=1.0\textwidth,clip=]{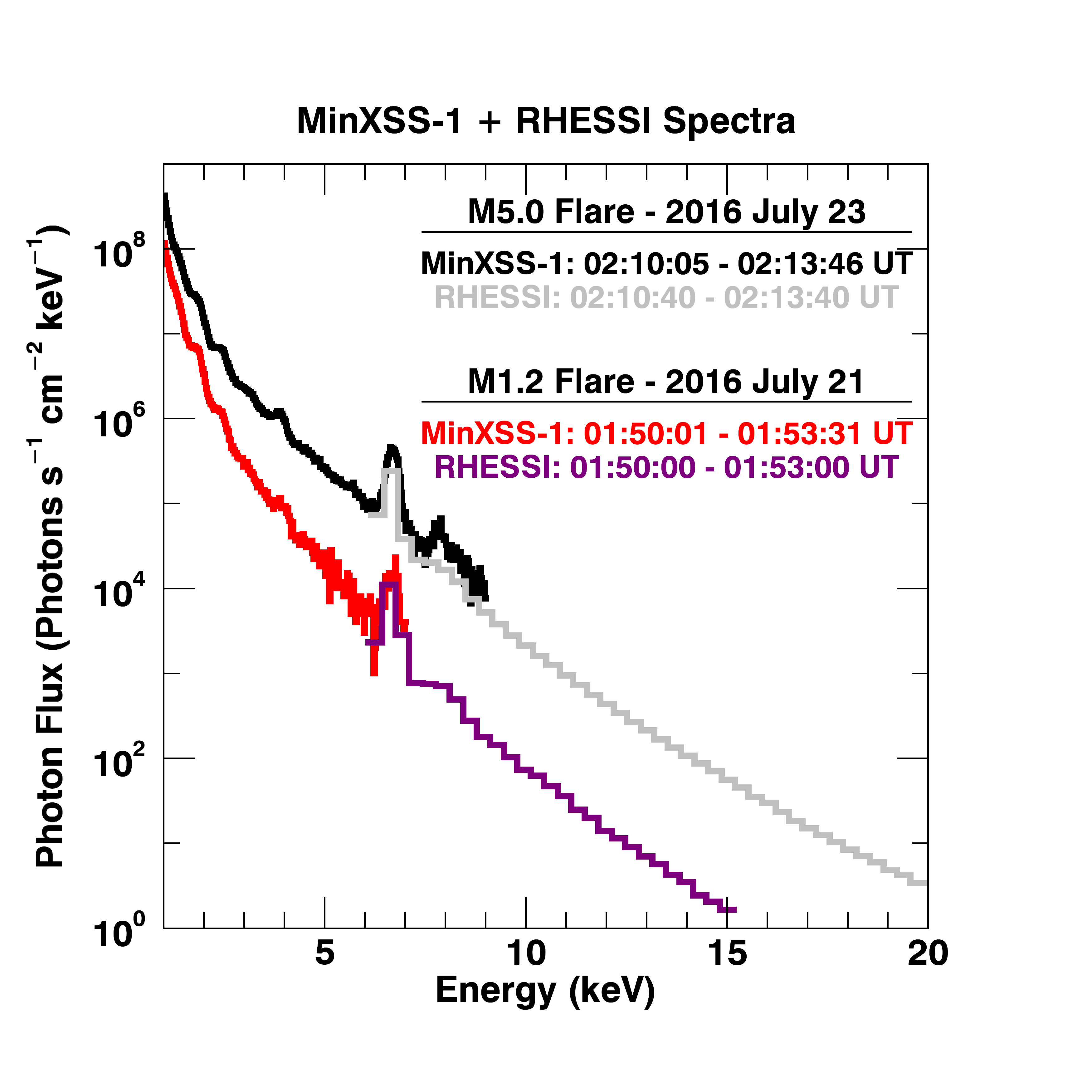}
              }
              \caption{MinXSS-1 M1.2 and M5.0 flare photon flux spectra with overlaid RHESSI spectra. These near simultaneous measurements provide complete spectral coverage from 1 keV to the minimum detected flux from RHESSI and spans eight orders of magnitude in flux. The main overlap between instruments for flares is near the 6.7 keV Fe complex. This comparison helps validate the MinXSS observations. 
                      }
   \label{Figure15}
   \end{figure}

The consistency between X123 and XP em loci provide further confidence that both instruments are performing nominally. The minimum of all the em loci for the MinXSS-1 X123 energy bins provides an upper limit to any multi-temperature and DEM fits. The energy must be spread over a range of temperatures and thus could not `touch' the em loci at any point. Thus em loci provide firm upper limits to the temperature distribution for a particular measurement from a particular instrument. Em loci are not a new tool, they have been used most notably in XRT, EIS, AIA, FOXSI, RHESSI and NuSTAR analysis to name a few. Em loci can be a valuable aid in MinXSS data analysis.


  \section{Summary} 
\label{S-Sumary} 
  
  The MinXSS CubeSats can provide the solar community with a new set of measurements that can augment current and future investigations of the solar corona. The SPS ancillary instrument, the X-ray detectors of XP and X123 have been characterized. The first version of MinXSS, MinXSS-1, has performed nominally over its mission at LEO.  This paper describes the MinXSS instrument suite, the X123 FOV sensitivity, X123 spectral resolution vs. photon energy, XP and X123 effective area curves, X123 detector response matrix, XP and X123 temperature response, the X123 linearity of response, GOES flux levels vs. MinXSS-1 X123 measured integrated counts, and MinXSS-2 X123 estimated counts, inferred temperature and emission measures from MinXSS spectra, and emission measure loci for the discussed data. These realizations further the notion that CubeSats can conduct significant targeted science. A summary of the main attributes of MinXSS are listed below.

\flushleft  
  \begin{enumerate}
\item  MinXSS-1 X123 has an effective solar flux energy range of 0.8 - 12 keV ($\sim$0.10 - 1.55 nm) with resolving power $\sim$40 at 5.9 keV, and with dead-time corrections applied accurate spectra up to low GOES X levels (but need to correct for pulse-pileup). 
\newline

\item  The MinXSS CubeSats X123 have a relatively higher spectral resolution over a fairly broad bandpass that allow inference of elemental abundance values for the elements Fe, Ca, Si, Mg, S, Ar and Ni, when there are sufficient counts at X123 energies for their respective line groups. The observed elemental abundance variation in this work:
\begin{itemize}
     \item clearly demonstrate a decrease in low FIP elements for flare-peak times vs. the pre-flare values for spectral fits with a single FIP-Bias multiplicative factor
     \item  display variance in the fractionation pattern among the low FIP elements when all elements are allowed to vary for GOES levels $>$ B1.
   \end{itemize} 

\item MinXSS-1 X123 flare measurements in this paper indicate that the hotter components in the flaring plasma for the C2.7, M1.2 and M5.0 flares bare peak temperatures near 15, 13 and 20 MK, respectively. 
\newline

\item MinXSS-1 X123 and XP plasma temperatures inferences are self consistent and single X123 temperature fits (1T-Free) are comparable with GOES XRS isothermal estimates between GOES $\sim$B1 - M5 levels (GOES XRS response becomes non-linear for lower flux levels).  However, these single temperature fits can not account for high energy X123 spectral counts nor are suitable fits for an entire flare spectrum alone.
\newline

\item MinXSS X123 can infer non-large-flaring Sun properties between 1.5 - 4 MK with high confidence, but limited capabilities for temperatures below 1.5 MK (due to limited sensitivity at lower energies, $<$ 1 keV). 
\newline

\item MinXSS X123 can only set upper limits on the emission measure and cannot definitively constrain the temperature values for `dimmer' plasma hotter than $\sim$5 MK during non-large-flaring times. This is due to the flattening nature of the temperature response above $\sim$4.5 MK for energy bins less than 3 keV and limited significant counts from energy bins greater than 3 keV. The latter is a consequence of the relatively small X123 aperture area (2.5 $\times$ 10$^{-4}$ cm$^{2}$). 
\newline
\end{enumerate}

  The anticipated mission length for MinXSS-1 was 1 year. The second version, MinXSS-2, has improved hardware, software, and detectors (see Section~\ref{Ss-Improvements}). The anticipated duration of the Sun-synchronous orbit for MinXSS-2 is 5 years, providing a possible total of 6 years of near continuous measurements. The MinXSS-1 Level-1 products are available on the MinXSS website: \url{http://lasp.colorado.edu/home/minxss/}.

 \subsection{Improvements of MinXSS-2} 
  \label{Ss-Improvements}
The MinXSS-2 X123 has a thinner Be window along with a lower noise preamplifier that will provide an improved low energy response, possibly extending the low energy limit to 0.6 keV. The newer preamplifier for the MinXSS-2 X123 Fast SDD detector allows for a better spectral resolution for the same nominal peaking time as the MinXSS-1 X123. We will be operating a shorter peaking time for the MinXSS-2 X123 that will allow for accurate spectra for a higher input flux. Preliminary estimates suggest GOES levels around X2 are the maximum that the MinXSS-1 X123 can handle before the spectra will need further processing to retain fidelity.


\appendix   

\section{SPS Architecture and Position Calculation} 
    \label{S-appendix_SPS}

\begin{figure}    
   \centerline{\includegraphics[width=1.0\textwidth,clip=]{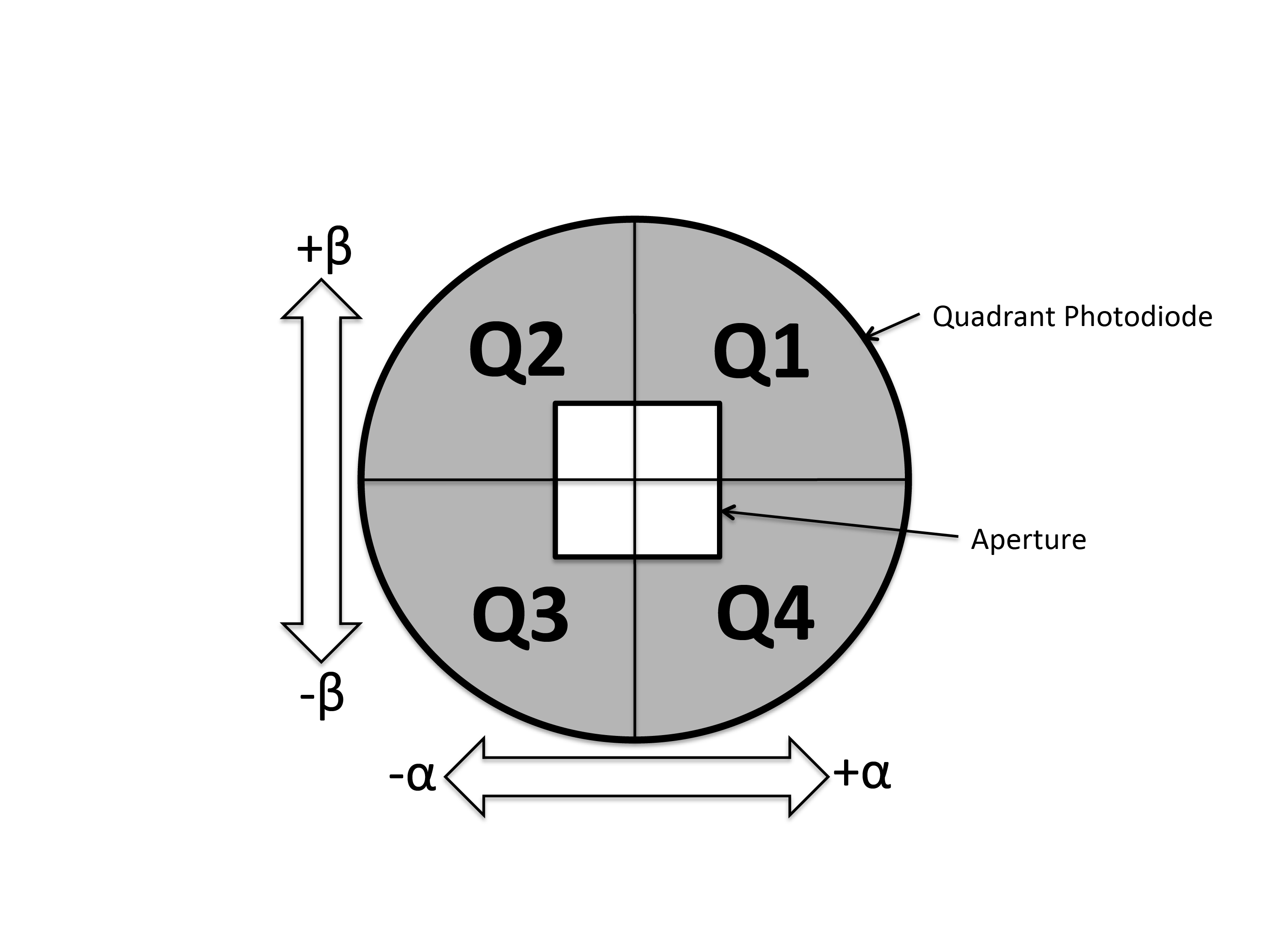}
              }
              \caption{SPS quad diode layout demonstrating the orientation of the $\alpha$ and $\beta$ angles. The SPS square aperture is 5 mm $\times$ 5 mm.  
                      }
   \label{FigureA1}
   \end{figure}

The SPS quad Si-photodiode detector assembly is used to calculate the solar position in the MinXSS FOV. SPS sits behind a ND7 filter (attenuation of a factor of 10$^{7})$ to limit the visible-light flux and eliminate all other electromagnetic contributions to the signal. SPS relies on the visible-light emission from the Sun to estimate the solar location within the FOV position by calculating $\alpha$ and $\beta$. Figure~\ref{FigureA1} shows the basic SPS layout of the quad diode assembly with a square 2 $\times$ 2 mm aperture. SPS has no focusing optics and thus its operating principle relies upon measuring the relative light detected on each of the four diodes Q1, Q2, Q3, and Q4 to calculate solar disc location. Equation~\ref{Eq-alpha} and Equation~\ref{Eq-beta} display the relation between the quad diode signal to the $\alpha$ and $\beta$ coordinates, which must be scaled by the known maximum angular deviation $\alpha_{max}$ and $\beta_{max}$. The maximum angular deviation is set by the geometrical layout of the aperture with respect to the quad diodes. For an aperture  of width, w, height, h and separation between the photodiode and aperture along the optical path, d, the maximum angles can be calculated from Equation~\ref{Eq-alpha_max} and Equation~\ref{Eq-beta_max}.
  
        \begin{equation}  \label{Eq-alpha} 
	\alpha  = \alpha_{max} \frac{(Q1 + Q4) - (Q2 + Q3)}{(Q1 + Q2 + Q3 + Q4)}
   \end{equation}
  
          \begin{equation}  \label{Eq-beta} 
	\beta  = \beta_{max} \frac{(Q1 + Q2) - (Q3 + Q4)}{(Q1 + Q2 + Q3 + Q4)}
   \end{equation}

           \begin{equation}  \label{Eq-alpha_max} 
	 \alpha_{max} = \alpha tan\bigg(\frac{w}{2d}\bigg)
   \end{equation}
   
              \begin{equation}  \label{Eq-beta_max} 
	 \beta_{max} = \beta tan\bigg(\frac{h}{2d}\bigg)
   \end{equation}
 
  
\begin{acks}
We would like to thank Micheal Klapetzky, Rick Kohnert, the entire MinXSS Team and the entire NIST SURF team for assistance in testing the MinXSS CubeSat. Support for C. S. Moore was provided through NASA Space Technology Research Fellowship (NSTRF) Program Grant $\#$NNX13AL35H. The MinXSS-1 CubeSat mission is supported by NASA Grant $\#$NNX14AN84G. A. Caspi was also partially supported by NASA grants NNX14AH54G and NNX15AK26G.
\end{acks}


\bibliographystyle{spr-mp-sola}
\bibliography{minxss_instrument_paper_final_resubmit2}  

\begin{thebibliography}{46}
\ifx\bisbn     \undefined \def\bisbn  #1{ISBN #1}\fi
\ifx\binits    \undefined \def\binits#1{#1}\fi
\ifx\bauthor   \undefined \def\bauthor#1{#1}\fi
\ifx\batitle   \undefined \def\batitle#1{#1}\fi
\ifx\bjtitle   \undefined \def\bjtitle#1{\textit{#1}}\fi
\ifx\bvolume   \undefined \def\bvolume#1{\textbf{#1}}\fi
\ifx\byear     \undefined \def\byear#1{#1}\fi
\ifx\bissue    \undefined \def\bissue#1{#1}\fi
\ifx\bfpage    \undefined \def\bfpage#1{#1}\fi
\ifx\blpage    \undefined \def\blpage #1{#1}\fi
\ifx\burl      \undefined \def\burl#1{\textsf{#1}}\fi
\ifx\href      \undefined \def\href#1#2{\textsf{#2}}\fi
\ifx\betal     \undefined \def\betal{\textit{et al.}}\fi
\ifx\bctitle   \undefined \def\bctitle#1{#1}\fi
\ifx\beditor   \undefined \def\beditor#1{#1}\fi
\ifx\bbtitle   \undefined \def\bbtitle#1{\textit{#1}}\fi
\ifx\bedition  \undefined \def\bedition#1{#1}\fi
\ifx\bseriesno \undefined \def\bseriesno#1{\textbf{#1}}\fi
\ifx\blocation \undefined \def\blocation#1{#1}\fi
\ifx\bsertitle \undefined \def\bsertitle#1{\textit{#1}}\fi
\ifx\bsnm      \undefined \def\bsnm#1{#1}\fi
\ifx\bsuffix   \undefined \def\bsuffix#1{#1}\fi
\ifx\bparticle \undefined \def\bparticle#1{#1}\fi
\ifx\barticle  \undefined \def\barticle#1{}\fi
\ifx\binstitute  \undefined \def\binstitute#1{#1}\fi
\ifx\bpublisher  \undefined \def\bpublisher#1{#1}\fi
\ifx\doiurl    \undefined
  \def\doiurl#1{\href{http://dx.doi.org/#1}{\textsf{DOI}}}\fi
\ifx\arxivurl  \undefined
  \def\arxivurl#1{\href{http://arxiv.org/abs/#1}{\textsf{arXiv}}}\fi
\ifx\adsurl    \undefined
  \def\adsurl#1{\href{http://adsabs.harvard.edu/abs/#1}{\textsf{ADS}}}\fi
\ifx\botherref \undefined \def\botherref#1{}\fi
\ifx\url       \undefined \def\url#1{\textsf{#1}}\fi
\ifx\bchapter  \undefined \def\bchapter#1{}\fi
\ifx\bbook     \undefined \def\bbook#1{}\fi
\ifx\bcomment  \undefined \def\bcomment#1{#1}\fi
\ifx\oauthor   \undefined \def\oauthor#1{#1}\fi
\ifx\citeauthoryear \undefined\def \citeauthoryear#1{#1}\fi
\ifx\endbibitem\undefined \def\endbibitem{}\fi
\ifx\bconflocation  \undefined \def\bconflocation#1{#1} \fi

\bibitem[\protect\citeauthoryear{{Brosius}, {Daw}, and
  {Rabin}}{2014}]{Brosius2014}
\begin{barticle}
\bauthor{\bsnm{{Brosius}}, \binits{J.W.}},
\bauthor{\bsnm{{Daw}}, \binits{A.N.}},
\bauthor{\bsnm{{Rabin}}, \binits{D.M.}}:
\byear{2014},
\batitle{{Pervasive Faint Fe XIX Emission from a Solar Active Region Observed
  with EUNIS-13: Evidence for Nanoflare Heating}}.
\bjtitle{\apj}
\bvolume{790},
\bfpage{112}.
\doiurl{10.1088/0004-637X/790/2/112}.
\adsurl{2014ApJ...790..112B}.
\end{barticle}
\endbibitem

\bibitem[\protect\citeauthoryear{{Caffau} \textit{et~al.}}{2011}]{Caffau2011}
\begin{barticle}
\bauthor{\bsnm{{Caffau}}, \binits{E.}},
\bauthor{\bsnm{{Ludwig}}, \binits{H.-G.}},
\bauthor{\bsnm{{Steffen}}, \binits{M.}},
\bauthor{\bsnm{{Freytag}}, \binits{B.}},
\bauthor{\bsnm{{Bonifacio}}, \binits{P.}}:
\byear{2011},
\batitle{{Solar Chemical Abundances Determined with a CO5BOLD 3D Model
  Atmosphere}}.
\bjtitle{\solphys}
\bvolume{268},
\bfpage{255}.
\doiurl{10.1007/s11207-010-9541-4}.
\adsurl{2011SoPh..268..255C}.
\end{barticle}
\endbibitem

\bibitem[\protect\citeauthoryear{{Caspi}, {Woods}, and
  {Warren}}{2015}]{Caspi2015}
\begin{barticle}
\bauthor{\bsnm{{Caspi}}, \binits{A.}},
\bauthor{\bsnm{{Woods}}, \binits{T.N.}},
\bauthor{\bsnm{{Warren}}, \binits{H.P.}}:
\byear{2015},
\batitle{{New Observations of the Solar 0.5-5 keV Soft X-Ray Spectrum}}.
\bjtitle{\apjl}
\bvolume{802},
\bfpage{L2}.
\doiurl{10.1088/2041-8205/802/1/L2}.
\adsurl{2015ApJ...802L...2C}.
\end{barticle}
\endbibitem

\bibitem[\protect\citeauthoryear{{Culhane} \textit{et~al.}}{2007}]{Culhane2007}
\begin{barticle}
\bauthor{\bsnm{{Culhane}}, \binits{J.L.}},
\bauthor{\bsnm{{Harra}}, \binits{L.K.}},
\bauthor{\bsnm{{James}}, \binits{A.M.}},
\bauthor{\bsnm{{Al-Janabi}}, \binits{K.}},
\bauthor{\bsnm{{Bradley}}, \binits{L.J.}},
\bauthor{\bsnm{{Chaudry}}, \binits{R.A.}},
\bauthor{\bsnm{{Rees}}, \binits{K.}},
\bauthor{\bsnm{{Tandy}}, \binits{J.A.}},
\bauthor{\bsnm{{Thomas}}, \binits{P.}},
\bauthor{\bsnm{{Whillock}}, \binits{M.C.R.}},
\bauthor{\bsnm{{Winter}}, \binits{B.}},
\bauthor{\bsnm{{Doschek}}, \binits{G.A.}},
\bauthor{\bsnm{{Korendyke}}, \binits{C.M.}},
\bauthor{\bsnm{{Brown}}, \binits{C.M.}},
\bauthor{\bsnm{{Myers}}, \binits{S.}},
\bauthor{\bsnm{{Mariska}}, \binits{J.}},
\bauthor{\bsnm{{Seely}}, \binits{J.}},
\bauthor{\bsnm{{Lang}}, \binits{J.}},
\bauthor{\bsnm{{Kent}}, \binits{B.J.}},
\bauthor{\bsnm{{Shaughnessy}}, \binits{B.M.}},
\bauthor{\bsnm{{Young}}, \binits{P.R.}},
\bauthor{\bsnm{{Simnett}}, \binits{G.M.}},
\bauthor{\bsnm{{Castelli}}, \binits{C.M.}},
\bauthor{\bsnm{{Mahmoud}}, \binits{S.}},
\bauthor{\bsnm{{Mapson-Menard}}, \binits{H.}},
\bauthor{\bsnm{{Probyn}}, \binits{B.J.}},
\bauthor{\bsnm{{Thomas}}, \binits{R.J.}},
\bauthor{\bsnm{{Davila}}, \binits{J.}},
\bauthor{\bsnm{{Dere}}, \binits{K.}},
\bauthor{\bsnm{{Windt}}, \binits{D.}},
\bauthor{\bsnm{{Shea}}, \binits{J.}},
\bauthor{\bsnm{{Hagood}}, \binits{R.}},
\bauthor{\bsnm{{Moye}}, \binits{R.}},
\bauthor{\bsnm{{Hara}}, \binits{H.}},
\bauthor{\bsnm{{Watanabe}}, \binits{T.}},
\bauthor{\bsnm{{Matsuzaki}}, \binits{K.}},
\bauthor{\bsnm{{Kosugi}}, \binits{T.}},
\bauthor{\bsnm{{Hansteen}}, \binits{V.}},
\bauthor{\bsnm{{Wikstol}}, \binits{{\O}.}}:
\byear{2007},
\batitle{{The EUV Imaging Spectrometer for Hinode}}.
\bjtitle{\solphys}
\bvolume{243},
\bfpage{19}.
\doiurl{10.1007/s01007-007-0293-1}.
\adsurl{2007SoPh..243...19C}.
\end{barticle}
\endbibitem

\bibitem[\protect\citeauthoryear{{De Pontieu}
  \textit{et~al.}}{2014}]{DePontieu2014}
\begin{barticle}
\bauthor{\bsnm{{De Pontieu}}, \binits{B.}},
\bauthor{\bsnm{{Title}}, \binits{A.M.}},
\bauthor{\bsnm{{Lemen}}, \binits{J.R.}},
\bauthor{\bsnm{{Kushner}}, \binits{G.D.}},
\bauthor{\bsnm{{Akin}}, \binits{D.J.}},
\bauthor{\bsnm{{Allard}}, \binits{B.}},
\bauthor{\bsnm{{Berger}}, \binits{T.}},
\bauthor{\bsnm{{Boerner}}, \binits{P.}},
\bauthor{\bsnm{{Cheung}}, \binits{M.}},
\bauthor{\bsnm{{Chou}}, \binits{C.}},
\bauthor{\bsnm{{Drake}}, \binits{J.F.}},
\bauthor{\bsnm{{Duncan}}, \binits{D.W.}},
\bauthor{\bsnm{{Freeland}}, \binits{S.}},
\bauthor{\bsnm{{Heyman}}, \binits{G.F.}},
\bauthor{\bsnm{{Hoffman}}, \binits{C.}},
\bauthor{\bsnm{{Hurlburt}}, \binits{N.E.}},
\bauthor{\bsnm{{Lindgren}}, \binits{R.W.}},
\bauthor{\bsnm{{Mathur}}, \binits{D.}},
\bauthor{\bsnm{{Rehse}}, \binits{R.}},
\bauthor{\bsnm{{Sabolish}}, \binits{D.}},
\bauthor{\bsnm{{Seguin}}, \binits{R.}},
\bauthor{\bsnm{{Schrijver}}, \binits{C.J.}},
\bauthor{\bsnm{{Tarbell}}, \binits{T.D.}},
\bauthor{\bsnm{{W{\"u}lser}}, \binits{J.-P.}},
\bauthor{\bsnm{{Wolfson}}, \binits{C.J.}},
\bauthor{\bsnm{{Yanari}}, \binits{C.}},
\bauthor{\bsnm{{Mudge}}, \binits{J.}},
\bauthor{\bsnm{{Nguyen-Phuc}}, \binits{N.}},
\bauthor{\bsnm{{Timmons}}, \binits{R.}},
\bauthor{\bsnm{{van Bezooijen}}, \binits{R.}},
\bauthor{\bsnm{{Weingrod}}, \binits{I.}},
\bauthor{\bsnm{{Brookner}}, \binits{R.}},
\bauthor{\bsnm{{Butcher}}, \binits{G.}},
\bauthor{\bsnm{{Dougherty}}, \binits{B.}},
\bauthor{\bsnm{{Eder}}, \binits{J.}},
\bauthor{\bsnm{{Knagenhjelm}}, \binits{V.}},
\bauthor{\bsnm{{Larsen}}, \binits{S.}},
\bauthor{\bsnm{{Mansir}}, \binits{D.}},
\bauthor{\bsnm{{Phan}}, \binits{L.}},
\bauthor{\bsnm{{Boyle}}, \binits{P.}},
\bauthor{\bsnm{{Cheimets}}, \binits{P.N.}},
\bauthor{\bsnm{{DeLuca}}, \binits{E.E.}},
\bauthor{\bsnm{{Golub}}, \binits{L.}},
\bauthor{\bsnm{{Gates}}, \binits{R.}},
\bauthor{\bsnm{{Hertz}}, \binits{E.}},
\bauthor{\bsnm{{McKillop}}, \binits{S.}},
\bauthor{\bsnm{{Park}}, \binits{S.}},
\bauthor{\bsnm{{Perry}}, \binits{T.}},
\bauthor{\bsnm{{Podgorski}}, \binits{W.A.}},
\bauthor{\bsnm{{Reeves}}, \binits{K.}},
\bauthor{\bsnm{{Saar}}, \binits{S.}},
\bauthor{\bsnm{{Testa}}, \binits{P.}},
\bauthor{\bsnm{{Tian}}, \binits{H.}},
\bauthor{\bsnm{{Weber}}, \binits{M.}},
\bauthor{\bsnm{{Dunn}}, \binits{C.}},
\bauthor{\bsnm{{Eccles}}, \binits{S.}},
\bauthor{\bsnm{{Jaeggli}}, \binits{S.A.}},
\bauthor{\bsnm{{Kankelborg}}, \binits{C.C.}},
\bauthor{\bsnm{{Mashburn}}, \binits{K.}},
\bauthor{\bsnm{{Pust}}, \binits{N.}},
\bauthor{\bsnm{{Springer}}, \binits{L.}},
\bauthor{\bsnm{{Carvalho}}, \binits{R.}},
\bauthor{\bsnm{{Kleint}}, \binits{L.}},
\bauthor{\bsnm{{Marmie}}, \binits{J.}},
\bauthor{\bsnm{{Mazmanian}}, \binits{E.}},
\bauthor{\bsnm{{Pereira}}, \binits{T.M.D.}},
\bauthor{\bsnm{{Sawyer}}, \binits{S.}},
\bauthor{\bsnm{{Strong}}, \binits{J.}},
\bauthor{\bsnm{{Worden}}, \binits{S.P.}},
\bauthor{\bsnm{{Carlsson}}, \binits{M.}},
\bauthor{\bsnm{{Hansteen}}, \binits{V.H.}},
\bauthor{\bsnm{{Leenaarts}}, \binits{J.}},
\bauthor{\bsnm{{Wiesmann}}, \binits{M.}},
\bauthor{\bsnm{{Aloise}}, \binits{J.}},
\bauthor{\bsnm{{Chu}}, \binits{K.-C.}},
\bauthor{\bsnm{{Bush}}, \binits{R.I.}},
\bauthor{\bsnm{{Scherrer}}, \binits{P.H.}},
\bauthor{\bsnm{{Brekke}}, \binits{P.}},
\bauthor{\bsnm{{Martinez-Sykora}}, \binits{J.}},
\bauthor{\bsnm{{Lites}}, \binits{B.W.}},
\bauthor{\bsnm{{McIntosh}}, \binits{S.W.}},
\bauthor{\bsnm{{Uitenbroek}}, \binits{H.}},
\bauthor{\bsnm{{Okamoto}}, \binits{T.J.}},
\bauthor{\bsnm{{Gummin}}, \binits{M.A.}},
\bauthor{\bsnm{{Auker}}, \binits{G.}},
\bauthor{\bsnm{{Jerram}}, \binits{P.}},
\bauthor{\bsnm{{Pool}}, \binits{P.}},
\bauthor{\bsnm{{Waltham}}, \binits{N.}}:
\byear{2014},
\batitle{{The Interface Region Imaging Spectrograph (IRIS)}}.
\bjtitle{\solphys}
\bvolume{289},
\bfpage{2733}.
\doiurl{10.1007/s11207-014-0485-y}.
\adsurl{2014SoPh..289.2733D}.
\end{barticle}
\endbibitem

\bibitem[\protect\citeauthoryear{{Del Zanna}}{2013}]{DelZanna2013}
\begin{barticle}
\bauthor{\bsnm{{Del Zanna}}, \binits{G.}}:
\byear{2013},
\batitle{{The multi-thermal emission in solar active regions}}.
\bjtitle{\aap}
\bvolume{558},
\bfpage{A73}.
\doiurl{10.1051/0004-6361/201321653}.
\adsurl{2013A\%26A...558A..73D}.
\end{barticle}
\endbibitem

\bibitem[\protect\citeauthoryear{{Del Zanna}
  \textit{et~al.}}{2015}]{DelZanna2015}
\begin{barticle}
\bauthor{\bsnm{{Del Zanna}}, \binits{G.}},
\bauthor{\bsnm{{Dere}}, \binits{K.P.}},
\bauthor{\bsnm{{Young}}, \binits{P.R.}},
\bauthor{\bsnm{{Landi}}, \binits{E.}},
\bauthor{\bsnm{{Mason}}, \binits{H.E.}}:
\byear{2015},
\batitle{{CHIANTI - An atomic database for emission lines. Version 8}}.
\bjtitle{\aap}
\bvolume{582},
\bfpage{A56}.
\doiurl{10.1051/0004-6361/201526827}.
\adsurl{2015A\%26A...582A..56D}.
\end{barticle}
\endbibitem

\bibitem[\protect\citeauthoryear{{Dennis} \textit{et~al.}}{2015}]{Dennis2015}
\begin{barticle}
\bauthor{\bsnm{{Dennis}}, \binits{B.R.}},
\bauthor{\bsnm{{Phillips}}, \binits{K.J.H.}},
\bauthor{\bsnm{{Schwartz}}, \binits{R.A.}},
\bauthor{\bsnm{{Tolbert}}, \binits{A.K.}},
\bauthor{\bsnm{{Starr}}, \binits{R.D.}},
\bauthor{\bsnm{{Nittler}}, \binits{L.R.}}:
\byear{2015},
\batitle{{Solar Flare Element Abundances from the Solar Assembly for X-Rays
  (SAX) on MESSENGER}}.
\bjtitle{\apj}
\bvolume{803},
\bfpage{67}.
\doiurl{10.1088/0004-637X/803/2/67}.
\adsurl{2015ApJ...803...67D}.
\end{barticle}
\endbibitem

\bibitem[\protect\citeauthoryear{{Doschek} and
  {Warren}}{2016}]{DoschekWarren2016}
\begin{barticle}
\bauthor{\bsnm{{Doschek}}, \binits{G.A.}},
\bauthor{\bsnm{{Warren}}, \binits{H.P.}}:
\byear{2016},
\batitle{{The Mysterious Case of the Solar Argon Abundance near Sunspots in
  Flares}}.
\bjtitle{\apj}
\bvolume{825},
\bfpage{36}.
\doiurl{10.3847/0004-637X/825/1/36}.
\adsurl{2016ApJ...825...36D}.
\end{barticle}
\endbibitem

\bibitem[\protect\citeauthoryear{{Doschek}, {Warren}, and
  {Feldman}}{2015}]{Doschek2015}
\begin{barticle}
\bauthor{\bsnm{{Doschek}}, \binits{G.A.}},
\bauthor{\bsnm{{Warren}}, \binits{H.P.}},
\bauthor{\bsnm{{Feldman}}, \binits{U.}}:
\byear{2015},
\batitle{{Anomalous Relative Ar/Ca Coronal Abundances Observed by the
  Hinode/EUV Imaging Spectrometer Near Sunspots}}.
\bjtitle{\apjl}
\bvolume{808},
\bfpage{L7}.
\doiurl{10.1088/2041-8205/808/1/L7}.
\adsurl{2015ApJ...808L...7D}.
\end{barticle}
\endbibitem

\bibitem[\protect\citeauthoryear{{Feldman}}{1992}]{Feldman1992}
\begin{barticle}
\bauthor{\bsnm{{Feldman}}, \binits{U.}}:
\byear{1992},
\batitle{{Elemental abundances in the upper solar atmosphere.}}
\bjtitle{\physscr}
\bvolume{46},
\bfpage{202}.
\doiurl{10.1088/0031-8949/46/3/002}.
\adsurl{1992PhyS...46..202F}.
\end{barticle}
\endbibitem

\bibitem[\protect\citeauthoryear{{Garcia}}{1994}]{Garcia1994}
\begin{barticle}
\bauthor{\bsnm{{Garcia}}, \binits{H.A.}}:
\byear{1994},
\batitle{{Temperature and emission measure from GOES soft X-ray measurements}}.
\bjtitle{\solphys}
\bvolume{154},
\bfpage{275}.
\doiurl{10.1007/BF00681100}.
\adsurl{1994SoPh..154..275G}.
\end{barticle}
\endbibitem

\bibitem[\protect\citeauthoryear{{Gburek} \textit{et~al.}}{2013}]{Gburek2013}
\begin{barticle}
\bauthor{\bsnm{{Gburek}}, \binits{S.}},
\bauthor{\bsnm{{Sylwester}}, \binits{J.}},
\bauthor{\bsnm{{Kowalinski}}, \binits{M.}},
\bauthor{\bsnm{{Bakala}}, \binits{J.}},
\bauthor{\bsnm{{Kordylewski}}, \binits{Z.}},
\bauthor{\bsnm{{Podgorski}}, \binits{P.}},
\bauthor{\bsnm{{Plocieniak}}, \binits{S.}},
\bauthor{\bsnm{{Siarkowski}}, \binits{M.}},
\bauthor{\bsnm{{Sylwester}}, \binits{B.}},
\bauthor{\bsnm{{Trzebinski}}, \binits{W.}},
\bauthor{\bsnm{{Kuzin}}, \binits{S.V.}},
\bauthor{\bsnm{{Pertsov}}, \binits{A.A.}},
\bauthor{\bsnm{{Kotov}}, \binits{Y.D.}},
\bauthor{\bsnm{{Farnik}}, \binits{F.}},
\bauthor{\bsnm{{Reale}}, \binits{F.}},
\bauthor{\bsnm{{Phillips}}, \binits{K.J.H.}}:
\byear{2013},
\batitle{{SphinX: The Solar Photometer in X-Rays}}.
\bjtitle{\solphys}
\bvolume{283},
\bfpage{631}.
\doiurl{10.1007/s11207-012-0201-8}.
\adsurl{2013SoPh..283..631G}.
\end{barticle}
\endbibitem

\bibitem[\protect\citeauthoryear{Golub and
  Pasachoff}{2010}]{golubpasachoff2010solar}
\begin{bbook}
\bauthor{\bsnm{Golub}, \binits{L.}},
\bauthor{\bsnm{Pasachoff}, \binits{J.M.}}:
\byear{2010},
\bbtitle{The solar corona},
\bsertitle{The Solar Corona},
\bpublisher{Cambridge University Press}, \blocation{???}
\bisbn{9780521882019}.
\burl{https://books.google.com/books?id=XaRH1ZxEWu4C}.
\end{bbook}
\endbibitem

\bibitem[\protect\citeauthoryear{{Golub} \textit{et~al.}}{2007}]{Golub2007}
\begin{barticle}
\bauthor{\bsnm{{Golub}}, \binits{L.}},
\bauthor{\bsnm{{Deluca}}, \binits{E.}},
\bauthor{\bsnm{{Austin}}, \binits{G.}},
\bauthor{\bsnm{{Bookbinder}}, \binits{J.}},
\bauthor{\bsnm{{Caldwell}}, \binits{D.}},
\bauthor{\bsnm{{Cheimets}}, \binits{P.}},
\bauthor{\bsnm{{Cirtain}}, \binits{J.}},
\bauthor{\bsnm{{Cosmo}}, \binits{M.}},
\bauthor{\bsnm{{Reid}}, \binits{P.}},
\bauthor{\bsnm{{Sette}}, \binits{A.}},
\bauthor{\bsnm{{Weber}}, \binits{M.}},
\bauthor{\bsnm{{Sakao}}, \binits{T.}},
\bauthor{\bsnm{{Kano}}, \binits{R.}},
\bauthor{\bsnm{{Shibasaki}}, \binits{K.}},
\bauthor{\bsnm{{Hara}}, \binits{H.}},
\bauthor{\bsnm{{Tsuneta}}, \binits{S.}},
\bauthor{\bsnm{{Kumagai}}, \binits{K.}},
\bauthor{\bsnm{{Tamura}}, \binits{T.}},
\bauthor{\bsnm{{Shimojo}}, \binits{M.}},
\bauthor{\bsnm{{McCracken}}, \binits{J.}},
\bauthor{\bsnm{{Carpenter}}, \binits{J.}},
\bauthor{\bsnm{{Haight}}, \binits{H.}},
\bauthor{\bsnm{{Siler}}, \binits{R.}},
\bauthor{\bsnm{{Wright}}, \binits{E.}},
\bauthor{\bsnm{{Tucker}}, \binits{J.}},
\bauthor{\bsnm{{Rutledge}}, \binits{H.}},
\bauthor{\bsnm{{Barbera}}, \binits{M.}},
\bauthor{\bsnm{{Peres}}, \binits{G.}},
\bauthor{\bsnm{{Varisco}}, \binits{S.}}:
\byear{2007},
\batitle{{The X-Ray Telescope (XRT) for the Hinode Mission}}.
\bjtitle{\solphys}
\bvolume{243},
\bfpage{63}.
\doiurl{10.1007/s11207-007-0182-1}.
\adsurl{2007SoPh..243...63G}.
\end{barticle}
\endbibitem

\bibitem[\protect\citeauthoryear{{Grefenstette}
  \textit{et~al.}}{2016}]{Grefenstette2016}
\begin{barticle}
\bauthor{\bsnm{{Grefenstette}}, \binits{B.W.}},
\bauthor{\bsnm{{Glesener}}, \binits{L.}},
\bauthor{\bsnm{{Krucker}}, \binits{S.}},
\bauthor{\bsnm{{Hudson}}, \binits{H.}},
\bauthor{\bsnm{{Hannah}}, \binits{I.G.}},
\bauthor{\bsnm{{Smith}}, \binits{D.M.}},
\bauthor{\bsnm{{Vogel}}, \binits{J.K.}},
\bauthor{\bsnm{{White}}, \binits{S.M.}},
\bauthor{\bsnm{{Madsen}}, \binits{K.K.}},
\bauthor{\bsnm{{Marsh}}, \binits{A.J.}},
\bauthor{\bsnm{{Caspi}}, \binits{A.}},
\bauthor{\bsnm{{Chen}}, \binits{B.}},
\bauthor{\bsnm{{Shih}}, \binits{A.}},
\bauthor{\bsnm{{Kuhar}}, \binits{M.}},
\bauthor{\bsnm{{Boggs}}, \binits{S.E.}},
\bauthor{\bsnm{{Christensen}}, \binits{F.E.}},
\bauthor{\bsnm{{Craig}}, \binits{W.W.}},
\bauthor{\bsnm{{Forster}}, \binits{K.}},
\bauthor{\bsnm{{Hailey}}, \binits{C.J.}},
\bauthor{\bsnm{{Harrison}}, \binits{F.A.}},
\bauthor{\bsnm{{Miyasaka}}, \binits{H.}},
\bauthor{\bsnm{{Stern}}, \binits{D.}},
\bauthor{\bsnm{{Zhang}}, \binits{W.W.}}:
\byear{2016},
\batitle{{The First Focused Hard X-ray Images of the Sun with NuSTAR}}.
\bjtitle{\apj}
\bvolume{826},
\bfpage{20}.
\doiurl{10.3847/0004-637X/826/1/20}.
\adsurl{2016ApJ...826...20G}.
\end{barticle}
\endbibitem

\bibitem[\protect\citeauthoryear{{Hannah} \textit{et~al.}}{2016}]{Hannah2016}
\begin{barticle}
\bauthor{\bsnm{{Hannah}}, \binits{I.G.}},
\bauthor{\bsnm{{Grefenstette}}, \binits{B.W.}},
\bauthor{\bsnm{{Smith}}, \binits{D.M.}},
\bauthor{\bsnm{{Glesener}}, \binits{L.}},
\bauthor{\bsnm{{Krucker}}, \binits{S.}},
\bauthor{\bsnm{{Hudson}}, \binits{H.S.}},
\bauthor{\bsnm{{Madsen}}, \binits{K.K.}},
\bauthor{\bsnm{{Marsh}}, \binits{A.}},
\bauthor{\bsnm{{White}}, \binits{S.M.}},
\bauthor{\bsnm{{Caspi}}, \binits{A.}},
\bauthor{\bsnm{{Shih}}, \binits{A.Y.}},
\bauthor{\bsnm{{Harrison}}, \binits{F.A.}},
\bauthor{\bsnm{{Stern}}, \binits{D.}},
\bauthor{\bsnm{{Boggs}}, \binits{S.E.}},
\bauthor{\bsnm{{Christensen}}, \binits{F.E.}},
\bauthor{\bsnm{{Craig}}, \binits{W.W.}},
\bauthor{\bsnm{{Hailey}}, \binits{C.J.}},
\bauthor{\bsnm{{Zhang}}, \binits{W.W.}}:
\byear{2016},
\batitle{{The First X-Ray Imaging Spectroscopy of Quiescent Solar Active
  Regions with NuSTAR}}.
\bjtitle{\apjl}
\bvolume{820},
\bfpage{L14}.
\doiurl{10.3847/2041-8205/820/1/L14}.
\adsurl{2016ApJ...820L..14H}.
\end{barticle}
\endbibitem

\bibitem[\protect\citeauthoryear{{Harrison}
  \textit{et~al.}}{2013}]{Harrison2013}
\begin{barticle}
\bauthor{\bsnm{{Harrison}}, \binits{F.A.}},
\bauthor{\bsnm{{Craig}}, \binits{W.W.}},
\bauthor{\bsnm{{Christensen}}, \binits{F.E.}},
\bauthor{\bsnm{{Hailey}}, \binits{C.J.}},
\bauthor{\bsnm{{Zhang}}, \binits{W.W.}},
\bauthor{\bsnm{{Boggs}}, \binits{S.E.}},
\bauthor{\bsnm{{Stern}}, \binits{D.}},
\bauthor{\bsnm{{Cook}}, \binits{W.R.}},
\bauthor{\bsnm{{Forster}}, \binits{K.}},
\bauthor{\bsnm{{Giommi}}, \binits{P.}},
\bauthor{\bsnm{{Grefenstette}}, \binits{B.W.}},
\bauthor{\bsnm{{Kim}}, \binits{Y.}},
\bauthor{\bsnm{{Kitaguchi}}, \binits{T.}},
\bauthor{\bsnm{{Koglin}}, \binits{J.E.}},
\bauthor{\bsnm{{Madsen}}, \binits{K.K.}},
\bauthor{\bsnm{{Mao}}, \binits{P.H.}},
\bauthor{\bsnm{{Miyasaka}}, \binits{H.}},
\bauthor{\bsnm{{Mori}}, \binits{K.}},
\bauthor{\bsnm{{Perri}}, \binits{M.}},
\bauthor{\bsnm{{Pivovaroff}}, \binits{M.J.}},
\bauthor{\bsnm{{Puccetti}}, \binits{S.}},
\bauthor{\bsnm{{Rana}}, \binits{V.R.}},
\bauthor{\bsnm{{Westergaard}}, \binits{N.J.}},
\bauthor{\bsnm{{Willis}}, \binits{J.}},
\bauthor{\bsnm{{Zoglauer}}, \binits{A.}},
\bauthor{\bsnm{{An}}, \binits{H.}},
\bauthor{\bsnm{{Bachetti}}, \binits{M.}},
\bauthor{\bsnm{{Barri{\`e}re}}, \binits{N.M.}},
\bauthor{\bsnm{{Bellm}}, \binits{E.C.}},
\bauthor{\bsnm{{Bhalerao}}, \binits{V.}},
\bauthor{\bsnm{{Brejnholt}}, \binits{N.F.}},
\bauthor{\bsnm{{Fuerst}}, \binits{F.}},
\bauthor{\bsnm{{Liebe}}, \binits{C.C.}},
\bauthor{\bsnm{{Markwardt}}, \binits{C.B.}},
\bauthor{\bsnm{{Nynka}}, \binits{M.}},
\bauthor{\bsnm{{Vogel}}, \binits{J.K.}},
\bauthor{\bsnm{{Walton}}, \binits{D.J.}},
\bauthor{\bsnm{{Wik}}, \binits{D.R.}},
\bauthor{\bsnm{{Alexander}}, \binits{D.M.}},
\bauthor{\bsnm{{Cominsky}}, \binits{L.R.}},
\bauthor{\bsnm{{Hornschemeier}}, \binits{A.E.}},
\bauthor{\bsnm{{Hornstrup}}, \binits{A.}},
\bauthor{\bsnm{{Kaspi}}, \binits{V.M.}},
\bauthor{\bsnm{{Madejski}}, \binits{G.M.}},
\bauthor{\bsnm{{Matt}}, \binits{G.}},
\bauthor{\bsnm{{Molendi}}, \binits{S.}},
\bauthor{\bsnm{{Smith}}, \binits{D.M.}},
\bauthor{\bsnm{{Tomsick}}, \binits{J.A.}},
\bauthor{\bsnm{{Ajello}}, \binits{M.}},
\bauthor{\bsnm{{Ballantyne}}, \binits{D.R.}},
\bauthor{\bsnm{{Balokovi{\'c}}}, \binits{M.}},
\bauthor{\bsnm{{Barret}}, \binits{D.}},
\bauthor{\bsnm{{Bauer}}, \binits{F.E.}},
\bauthor{\bsnm{{Blandford}}, \binits{R.D.}},
\bauthor{\bsnm{{Brandt}}, \binits{W.N.}},
\bauthor{\bsnm{{Brenneman}}, \binits{L.W.}},
\bauthor{\bsnm{{Chiang}}, \binits{J.}},
\bauthor{\bsnm{{Chakrabarty}}, \binits{D.}},
\bauthor{\bsnm{{Chenevez}}, \binits{J.}},
\bauthor{\bsnm{{Comastri}}, \binits{A.}},
\bauthor{\bsnm{{Dufour}}, \binits{F.}},
\bauthor{\bsnm{{Elvis}}, \binits{M.}},
\bauthor{\bsnm{{Fabian}}, \binits{A.C.}},
\bauthor{\bsnm{{Farrah}}, \binits{D.}},
\bauthor{\bsnm{{Fryer}}, \binits{C.L.}},
\bauthor{\bsnm{{Gotthelf}}, \binits{E.V.}},
\bauthor{\bsnm{{Grindlay}}, \binits{J.E.}},
\bauthor{\bsnm{{Helfand}}, \binits{D.J.}},
\bauthor{\bsnm{{Krivonos}}, \binits{R.}},
\bauthor{\bsnm{{Meier}}, \binits{D.L.}},
\bauthor{\bsnm{{Miller}}, \binits{J.M.}},
\bauthor{\bsnm{{Natalucci}}, \binits{L.}},
\bauthor{\bsnm{{Ogle}}, \binits{P.}},
\bauthor{\bsnm{{Ofek}}, \binits{E.O.}},
\bauthor{\bsnm{{Ptak}}, \binits{A.}},
\bauthor{\bsnm{{Reynolds}}, \binits{S.P.}},
\bauthor{\bsnm{{Rigby}}, \binits{J.R.}},
\bauthor{\bsnm{{Tagliaferri}}, \binits{G.}},
\bauthor{\bsnm{{Thorsett}}, \binits{S.E.}},
\bauthor{\bsnm{{Treister}}, \binits{E.}},
\bauthor{\bsnm{{Urry}}, \binits{C.M.}}:
\byear{2013},
\batitle{{The Nuclear Spectroscopic Telescope Array (NuSTAR) High-energy X-Ray
  Mission}}.
\bjtitle{\apj}
\bvolume{770},
\bfpage{103}.
\doiurl{10.1088/0004-637X/770/2/103}.
\adsurl{2013ApJ...770..103H}.
\end{barticle}
\endbibitem

\bibitem[\protect\citeauthoryear{{Hill} \textit{et~al.}}{2005}]{Hill2005}
\begin{barticle}
\bauthor{\bsnm{{Hill}}, \binits{S.M.}},
\bauthor{\bsnm{{Pizzo}}, \binits{V.J.}},
\bauthor{\bsnm{{Balch}}, \binits{C.C.}},
\bauthor{\bsnm{{Biesecker}}, \binits{D.A.}},
\bauthor{\bsnm{{Bornmann}}, \binits{P.}},
\bauthor{\bsnm{{Hildner}}, \binits{E.}},
\bauthor{\bsnm{{Lewis}}, \binits{L.D.}},
\bauthor{\bsnm{{Grubb}}, \binits{R.N.}},
\bauthor{\bsnm{{Husler}}, \binits{M.P.}},
\bauthor{\bsnm{{Prendergast}}, \binits{K.}},
\bauthor{\bsnm{{Vickroy}}, \binits{J.}},
\bauthor{\bsnm{{Greer}}, \binits{S.}},
\bauthor{\bsnm{{Defoor}}, \binits{T.}},
\bauthor{\bsnm{{Wilkinson}}, \binits{D.C.}},
\bauthor{\bsnm{{Hooker}}, \binits{R.}},
\bauthor{\bsnm{{Mulligan}}, \binits{P.}},
\bauthor{\bsnm{{Chipman}}, \binits{E.}},
\bauthor{\bsnm{{Bysal}}, \binits{H.}},
\bauthor{\bsnm{{Douglas}}, \binits{J.P.}},
\bauthor{\bsnm{{Reynolds}}, \binits{R.}},
\bauthor{\bsnm{{Davis}}, \binits{J.M.}},
\bauthor{\bsnm{{Wallace}}, \binits{K.S.}},
\bauthor{\bsnm{{Russell}}, \binits{K.}},
\bauthor{\bsnm{{Freestone}}, \binits{K.}},
\bauthor{\bsnm{{Bagdigian}}, \binits{D.}},
\bauthor{\bsnm{{Page}}, \binits{T.}},
\bauthor{\bsnm{{Kerns}}, \binits{S.}},
\bauthor{\bsnm{{Hoffman}}, \binits{R.}},
\bauthor{\bsnm{{Cauffman}}, \binits{S.A.}},
\bauthor{\bsnm{{Davis}}, \binits{M.A.}},
\bauthor{\bsnm{{Studer}}, \binits{R.}},
\bauthor{\bsnm{{Berthiaume}}, \binits{F.E.}},
\bauthor{\bsnm{{Saha}}, \binits{T.T.}},
\bauthor{\bsnm{{Berthiume}}, \binits{G.D.}},
\bauthor{\bsnm{{Farthing}}, \binits{H.}},
\bauthor{\bsnm{{Zimmermann}}, \binits{F.}}:
\byear{2005},
\batitle{{The NOAA Goes-12 Solar X-Ray Imager (SXI) 1. Instrument, Operations,
  and Data}}.
\bjtitle{\solphys}
\bvolume{226},
\bfpage{255}.
\doiurl{10.1007/s11207-005-7416-x}.
\adsurl{2005SoPh..226..255H}.
\end{barticle}
\endbibitem

\bibitem[\protect\citeauthoryear{{Ishikawa}
  \textit{et~al.}}{2014}]{Ishikawa2014}
\begin{barticle}
\bauthor{\bsnm{{Ishikawa}}, \binits{S.-n.}},
\bauthor{\bsnm{{Glesener}}, \binits{L.}},
\bauthor{\bsnm{{Christe}}, \binits{S.}},
\bauthor{\bsnm{{Ishibashi}}, \binits{K.}},
\bauthor{\bsnm{{Brooks}}, \binits{D.H.}},
\bauthor{\bsnm{{Williams}}, \binits{D.R.}},
\bauthor{\bsnm{{Shimojo}}, \binits{M.}},
\bauthor{\bsnm{{Sako}}, \binits{N.}},
\bauthor{\bsnm{{Krucker}}, \binits{S.}}:
\byear{2014},
\batitle{{Constraining hot plasma in a non-flaring solar active region with
  FOXSI hard X-ray observations}}.
\bjtitle{\pasj}
\bvolume{66},
\bfpage{S15}.
\doiurl{10.1093/pasj/psu090}.
\adsurl{2014PASJ...66S..15I}.
\end{barticle}
\endbibitem

\bibitem[\protect\citeauthoryear{{Ishikawa}
  \textit{et~al.}}{2017}]{Ishikawa2017}
\begin{barticle}
\bauthor{\bsnm{{Ishikawa}}, \binits{S.-n.}},
\bauthor{\bsnm{{Glesener}}, \binits{L.}},
\bauthor{\bsnm{{Krucker}}, \binits{S.}},
\bauthor{\bsnm{{Christe}}, \binits{S.}},
\bauthor{\bsnm{{Buitrago-Casas}}, \binits{J.C.}},
\bauthor{\bsnm{{Narukage}}, \binits{N.}},
\bauthor{\bsnm{{Vievering}}, \binits{J.}}:
\byear{2017},
\batitle{{Detection of nanoflare-heated plasma in the solar corona by the
  FOXSI-2 sounding rocket}}.
\bjtitle{Nature Astronomy}
\bvolume{1},
\bfpage{771}.
\doiurl{10.1038/s41550-017-0269-z}.
\adsurl{2017NatAs...1..771I}.
\end{barticle}
\endbibitem

\bibitem[\protect\citeauthoryear{Knoll}{2010}]{knoll2010}
\begin{bbook}
\bauthor{\bsnm{Knoll}, \binits{G.F.}}:
\byear{2010},
\bbtitle{Radiation detection and measurement},
\bpublisher{John Wiley \& Sons}, \blocation{???}
\bisbn{9780470131480}.
\burl{https://books.google.com/books?id=4vTJ7UDel5IC}.
\end{bbook}
\endbibitem

\bibitem[\protect\citeauthoryear{{Kobayashi}
  \textit{et~al.}}{2011}]{Kobayashi2011}
\begin{bchapter}
\bauthor{\bsnm{{Kobayashi}}, \binits{K.}},
\bauthor{\bsnm{{Cirtain}}, \binits{J.}},
\bauthor{\bsnm{{Golub}}, \binits{L.}},
\bauthor{\bsnm{{Winebarger}}, \binits{A.}},
\bauthor{\bsnm{{Hertz}}, \binits{E.}},
\bauthor{\bsnm{{Cheimets}}, \binits{P.}},
\bauthor{\bsnm{{Caldwell}}, \binits{D.}},
\bauthor{\bsnm{{Korreck}}, \binits{K.}},
\bauthor{\bsnm{{Robinson}}, \binits{B.}},
\bauthor{\bsnm{{Reardon}}, \binits{P.}},
\bauthor{\bsnm{{Kester}}, \binits{T.}},
\bauthor{\bsnm{{Griffith}}, \binits{C.}},
\bauthor{\bsnm{{Young}}, \binits{M.}}:
\byear{2011},
\bctitle{{The Marshall Grazing Incidence X-ray Spectrograph (MaGIXS)}}.
In: \bbtitle{Society of Photo-Optical Instrumentation Engineers (SPIE)
  Conference Series},
\bsertitle{\procspie}
\bseriesno{8147},
\bfpage{81471M}.
\doiurl{10.1117/12.894071}.
\adsurl{2011SPIE.8147E..1MK}.
\end{bchapter}
\endbibitem

\bibitem[\protect\citeauthoryear{{Kosugi} \textit{et~al.}}{2007}]{Kosugi2007}
\begin{barticle}
\bauthor{\bsnm{{Kosugi}}, \binits{T.}},
\bauthor{\bsnm{{Matsuzaki}}, \binits{K.}},
\bauthor{\bsnm{{Sakao}}, \binits{T.}},
\bauthor{\bsnm{{Shimizu}}, \binits{T.}},
\bauthor{\bsnm{{Sone}}, \binits{Y.}},
\bauthor{\bsnm{{Tachikawa}}, \binits{S.}},
\bauthor{\bsnm{{Hashimoto}}, \binits{T.}},
\bauthor{\bsnm{{Minesugi}}, \binits{K.}},
\bauthor{\bsnm{{Ohnishi}}, \binits{A.}},
\bauthor{\bsnm{{Yamada}}, \binits{T.}},
\bauthor{\bsnm{{Tsuneta}}, \binits{S.}},
\bauthor{\bsnm{{Hara}}, \binits{H.}},
\bauthor{\bsnm{{Ichimoto}}, \binits{K.}},
\bauthor{\bsnm{{Suematsu}}, \binits{Y.}},
\bauthor{\bsnm{{Shimojo}}, \binits{M.}},
\bauthor{\bsnm{{Watanabe}}, \binits{T.}},
\bauthor{\bsnm{{Shimada}}, \binits{S.}},
\bauthor{\bsnm{{Davis}}, \binits{J.M.}},
\bauthor{\bsnm{{Hill}}, \binits{L.D.}},
\bauthor{\bsnm{{Owens}}, \binits{J.K.}},
\bauthor{\bsnm{{Title}}, \binits{A.M.}},
\bauthor{\bsnm{{Culhane}}, \binits{J.L.}},
\bauthor{\bsnm{{Harra}}, \binits{L.K.}},
\bauthor{\bsnm{{Doschek}}, \binits{G.A.}},
\bauthor{\bsnm{{Golub}}, \binits{L.}}:
\byear{2007},
\batitle{{The Hinode (Solar-B) Mission: An Overview}}.
\bjtitle{\solphys}
\bvolume{243},
\bfpage{3}.
\doiurl{10.1007/s11207-007-9014-6}.
\adsurl{2007SoPh..243....3K}.
\end{barticle}
\endbibitem

\bibitem[\protect\citeauthoryear{{Laming}}{2015}]{Laming2015}
\begin{barticle}
\bauthor{\bsnm{{Laming}}, \binits{J.M.}}:
\byear{2015},
\batitle{{The FIP and Inverse FIP Effects in Solar and Stellar Coronae}}.
\bjtitle{Living Reviews in Solar Physics}
\bvolume{12},
\bfpage{2}.
\doiurl{10.1007/lrsp-2015-2}.
\adsurl{2015LRSP...12....2L}.
\end{barticle}
\endbibitem

\bibitem[\protect\citeauthoryear{{Landi} and {Klimchuk}}{2010}]{Landi2010}
\begin{barticle}
\bauthor{\bsnm{{Landi}}, \binits{E.}},
\bauthor{\bsnm{{Klimchuk}}, \binits{J.A.}}:
\byear{2010},
\batitle{{On the Isothermality of Solar Plasmas}}.
\bjtitle{\apj}
\bvolume{723},
\bfpage{320}.
\doiurl{10.1088/0004-637X/723/1/320}.
\adsurl{2010ApJ...723..320L}.
\end{barticle}
\endbibitem

\bibitem[\protect\citeauthoryear{{Lin} \textit{et~al.}}{2002}]{Lin2002}
\begin{barticle}
\bauthor{\bsnm{{Lin}}, \binits{R.P.}},
\bauthor{\bsnm{{Dennis}}, \binits{B.R.}},
\bauthor{\bsnm{{Hurford}}, \binits{G.J.}},
\bauthor{\bsnm{{Smith}}, \binits{D.M.}},
\bauthor{\bsnm{{Zehnder}}, \binits{A.}},
\bauthor{\bsnm{{Harvey}}, \binits{P.R.}},
\bauthor{\bsnm{{Curtis}}, \binits{D.W.}},
\bauthor{\bsnm{{Pankow}}, \binits{D.}},
\bauthor{\bsnm{{Turin}}, \binits{P.}},
\bauthor{\bsnm{{Bester}}, \binits{M.}},
\bauthor{\bsnm{{Csillaghy}}, \binits{A.}},
\bauthor{\bsnm{{Lewis}}, \binits{M.}},
\bauthor{\bsnm{{Madden}}, \binits{N.}},
\bauthor{\bsnm{{van Beek}}, \binits{H.F.}},
\bauthor{\bsnm{{Appleby}}, \binits{M.}},
\bauthor{\bsnm{{Raudorf}}, \binits{T.}},
\bauthor{\bsnm{{McTiernan}}, \binits{J.}},
\bauthor{\bsnm{{Ramaty}}, \binits{R.}},
\bauthor{\bsnm{{Schmahl}}, \binits{E.}},
\bauthor{\bsnm{{Schwartz}}, \binits{R.}},
\bauthor{\bsnm{{Krucker}}, \binits{S.}},
\bauthor{\bsnm{{Abiad}}, \binits{R.}},
\bauthor{\bsnm{{Quinn}}, \binits{T.}},
\bauthor{\bsnm{{Berg}}, \binits{P.}},
\bauthor{\bsnm{{Hashii}}, \binits{M.}},
\bauthor{\bsnm{{Sterling}}, \binits{R.}},
\bauthor{\bsnm{{Jackson}}, \binits{R.}},
\bauthor{\bsnm{{Pratt}}, \binits{R.}},
\bauthor{\bsnm{{Campbell}}, \binits{R.D.}},
\bauthor{\bsnm{{Malone}}, \binits{D.}},
\bauthor{\bsnm{{Landis}}, \binits{D.}},
\bauthor{\bsnm{{Barrington-Leigh}}, \binits{C.P.}},
\bauthor{\bsnm{{Slassi-Sennou}}, \binits{S.}},
\bauthor{\bsnm{{Cork}}, \binits{C.}},
\bauthor{\bsnm{{Clark}}, \binits{D.}},
\bauthor{\bsnm{{Amato}}, \binits{D.}},
\bauthor{\bsnm{{Orwig}}, \binits{L.}},
\bauthor{\bsnm{{Boyle}}, \binits{R.}},
\bauthor{\bsnm{{Banks}}, \binits{I.S.}},
\bauthor{\bsnm{{Shirey}}, \binits{K.}},
\bauthor{\bsnm{{Tolbert}}, \binits{A.K.}},
\bauthor{\bsnm{{Zarro}}, \binits{D.}},
\bauthor{\bsnm{{Snow}}, \binits{F.}},
\bauthor{\bsnm{{Thomsen}}, \binits{K.}},
\bauthor{\bsnm{{Henneck}}, \binits{R.}},
\bauthor{\bsnm{{McHedlishvili}}, \binits{A.}},
\bauthor{\bsnm{{Ming}}, \binits{P.}},
\bauthor{\bsnm{{Fivian}}, \binits{M.}},
\bauthor{\bsnm{{Jordan}}, \binits{J.}},
\bauthor{\bsnm{{Wanner}}, \binits{R.}},
\bauthor{\bsnm{{Crubb}}, \binits{J.}},
\bauthor{\bsnm{{Preble}}, \binits{J.}},
\bauthor{\bsnm{{Matranga}}, \binits{M.}},
\bauthor{\bsnm{{Benz}}, \binits{A.}},
\bauthor{\bsnm{{Hudson}}, \binits{H.}},
\bauthor{\bsnm{{Canfield}}, \binits{R.C.}},
\bauthor{\bsnm{{Holman}}, \binits{G.D.}},
\bauthor{\bsnm{{Crannell}}, \binits{C.}},
\bauthor{\bsnm{{Kosugi}}, \binits{T.}},
\bauthor{\bsnm{{Emslie}}, \binits{A.G.}},
\bauthor{\bsnm{{Vilmer}}, \binits{N.}},
\bauthor{\bsnm{{Brown}}, \binits{J.C.}},
\bauthor{\bsnm{{Johns-Krull}}, \binits{C.}},
\bauthor{\bsnm{{Aschwanden}}, \binits{M.}},
\bauthor{\bsnm{{Metcalf}}, \binits{T.}},
\bauthor{\bsnm{{Conway}}, \binits{A.}}:
\byear{2002},
\batitle{{The Reuven Ramaty High-Energy Solar Spectroscopic Imager (RHESSI)}}.
\bjtitle{\solphys}
\bvolume{210},
\bfpage{3}.
\doiurl{10.1023/A:1022428818870}.
\adsurl{2002SoPh..210....3L}.
\end{barticle}
\endbibitem

\bibitem[\protect\citeauthoryear{{Mason} \textit{et~al.}}{2016}]{Mason2016}
\begin{barticle}
\bauthor{\bsnm{{Mason}}, \binits{J.P.}},
\bauthor{\bsnm{{Woods}}, \binits{T.N.}},
\bauthor{\bsnm{{Caspi}}, \binits{A.}},
\bauthor{\bsnm{{Chamberlin}}, \binits{P.C.}},
\bauthor{\bsnm{{Moore}}, \binits{C.}},
\bauthor{\bsnm{{Jones}}, \binits{A.}},
\bauthor{\bsnm{{Kohnert}}, \binits{R.}},
\bauthor{\bsnm{{Li}}, \binits{X.}},
\bauthor{\bsnm{{Palo}}, \binits{S.}},
\bauthor{\bsnm{{Solomon}}, \binits{S.C.}}:
\byear{2016},
\batitle{{Miniature X-Ray Solar Spectrometer: A Science-Oriented, University 3U
  CubeSat}}.
\bjtitle{Journal of Spacecraft and Rockets}
\bvolume{53},
\bfpage{328}.
\doiurl{10.2514/1.A33351}.
\adsurl{2016JSpRo..53..328M}.
\end{barticle}
\endbibitem

\bibitem[\protect\citeauthoryear{{Miceli} \textit{et~al.}}{2012}]{Miceli2012}
\begin{barticle}
\bauthor{\bsnm{{Miceli}}, \binits{M.}},
\bauthor{\bsnm{{Reale}}, \binits{F.}},
\bauthor{\bsnm{{Gburek}}, \binits{S.}},
\bauthor{\bsnm{{Terzo}}, \binits{S.}},
\bauthor{\bsnm{{Barbera}}, \binits{M.}},
\bauthor{\bsnm{{Collura}}, \binits{A.}},
\bauthor{\bsnm{{Sylwester}}, \binits{J.}},
\bauthor{\bsnm{{Kowalinski}}, \binits{M.}},
\bauthor{\bsnm{{Podgorski}}, \binits{P.}},
\bauthor{\bsnm{{Gryciuk}}, \binits{M.}}:
\byear{2012},
\batitle{{X-ray emitting hot plasma in solar active regions observed by the
  SphinX spectrometer}}.
\bjtitle{\aap}
\bvolume{544},
\bfpage{A139}.
\doiurl{10.1051/0004-6361/201219670}.
\adsurl{2012A\%26A...544A.139M}.
\end{barticle}
\endbibitem

\bibitem[\protect\citeauthoryear{{Moore} \textit{et~al.}}{2016}]{Moore2016}
\begin{bchapter}
\bauthor{\bsnm{{Moore}}, \binits{C.S.}},
\bauthor{\bsnm{{Woods}}, \binits{T.N.}},
\bauthor{\bsnm{{Caspi}}, \binits{A.}},
\bauthor{\bsnm{{Mason}}, \binits{J.P.}}:
\byear{2016},
\bctitle{{The Miniature X-ray Solar Spectrometer (MinXSS) CubeSats:
  spectrometer characterization techniques, spectrometer capabilities, and
  solar science objectives}}.
In: \bbtitle{Space Telescopes and Instrumentation 2016: Ultraviolet to Gamma
  Ray},
\bsertitle{\procspie}
\bseriesno{9905},
\bfpage{990509}.
\doiurl{10.1117/12.2231945}.
\adsurl{2016SPIE.9905E..09M}.
\end{bchapter}
\endbibitem

\bibitem[\protect\citeauthoryear{{Pesnell}, {Thompson}, and
  {Chamberlin}}{2012}]{Pesnell2012}
\begin{barticle}
\bauthor{\bsnm{{Pesnell}}, \binits{W.D.}},
\bauthor{\bsnm{{Thompson}}, \binits{B.J.}},
\bauthor{\bsnm{{Chamberlin}}, \binits{P.C.}}:
\byear{2012},
\batitle{{The Solar Dynamics Observatory (SDO)}}.
\bjtitle{\solphys}
\bvolume{275},
\bfpage{3}.
\doiurl{10.1007/s11207-011-9841-3}.
\adsurl{2012SoPh..275....3P}.
\end{barticle}
\endbibitem

\bibitem[\protect\citeauthoryear{{Rayet}}{1868}]{Rayet1868}
\begin{botherref}
\oauthor{\bsnm{{Rayet}}, \binits{G.A.P.}}:
1868,
{ }.
\textit{Compt. Rend. 67, 757}
\textbf{67}.
\end{botherref}
\endbibitem

\bibitem[\protect\citeauthoryear{{Reale} \textit{et~al.}}{2009}]{Reale2009}
\begin{barticle}
\bauthor{\bsnm{{Reale}}, \binits{F.}},
\bauthor{\bsnm{{Testa}}, \binits{P.}},
\bauthor{\bsnm{{Klimchuk}}, \binits{J.A.}},
\bauthor{\bsnm{{Parenti}}, \binits{S.}}:
\byear{2009},
\batitle{{Evidence of Widespread Hot Plasma in a Nonflaring Coronal Active
  Region from Hinode/X-Ray Telescope}}.
\bjtitle{\apj}
\bvolume{698},
\bfpage{756}.
\doiurl{10.1088/0004-637X/698/1/756}.
\adsurl{2009ApJ...698..756R}.
\end{barticle}
\endbibitem

\bibitem[\protect\citeauthoryear{Redus, Huber, and Sperry}{2008}]{Redus2008}
\begin{bchapter}
\bauthor{\bsnm{Redus}, \binits{R.H.}},
\bauthor{\bsnm{Huber}, \binits{A.C.}},
\bauthor{\bsnm{Sperry}, \binits{D.J.}}:
\byear{2008},
\bctitle{Dead time correction in the dp5 digital pulse processor}.
In: \bbtitle{2008 IEEE Nuclear Science Symposium Conference Record},
\bfpage{3416}.
\doiurl{10.1109/NSSMIC.2008.4775075}.
\end{bchapter}
\endbibitem

\bibitem[\protect\citeauthoryear{{Schlemm} \textit{et~al.}}{2007}]{Schlemm2007}
\begin{barticle}
\bauthor{\bsnm{{Schlemm}}, \binits{C.E.}},
\bauthor{\bsnm{{Starr}}, \binits{R.D.}},
\bauthor{\bsnm{{Ho}}, \binits{G.C.}},
\bauthor{\bsnm{{Bechtold}}, \binits{K.E.}},
\bauthor{\bsnm{{Hamilton}}, \binits{S.A.}},
\bauthor{\bsnm{{Boldt}}, \binits{J.D.}},
\bauthor{\bsnm{{Boynton}}, \binits{W.V.}},
\bauthor{\bsnm{{Bradley}}, \binits{W.}},
\bauthor{\bsnm{{Fraeman}}, \binits{M.E.}},
\bauthor{\bsnm{{Gold}}, \binits{R.E.}},
\bauthor{\bsnm{{Goldsten}}, \binits{J.O.}},
\bauthor{\bsnm{{Hayes}}, \binits{J.R.}},
\bauthor{\bsnm{{Jaskulek}}, \binits{S.E.}},
\bauthor{\bsnm{{Rossano}}, \binits{E.}},
\bauthor{\bsnm{{Rumpf}}, \binits{R.A.}},
\bauthor{\bsnm{{Schaefer}}, \binits{E.D.}},
\bauthor{\bsnm{{Strohbehn}}, \binits{K.}},
\bauthor{\bsnm{{Shelton}}, \binits{R.G.}},
\bauthor{\bsnm{{Thompson}}, \binits{R.E.}},
\bauthor{\bsnm{{Trombka}}, \binits{J.I.}},
\bauthor{\bsnm{{Williams}}, \binits{B.D.}}:
\byear{2007},
\batitle{{The X-Ray Spectrometer on the MESSENGER Spacecraft}}.
\bjtitle{\ssr}
\bvolume{131},
\bfpage{393}.
\doiurl{10.1007/s11214-007-9248-5}.
\adsurl{2007SSRv..131..393S}.
\end{barticle}
\endbibitem

\bibitem[\protect\citeauthoryear{{Schmelz}
  \textit{et~al.}}{2009a}]{Schmelz2009a}
\begin{barticle}
\bauthor{\bsnm{{Schmelz}}, \binits{J.T.}},
\bauthor{\bsnm{{Saar}}, \binits{S.H.}},
\bauthor{\bsnm{{DeLuca}}, \binits{E.E.}},
\bauthor{\bsnm{{Golub}}, \binits{L.}},
\bauthor{\bsnm{{Kashyap}}, \binits{V.L.}},
\bauthor{\bsnm{{Weber}}, \binits{M.A.}},
\bauthor{\bsnm{{Klimchuk}}, \binits{J.A.}}:
\byear{2009}a,
\batitle{{Hinode X-Ray Telescope Detection of Hot Emission from Quiescent
  Active Regions: A Nanoflare Signature?}}
\bjtitle{\apjl}
\bvolume{693},
\bfpage{L131}.
\doiurl{10.1088/0004-637X/693/2/L131}.
\adsurl{2009ApJ...693L.131S}.
\end{barticle}
\endbibitem

\bibitem[\protect\citeauthoryear{{Schmelz}
  \textit{et~al.}}{2009b}]{Schmelz2009b}
\begin{barticle}
\bauthor{\bsnm{{Schmelz}}, \binits{J.T.}},
\bauthor{\bsnm{{Kashyap}}, \binits{V.L.}},
\bauthor{\bsnm{{Saar}}, \binits{S.H.}},
\bauthor{\bsnm{{Dennis}}, \binits{B.R.}},
\bauthor{\bsnm{{Grigis}}, \binits{P.C.}},
\bauthor{\bsnm{{Lin}}, \binits{L.}},
\bauthor{\bsnm{{De Luca}}, \binits{E.E.}},
\bauthor{\bsnm{{Holman}}, \binits{G.D.}},
\bauthor{\bsnm{{Golub}}, \binits{L.}},
\bauthor{\bsnm{{Weber}}, \binits{M.A.}}:
\byear{2009}b,
\batitle{{Some Like It Hot: Coronal Heating Observations from Hinode X-ray
  Telescope and RHESSI}}.
\bjtitle{\apj}
\bvolume{704},
\bfpage{863}.
\doiurl{10.1088/0004-637X/704/1/863}.
\adsurl{2009ApJ...704..863S}.
\end{barticle}
\endbibitem

\bibitem[\protect\citeauthoryear{{Schmelz} \textit{et~al.}}{2012}]{Schmelz2012}
\begin{barticle}
\bauthor{\bsnm{{Schmelz}}, \binits{J.T.}},
\bauthor{\bsnm{{Reames}}, \binits{D.V.}},
\bauthor{\bsnm{{von Steiger}}, \binits{R.}},
\bauthor{\bsnm{{Basu}}, \binits{S.}}:
\byear{2012},
\batitle{{Composition of the Solar Corona, Solar Wind, and Solar Energetic
  Particles}}.
\bjtitle{\apj}
\bvolume{755},
\bfpage{33}.
\doiurl{10.1088/0004-637X/755/1/33}.
\adsurl{2012ApJ...755...33S}.
\end{barticle}
\endbibitem

\bibitem[\protect\citeauthoryear{{Schmelz} \textit{et~al.}}{2015}]{Schmelz2015}
\begin{barticle}
\bauthor{\bsnm{{Schmelz}}, \binits{J.T.}},
\bauthor{\bsnm{{Asgari-Targhi}}, \binits{M.}},
\bauthor{\bsnm{{Christian}}, \binits{G.M.}},
\bauthor{\bsnm{{Dhaliwal}}, \binits{R.S.}},
\bauthor{\bsnm{{Pathak}}, \binits{S.}}:
\byear{2015},
\batitle{{Hot Plasma from Solar Active Region Cores: a Test of AC and DC
  Coronal Heating Models?}}
\bjtitle{\apj}
\bvolume{806},
\bfpage{232}.
\doiurl{10.1088/0004-637X/806/2/232}.
\adsurl{2015ApJ...806..232S}.
\end{barticle}
\endbibitem

\bibitem[\protect\citeauthoryear{{Seechi}}{1875}]{Seechi1875}
\begin{botherref}
\oauthor{\bsnm{{Seechi}}, \binits{A.}}:
1875,
{ }.
\textit{Le Soleil, 2nd edn., Gauthier-Villars, Paris}.
\end{botherref}
\endbibitem

\bibitem[\protect\citeauthoryear{{Tsuneta} \textit{et~al.}}{2008}]{Tsuneta2008}
\begin{barticle}
\bauthor{\bsnm{{Tsuneta}}, \binits{S.}},
\bauthor{\bsnm{{Ichimoto}}, \binits{K.}},
\bauthor{\bsnm{{Katsukawa}}, \binits{Y.}},
\bauthor{\bsnm{{Nagata}}, \binits{S.}},
\bauthor{\bsnm{{Otsubo}}, \binits{M.}},
\bauthor{\bsnm{{Shimizu}}, \binits{T.}},
\bauthor{\bsnm{{Suematsu}}, \binits{Y.}},
\bauthor{\bsnm{{Nakagiri}}, \binits{M.}},
\bauthor{\bsnm{{Noguchi}}, \binits{M.}},
\bauthor{\bsnm{{Tarbell}}, \binits{T.}},
\bauthor{\bsnm{{Title}}, \binits{A.}},
\bauthor{\bsnm{{Shine}}, \binits{R.}},
\bauthor{\bsnm{{Rosenberg}}, \binits{W.}},
\bauthor{\bsnm{{Hoffmann}}, \binits{C.}},
\bauthor{\bsnm{{Jurcevich}}, \binits{B.}},
\bauthor{\bsnm{{Kushner}}, \binits{G.}},
\bauthor{\bsnm{{Levay}}, \binits{M.}},
\bauthor{\bsnm{{Lites}}, \binits{B.}},
\bauthor{\bsnm{{Elmore}}, \binits{D.}},
\bauthor{\bsnm{{Matsushita}}, \binits{T.}},
\bauthor{\bsnm{{Kawaguchi}}, \binits{N.}},
\bauthor{\bsnm{{Saito}}, \binits{H.}},
\bauthor{\bsnm{{Mikami}}, \binits{I.}},
\bauthor{\bsnm{{Hill}}, \binits{L.D.}},
\bauthor{\bsnm{{Owens}}, \binits{J.K.}}:
\byear{2008},
\batitle{{The Solar Optical Telescope for the Hinode Mission: An Overview}}.
\bjtitle{\solphys}
\bvolume{249},
\bfpage{167}.
\doiurl{10.1007/s11207-008-9174-z}.
\adsurl{2008SoPh..249..167T}.
\end{barticle}
\endbibitem

\bibitem[\protect\citeauthoryear{{Warren} and {Brooks}}{2009}]{Warren2009}
\begin{barticle}
\bauthor{\bsnm{{Warren}}, \binits{H.P.}},
\bauthor{\bsnm{{Brooks}}, \binits{D.H.}}:
\byear{2009},
\batitle{{The Temperature and Density Structure of the Solar Corona. I.
  Observations of the Quiet Sun with the EUV Imaging Spectrometer on Hinode}}.
\bjtitle{\apj}
\bvolume{700},
\bfpage{762}.
\doiurl{10.1088/0004-637X/700/1/762}.
\adsurl{2009ApJ...700..762W}.
\end{barticle}
\endbibitem

\bibitem[\protect\citeauthoryear{{White}, {Thomas}, and
  {Schwartz}}{2005}]{White2005}
\begin{barticle}
\bauthor{\bsnm{{White}}, \binits{S.M.}},
\bauthor{\bsnm{{Thomas}}, \binits{R.J.}},
\bauthor{\bsnm{{Schwartz}}, \binits{R.A.}}:
\byear{2005},
\batitle{{Updated Expressions for Determining Temperatures and Emission
  Measures from Goes Soft X-Ray Measurements}}.
\bjtitle{\solphys}
\bvolume{227},
\bfpage{231}.
\doiurl{10.1007/s11207-005-2445-z}.
\adsurl{2005SoPh..227..231W}.
\end{barticle}
\endbibitem

\bibitem[\protect\citeauthoryear{{Woods} \textit{et~al.}}{2012}]{Woods2012}
\begin{barticle}
\bauthor{\bsnm{{Woods}}, \binits{T.N.}},
\bauthor{\bsnm{{Eparvier}}, \binits{F.G.}},
\bauthor{\bsnm{{Hock}}, \binits{R.}},
\bauthor{\bsnm{{Jones}}, \binits{A.R.}},
\bauthor{\bsnm{{Woodraska}}, \binits{D.}},
\bauthor{\bsnm{{Judge}}, \binits{D.}},
\bauthor{\bsnm{{Didkovsky}}, \binits{L.}},
\bauthor{\bsnm{{Lean}}, \binits{J.}},
\bauthor{\bsnm{{Mariska}}, \binits{J.}},
\bauthor{\bsnm{{Warren}}, \binits{H.}},
\bauthor{\bsnm{{McMullin}}, \binits{D.}},
\bauthor{\bsnm{{Chamberlin}}, \binits{P.}},
\bauthor{\bsnm{{Berthiaume}}, \binits{G.}},
\bauthor{\bsnm{{Bailey}}, \binits{S.}},
\bauthor{\bsnm{{Fuller-Rowell}}, \binits{T.}},
\bauthor{\bsnm{{Sojka}}, \binits{J.}},
\bauthor{\bsnm{{Tobiska}}, \binits{W.K.}},
\bauthor{\bsnm{{Viereck}}, \binits{R.}}:
\byear{2012},
\batitle{{Extreme Ultraviolet Variability Experiment (EVE) on the Solar
  Dynamics Observatory (SDO): Overview of Science Objectives, Instrument
  Design, Data Products, and Model Developments}}.
\bjtitle{\solphys}
\bvolume{275},
\bfpage{115}.
\doiurl{10.1007/s11207-009-9487-6}.
\adsurl{2012SoPh..275..115W}.
\end{barticle}
\endbibitem

\bibitem[\protect\citeauthoryear{{Woods} \textit{et~al.}}{2017}]{Woods2017}
\begin{barticle}
\bauthor{\bsnm{{Woods}}, \binits{T.N.}},
\bauthor{\bsnm{{Caspi}}, \binits{A.}},
\bauthor{\bsnm{{Chamberlin}}, \binits{P.C.}},
\bauthor{\bsnm{{Jones}}, \binits{A.}},
\bauthor{\bsnm{{Kohnert}}, \binits{R.}},
\bauthor{\bsnm{{Mason}}, \binits{J.P.}},
\bauthor{\bsnm{{Moore}}, \binits{C.S.}},
\bauthor{\bsnm{{Palo}}, \binits{S.}},
\bauthor{\bsnm{{Rouleau}}, \binits{C.}},
\bauthor{\bsnm{{Solomon}}, \binits{S.C.}},
\bauthor{\bsnm{{Machol}}, \binits{J.}},
\bauthor{\bsnm{{Viereck}}, \binits{R.}}:
\byear{2017},
\batitle{{New Solar Irradiance Measurements from the Miniature X-Ray Solar
  Spectrometer CubeSat}}.
\bjtitle{\apj}
\bvolume{835},
\bfpage{122}.
\doiurl{10.3847/1538-4357/835/2/122}.
\adsurl{2017ApJ...835..122W}.
\end{barticle}
\endbibitem

\bibitem[\protect\citeauthoryear{{Young} \textit{et~al.}}{2016}]{Young2016}
\begin{barticle}
\bauthor{\bsnm{{Young}}, \binits{P.R.}},
\bauthor{\bsnm{{Dere}}, \binits{K.P.}},
\bauthor{\bsnm{{Landi}}, \binits{E.}},
\bauthor{\bsnm{{Del Zanna}}, \binits{G.}},
\bauthor{\bsnm{{Mason}}, \binits{H.E.}}:
\byear{2016},
\batitle{{The CHIANTI atomic database}}.
\bjtitle{Journal of Physics B Atomic Molecular Physics}
\bvolume{49}(\bissue{7}),
\bfpage{074009}.
\doiurl{10.1088/0953-4075/49/7/074009}.
\adsurl{2016JPhB...49g4009Y}.
\end{barticle}
\endbibitem

\end{thebibliography}

\IfFileExists{\jobname.bbl}{} {\typeout{}
\typeout{****************************************************}
\typeout{****************************************************}
\typeout{** Please run "bibtex \jobname" to obtain} \typeout{**
the bibliography and then re-run LaTeX} \typeout{** twice to fix
the references !}
\typeout{****************************************************}
\typeout{****************************************************}
\typeout{}}

\end{article} 

\end{document}